\RequirePackage[mathlines]{lineno}
\documentclass[aps,prl,twocolumn,superscriptaddress,showpacs,preprintnumbers,amsmath,amssymb,nofootinbib]{revtex4}

\usepackage{graphicx} 
\usepackage{dcolumn}  
\usepackage{ltablex}
\usepackage[english]{babel}
\usepackage{epsfig}
\usepackage{color} 
\usepackage{amsmath}    
\usepackage{subfigure}  
\usepackage{hyperref}   
\usepackage{xcolor}
\usepackage{verbatim}   
\usepackage{longtable}
\graphicspath{{ps}}

\newcommand{\fa}{{\it a})}
\newcommand{\fb}{{\it b})}
\newcommand{\fc}{{\it c})}
\newcommand{\fd}{{\it d})}

\newcommand{\ffa}{{\it a}}
\newcommand{\ffb}{{\it b}}

\newcommand{\bA}{\ensuremath{\text{\bf {A}}}}
\newcommand{\bB}{\ensuremath{\text{\bf {B}}}}
\newcommand{\bC}{\ensuremath{\text{\bf {C}}}}
\newcommand{\bD}{\ensuremath{\text{\bf {D}}}}

\newcommand{\Br}{\ensuremath{\mathcal{B}}}
\newcommand{\ecm}{\ensuremath{E_{\rm cm}}}

\newcommand{\ee}{\ensuremath{e^+ e^-}}
\newcommand{\gisr}{\ensuremath{\gamma_{\text{ISR}}}}
\newcommand{\pis}{\ensuremath{\pi_{\text{slow}}}}
\newcommand{\pmis}{\ensuremath{\pi_{\text{miss}}}}
\newcommand{\pif}{\ensuremath{\pi_{\text{fast}}}}

\newcommand{\mrec}{\ensuremath{M_{\text{rec}}}}
\newcommand{\dmrec}{\ensuremath{\Delta \mrec}}
\newcommand{\dmrecf}{\ensuremath{\Delta \mrec^{\text{fit}}}}

\newcommand{\Ufo}{\ensuremath{\Upsilon(4S)}}
\newcommand{\Ufi}{\ensuremath{\Upsilon(5S)}}
\newcommand{\jpsi}{\,\ensuremath{J/\psi}}
\newcommand{\pts}{\,\ensuremath{\psi(2S)}}
\newcommand{\pp}{\,\ensuremath{\pi^+ \pi^-}}
\newcommand{\ks}{\,\ensuremath{K^0_S}}
\newcommand{\kl}{\,\ensuremath{K^0_L}}

\newcommand{\Dp}{\ensuremath{D^{+}}}

\newcommand{\Dap}{\ensuremath{D^{(*)+}}}

\newcommand{\Dst}{\ensuremath{D^{*}}}
\newcommand{\Dstp}{\ensuremath{D^{*+}}}
\newcommand{\Dstm}{\ensuremath{D^{*-}}}

\newcommand{\DDa}{\ensuremath{D^{(*)+}D^{*-}}}
\newcommand{\DDst}{\ensuremath{D^{+}D^{*-}}}
\newcommand{\DstDst}{\ensuremath{D^{*+}D^{*-}}}

\newcommand{\Ma}{\ensuremath{M(D^{(*)+}D^{*-})}}
\newcommand{\Md}{\ensuremath{M(D^{+}D^{*-})}}
\newcommand{\Mdst}{\ensuremath{M(D^{*+}D^{*-})}}

\newcommand{\ifb}{\ensuremath{\,\mathrm{fb}^{-1}}}
\newcommand{\gevc}{\,{\ensuremath{\mathrm{\mbox{GeV}}/c^2}}}
\newcommand{\gev}{\,{\ensuremath{\mathrm{\mbox{GeV}}}}}
\newcommand{\mevc}{\,{\ensuremath{\mathrm{\mbox{MeV}}/c^2}}}
\newcommand{\mev}{\,{\ensuremath{\mathrm{\mbox{MeV}}}}}

\addto\captionsenglish{}

\begin{document}

\vspace*{2.5\baselineskip}

\preprint{\vbox{ \hbox{   }
	          	\hbox{Belle Preprint 2017-15}
                         \hbox{KEK Preprint 2017-16}
}}

\title{\quad\\[1.0cm] Angular analysis of the $\boldmath {e^+ e^- \to
    D^{(*) \pm} D^{* \mp}}$ process near the open charm threshold
  using initial-state radiation}

\noaffiliation
\affiliation{University of the Basque Country UPV/EHU, 48080 Bilbao}
\affiliation{Beihang University, Beijing 100191}
\affiliation{Budker Institute of Nuclear Physics SB RAS, Novosibirsk 630090}
\affiliation{Faculty of Mathematics and Physics, Charles University, 121 16 Prague}
\affiliation{University of Cincinnati, Cincinnati, Ohio 45221}
\affiliation{Deutsches Elektronen--Synchrotron, 22607 Hamburg}
\affiliation{University of Florida, Gainesville, Florida 32611}
\affiliation{Justus-Liebig-Universit\"at Gie\ss{}en, 35392 Gie\ss{}en}
\affiliation{SOKENDAI (The Graduate University for Advanced Studies), Hayama 240-0193}
\affiliation{Gyeongsang National University, Chinju 660-701}
\affiliation{Hanyang University, Seoul 133-791}
\affiliation{University of Hawaii, Honolulu, Hawaii 96822}
\affiliation{High Energy Accelerator Research Organization (KEK), Tsukuba 305-0801}
\affiliation{J-PARC Branch, KEK Theory Center, High Energy Accelerator Research Organization (KEK), Tsukuba 305-0801}
\affiliation{IKERBASQUE, Basque Foundation for Science, 48013 Bilbao}
\affiliation{Indian Institute of Technology Bhubaneswar, Satya Nagar 751007}
\affiliation{Indian Institute of Technology Guwahati, Assam 781039}
\affiliation{Indian Institute of Technology Madras, Chennai 600036}
\affiliation{Indiana University, Bloomington, Indiana 47408}
\affiliation{Institute of High Energy Physics, Chinese Academy of Sciences, Beijing 100049}
\affiliation{Institute of High Energy Physics, Vienna 1050}
\affiliation{Institute for High Energy Physics, Protvino 142281}
\affiliation{INFN - Sezione di Napoli, 80126 Napoli}
\affiliation{INFN - Sezione di Torino, 10125 Torino}
\affiliation{Advanced Science Research Center, Japan Atomic Energy Agency, Naka 319-1195}
\affiliation{J. Stefan Institute, 1000 Ljubljana}
\affiliation{Kanagawa University, Yokohama 221-8686}
\affiliation{Institut f\"ur Experimentelle Kernphysik, Karlsruher Institut f\"ur Technologie, 76131 Karlsruhe}
\affiliation{Kennesaw State University, Kennesaw, Georgia 30144}
\affiliation{King Abdulaziz City for Science and Technology, Riyadh 11442}
\affiliation{Department of Physics, Faculty of Science, King Abdulaziz University, Jeddah 21589}
\affiliation{Korea Institute of Science and Technology Information, Daejeon 305-806}
\affiliation{Korea University, Seoul 136-713}
\affiliation{Kyoto University, Kyoto 606-8502}
\affiliation{Kyungpook National University, Daegu 702-701}
\affiliation{\'Ecole Polytechnique F\'ed\'erale de Lausanne (EPFL), Lausanne 1015}
\affiliation{P.N. Lebedev Physical Institute of the Russian Academy of Sciences, Moscow 119991}
\affiliation{Faculty of Mathematics and Physics, University of Ljubljana, 1000 Ljubljana}
\affiliation{Ludwig Maximilians University, 80539 Munich}
\affiliation{Luther College, Decorah, Iowa 52101}
\affiliation{University of Malaya, 50603 Kuala Lumpur}
\affiliation{University of Maribor, 2000 Maribor}
\affiliation{Max-Planck-Institut f\"ur Physik, 80805 M\"unchen}
\affiliation{School of Physics, University of Melbourne, Victoria 3010}
\affiliation{University of Miyazaki, Miyazaki 889-2192}
\affiliation{Moscow Physical Engineering Institute, Moscow 115409}
\affiliation{Moscow Institute of Physics and Technology, Moscow Region 141700}
\affiliation{Graduate School of Science, Nagoya University, Nagoya 464-8602}
\affiliation{Nara Women's University, Nara 630-8506}
\affiliation{National Central University, Chung-li 32054}
\affiliation{National United University, Miao Li 36003}
\affiliation{Department of Physics, National Taiwan University, Taipei 10617}
\affiliation{H. Niewodniczanski Institute of Nuclear Physics, Krakow 31-342}
\affiliation{Nippon Dental University, Niigata 951-8580}
\affiliation{Niigata University, Niigata 950-2181}
\affiliation{Novosibirsk State University, Novosibirsk 630090}
\affiliation{Osaka University, Osaka 565-0871}
\affiliation{Pacific Northwest National Laboratory, Richland, Washington 99352}
\affiliation{University of Pittsburgh, Pittsburgh, Pennsylvania 15260}
\affiliation{Theoretical Research Division, Nishina Center, RIKEN, Saitama 351-0198}
\affiliation{University of Science and Technology of China, Hefei 230026}
\affiliation{Seoul National University, Seoul 151-742}
\affiliation{Showa Pharmaceutical University, Tokyo 194-8543}
\affiliation{Soongsil University, Seoul 156-743}
\affiliation{Sungkyunkwan University, Suwon 440-746}
\affiliation{Department of Physics, Faculty of Science, University of Tabuk, Tabuk 71451}
\affiliation{Tata Institute of Fundamental Research, Mumbai 400005}
\affiliation{Excellence Cluster Universe, Technische Universit\"at M\"unchen, 85748 Garching}
\affiliation{Department of Physics, Technische Universit\"at M\"unchen, 85748 Garching}
\affiliation{Toho University, Funabashi 274-8510}
\affiliation{Department of Physics, Tohoku University, Sendai 980-8578}
\affiliation{Earthquake Research Institute, University of Tokyo, Tokyo 113-0032}
\affiliation{Department of Physics, University of Tokyo, Tokyo 113-0033}
\affiliation{Tokyo Institute of Technology, Tokyo 152-8550}
\affiliation{Tokyo Metropolitan University, Tokyo 192-0397}
\affiliation{Virginia Polytechnic Institute and State University, Blacksburg, Virginia 24061}
\affiliation{Wayne State University, Detroit, Michigan 48202}
\affiliation{Yamagata University, Yamagata 990-8560}
\affiliation{Yonsei University, Seoul 120-749}
  \author{V.~Zhukova}\affiliation{P.N. Lebedev Physical Institute of the Russian Academy of Sciences, Moscow 119991}\affiliation{Moscow Physical Engineering Institute, Moscow 115409} 
 \author{G.~Pakhlova}\affiliation{P.N. Lebedev Physical Institute of the Russian Academy of Sciences, Moscow 119991}\affiliation{Moscow Institute of Physics and Technology, Moscow Region 141700} 
 \author{P.~Pakhlov}\affiliation{P.N. Lebedev Physical Institute of the Russian Academy of Sciences, Moscow 119991}\affiliation{Moscow Physical Engineering Institute, Moscow 115409} 
 \author{I.~Adachi}\affiliation{High Energy Accelerator Research Organization (KEK), Tsukuba 305-0801}\affiliation{SOKENDAI (The Graduate University for Advanced Studies), Hayama 240-0193} 
  \author{H.~Aihara}\affiliation{Department of Physics, University of Tokyo, Tokyo 113-0033} 
  \author{S.~Al~Said}\affiliation{Department of Physics, Faculty of Science, University of Tabuk, Tabuk 71451}\affiliation{Department of Physics, Faculty of Science, King Abdulaziz University, Jeddah 21589} 
  \author{D.~M.~Asner}\affiliation{Pacific Northwest National Laboratory, Richland, Washington 99352} 
  \author{V.~Aulchenko}\affiliation{Budker Institute of Nuclear Physics SB RAS, Novosibirsk 630090}\affiliation{Novosibirsk State University, Novosibirsk 630090} 
  \author{T.~Aushev}\affiliation{Moscow Institute of Physics and Technology, Moscow Region 141700} 
  \author{R.~Ayad}\affiliation{Department of Physics, Faculty of Science, University of Tabuk, Tabuk 71451} 
% ...
\collaboration{The Belle Collaboration}
 

\begin{abstract}
We report a new measurement of the exclusive $\ee \to D^{(*) \pm} D^{*
  \mp}$ cross sections as a function of the center-of-mass energy from the
$D^{(*) \pm} D^{* \mp}$ threshold through $\sqrt{s}=6.0\gev$ with
initial-state radiation. The analysis is based on a data sample
collected with the Belle detector with an integrated luminosity of
$951\ifb$. The accuracy of the cross section measurement is increased
by a factor of two over the first Belle study. We perform the
first angular analysis of the $\ee \to D^{* \pm} D^{* \mp}$ process
and decompose this exclusive cross section into three components
corresponding to the $D^*$ helicities.

\end{abstract}


\maketitle

\tighten

{\renewcommand{\thefootnote}{\fnsymbol{footnote}}}
\setcounter{footnote}{0}

\section{Introduction} 

Parameters of the vector charmonia ($\psi$'s) lying above the
open-charm threshold were obtained from the total \ee\ cross section
to hadronic final states in measurements by MARK-I, DELCO, DASP,
MARK-II, Crystal Ball, and BES~\cite{pdg}. In 2008, BES remodeled
their data, including in particular the interference and the relative
phases but using model predictions for the $\psi$ decays to two-body
open-charm final states~\cite{bes:fit}. As a result, the obtained
parameters of the $\psi$ states cannot be regarded as universal and
model-indepenent. Notably, BES did not extract the parameters of other
vector charmoniumlike resonances, such as Y(4008), Y(4260), Y(4360)
and Y(4660), which were observed by
BaBar~\cite{babar_y1,babar_y2,babar_y3} and
Belle~\cite{belle_y1,belle_y2,belle_y3,belle_y4,belle_y5} in
\ee\ annihilation. Y(4008) and Y(4260) decay into the $\jpsi\ \pi^+
\pi^-$ final state while Y(4360) and Y(4660) decay into $\pts\ \pi^+
\pi^-$. The Y states do not appear as peaks either in the total
hadronic cross section nor in the exclusive \ee\ cross sections to
open-charm final states that were measured later. In addition, there
exist predictions in the literature that some of the Y states could
manifest themselves as coupled-channel effects~\cite{y_predict}.

A detailed study of the exclusive \ee\ cross sections to open-charm
final states could help establish parameters of the vector charmonia
and charmoniumlike states in a model-independent way and, therefore,
to shed light on the nature of the Y-family. Such exclusive \ee\ cross
sections to various open-charm final states were first measured at
$B$-factories by
Belle~\cite{belle:dd,belle:ddst,belle:ddpi,belle:lala,belle:ddstpi,belle:dsds}
and BaBar~\cite{babar:dd,babar:ddst,babar:dsds} using the
initial-state radiation (ISR) processes, and by CLEO~\cite{cleo:cs}
using an energy scan in the range $\sqrt{s}=3.97-4.26\gev$. Belle has
demonstrated that the sum of the two-body ($D^{(*)} \bar{D}{}^{(*)}$,
$D_s^{(*)+} D_s^{(*)-}$, $\Lambda_c^+\Lambda_c^-$) and three-body $D
\bar{D}{}^{(*)} \pi$ cross sections almost saturates the total
hadronic cross section after the subtraction of the $u$-, $d$-, and
$s$-continuum in the region $\sqrt{s}\leq
5\gev$~\cite{Bevan:2014iga}. The main contribution to the inclusive
open-charm cross section comes from the $D \bar{D}$, $D \bar{D}{}^*$,
and $D^* \bar{D}{}^*$ final states.

The first attempt to extract the parameters of the $\psi$ states (in
particular, their couplings to the open-charm channels) from a
combined coupled-channel fit for all exclusive open-charm cross
sections was performed in Ref.~\cite{Uglov:2016orr}. At the time,
although the suggested approach provides a good overall description of
the line shapes, reliable conclusions could not be made because of the
limited statistical accuracy of the data and because of the absense of
experimental information on each of the three helicity amplitudes of
the $D^{*} \bar{D}{}^{*}$ system. For further details, see
Ref.~\cite{Uglov:2016orr} and references therein.

Here, we report a new measurement of the exclusive $\ee \to \DDa$
cross sections as a function of the center-of-mass energy near the
$\DDa$ threshold in the initial-state radiation
processes\footnote{Charge-conjugate modes are included throughout
  this paper.}. Compared to Ref.~\cite{belle:ddst}, the larger data set,
the improved track reconstruction, and the additional modes used in
the $D$ and $D^*$ reconstruction allow one to obtain more precise
determination of these cross sections.  We also perform the first
angular analysis of the $\ee \to D^{* \pm} D^{* \mp}$ processes and
explicitly decompose these exclusive cross sections into the three
components corresponding to the $D^*$'s helicities. Employing a
coupled-channel technique --- following the lines of
Ref.~\cite{Uglov:2016orr}, for example---a future study could use
these results to extract the parameters of the vector charmonium
states.

\section{Data sample and Belle detector}

The analysis reported in this work is based on the data sample with an
integrated luminosity of $951\ifb$ collected with the Belle detector
\cite{Belle} at the KEKB asymmetric-energy \ee\ collider near the
energies of the \Ufo\ and \Ufi\ resonances \cite{KEKB}.

The Belle detector is a large-solid-angle magnetic spectrometer that
consists of a silicon vertex detector (SVD), a 50-layer central drift
chamber (CDC), an array of aerogel threshold Cherenkov counters (ACC),
a barrel-like arrangement of time-of-flight scintillation counters
(TOF), and an electromagnetic calorimeter (ECL) composed of CsI(Tl)
crystals located inside a superconducting solenoid coil that provides
a 1.5 T magnetic field.  An iron flux-return located outside of the
coil is instrumented to detect\kl\ mesons and to identify muons
(KLM). A detailed description of the detector can be found, for
example, in Ref.~\cite{Belle}.

\section{Method}

To select the $\ee \to \DDa \gisr$ signal, we use the method described
in Ref.~\cite{belle:ddst}. We require full reconstruction of only one
of the \Dstp\ (for $\ee \to \DstDst$) or \Dp\ mesons (for $\ee \to
\DDst$), the $\gisr$ photon, and the slow pion from the other
\Dstm. The partial \Dstm\ reconstruction without reconstruction of the
$\bar{D}^0$ daughter of the \Dstm\ increases the overall efficiency by
a factor of $\sim 20$ for $\ee \to \DstDst$ and $\sim 10$ for $\ee \to
\DDst$, while suppressing the backgrounds enough to be able subtract
them reliably using the data. Unlike the usual method for
reconstruction of ISR processes, where the hadronic final state is
fully reconstructed and \gisr\ is inferred from the spectrum of masses
recoiling against the hadronic system, we require here that the
\gisr\ to be reconstructed. This requirement does not significantly
decrease the efficiency as the slow pion from $D^{*\pm}$ decay has a
low reconstruction efficiency when \gisr\ is outside the detector
acceptance (because of the very low transverse momentum of \gisr\ in this
case).

For the signal candidates, the spectrum of the mass recoiling against
the $D^{(*+)} \gisr$ system,~\footnote{Hereinafter, the speed of light
  is set to unity for convenience.}
\begin{linenomath*}
\begin{equation}
\mrec(\Dap\!\gisr)=\sqrt{(\ecm\!- \!E_{D^{(*)+}
    \gamma_{\text{ISR}}})^2 - p^2_{D^{(*)+} \gamma_{\text{ISR}}}},
\end{equation}
\end{linenomath*}
peaks at the \Dstm\ mass. Here, \ecm\ is the \ee\ center-of-mass
energy, and $E_{\Dap \gisr}$ and $p_{\Dap \gisr}$ are the
center-of-mass energy, and momentum, respectively, of the $\Dap \gisr$
system. According to a Monte Carlo (MC) study, this peak is wide
($\sigma \sim 150\mevc$) and asymmetric due to the asymmetric photon
energy resolution function and higher-order ISR corrections ({\it
  i.e.}, more than one $\gisr$ in the event). Because of poor
$\mrec(\Dap \gisr)$ resolution, the signals from $D \bar{D}$,
$D\bar{D}{}^*$, and $D^* \bar{D}{}^*$ overlap strongly overlap; hence,
one cannot distinguish these processes using this measure alone. For
illustration, Fig.~\ref{rmdg} shows the MC $\mrec(\Dstp \gisr)$
spectra for the $\ee \to \DstDst \gisr$ and $\ee \to \DDst \gisr$
processes.
\begin{figure}[h!]
\centering \includegraphics[width=.37\textwidth]{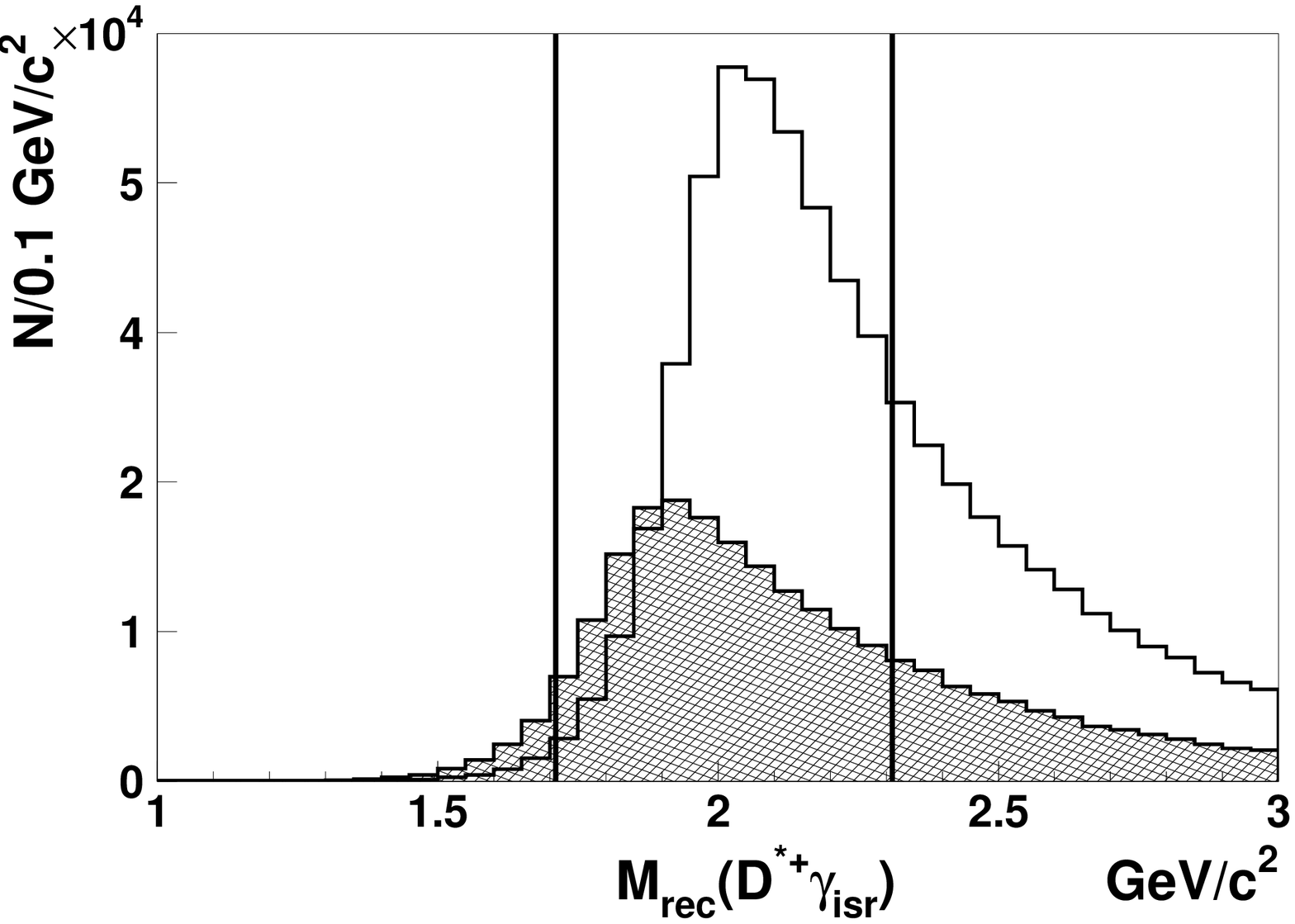}
\caption{MC simulation $\mrec(\Dstp \gisr)$ spectra for $\ee \to
  \DstDst \gisr$ (histogram) and $\ee \to \DDst \gisr$ (hatched
  histogram). The signal window is shown by vertical lines. }
\label{rmdg}
\end{figure}

\begin{figure}[h!]
\centering \includegraphics[width=0.37\textwidth]{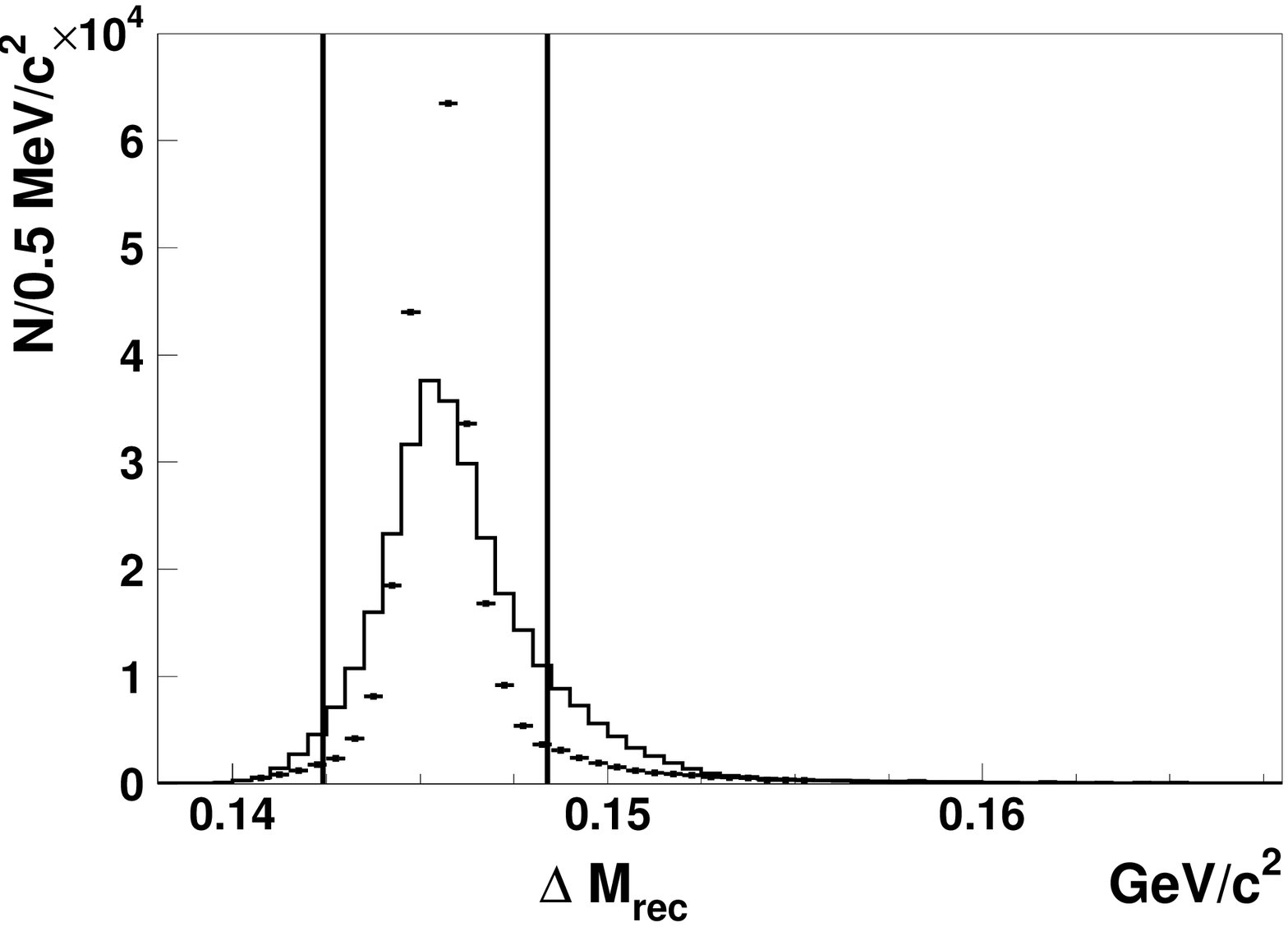}
\caption{The $\dmrec$ distributions before (histogram) and after
  (points with errors) refit procedure for the MC events of $\ee \to
  \DstDst \gisr$. The signal window lies between vertical lines.}
\label{rmd}
\end{figure}

To resolve this, we use the information provided by the slow pion
from the unreconstructed \Dstm\ meson. The distribution of the
difference between the masses recoiling against the $\Dap \gisr$ and
$\Dap \pis^{-} \gisr$, termed the recoil-mass difference,
\begin{linenomath*}
\begin{equation} 
\Delta \mrec \! = \! \mrec (\Dap \gisr) \! - \! \mrec(\Dap \pis^- \gisr),
\end{equation}
\end{linenomath*}
has a narrow peak for the signal process around the $m_{\Dstp} -
m_{D^0}$ mass difference (Fig.~\ref{rmd}, histogram). The resolution
of this peak is below 2\mevc\ as the uncertainty of the
\gisr\ momentum is mostly canceled out for this variable. Thus, the
existence of a partially reconstructed \Dstm\ in the event is
identified by the presence of this peak. The method does not exclude
contributions in the \dmrec\ signal window from processes with extra
neutrals in the final state ({\it e.g.}, $\ee \to \DDa
\pi^0$). However, this background is suppressed and its residual
contribution can be reliably determined using the data, as discussed
in the data analysis section.

To measure the exclusive cross sections as a function of $\sqrt{s}$,
one needs to obtain the \DDa\ mass spectrum despite one of $D^*$
mesons being unreconstructed. In the absence of higher-order QED
processes, the \DDa\ mass corresponds to the mass recoiling against
the single ISR photon: $\Ma \equiv \mrec(\gisr)$. However, the poor
$\mrec(\gisr)$ resolution ($\sigma \sim 120\mevc$) precludes the study
of relatively narrow charmonium states in the \DDa\ mass spectra. To
improve the $\mrec(\gisr)$ resolution, we refit the recoil mass
against the $\Dap \gisr$ system, constrained to the \Dstm\ mass. This
procedure utilizes the well-measured momentum of the reconstructed
\Dap\ meson and the signal kinematics to better determine the momentum
of the ISR photon. It works well even in case of a second ISR photon,
as checked with the MC. As a result, the $\mrec(\gisr)$ resolution is
drastically improved: near the threshold, the resolution is better
than $3\mevc$, and smoothly increases to $15\mevc$ at $\sqrt{s}\sim
6\gev$ (Fig.~\ref{res}). The resolution of the recoil-mass difference
after refit, \dmrecf, improves by a factor of $\sim 2$
(Fig.~\ref{rmd}, points with error bars); this is exploited for
more effective suppression of the combinatorial background.
\begin{figure}[h!]
\centering \includegraphics[width=0.37\textwidth]{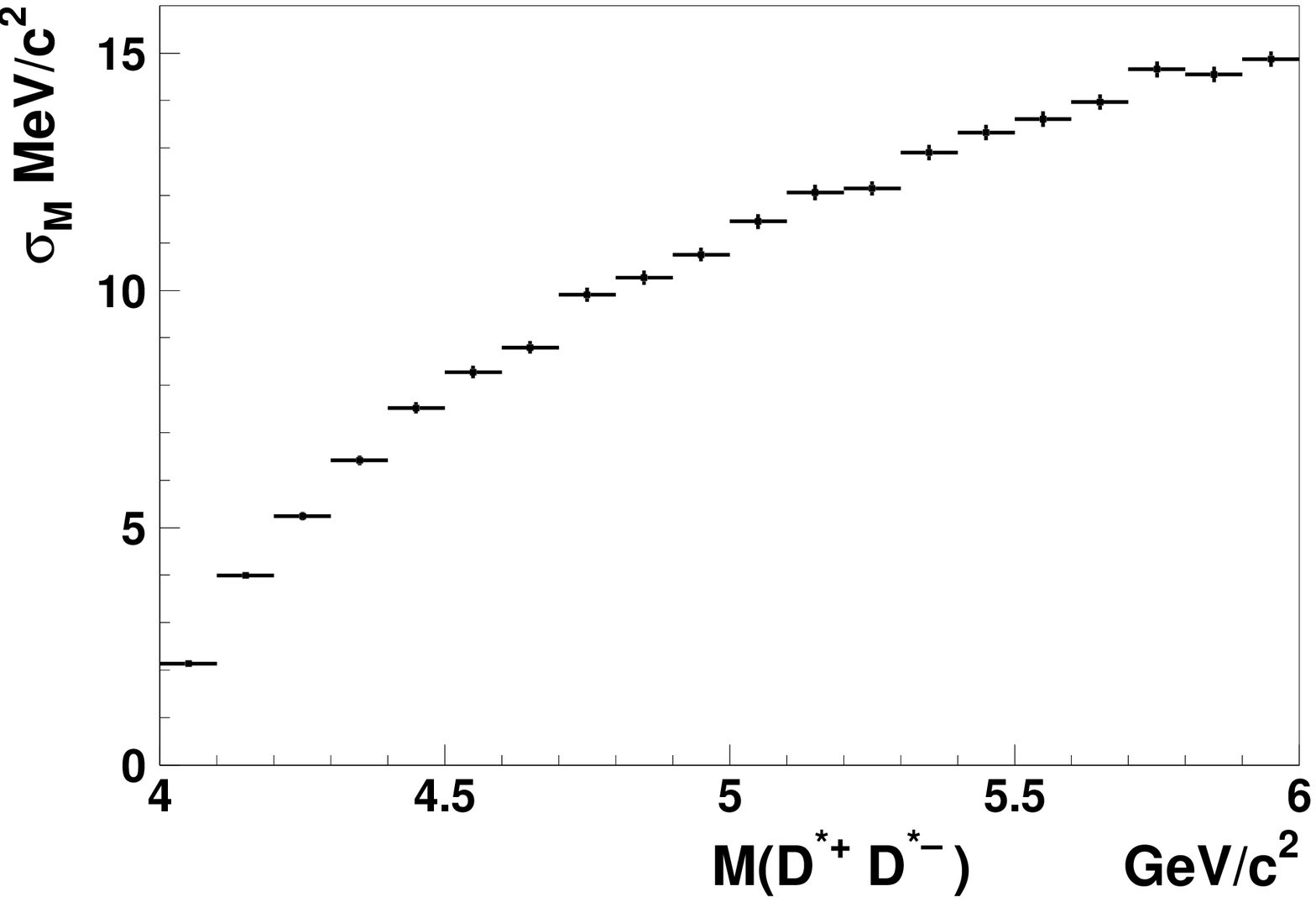}
\caption{The \DstDst\ mass resolution after the refit procedure as a
  function of \Mdst\ for the process $\ee \to \DstDst \gisr$.}
\label{res}
\end{figure}

\section{Monte Carlo study and calibration of $\gisr$ energy}

The simulation of the signal and background processes up to the
second-order ISR corrections and vacuum-polarization corrections is
performed using the {\tt PHOKHARA} MC
generator~\cite{Rodrigo:2002hk}. In the signal-MC samples, the
\DDa\ characteristics (mass spectrum and angular distributions) are
tuned to those measured in the data. As the measured distributions are
extracted from the data using the MC simulation, this tuning is
repeated until the difference between successive iterations is
negligibly small.

To improve the accuracy of the MC simulation, we calibrate the photon
energy. For soft and medium energy photons, this is done using $\pi^0
\to \gamma \gamma$ ~\cite{belle:pi0}. For energetic ($\sim 4 \gev$)
photons, where there are few energetic $\pi^0$ in the data, we select
a clean sample of fully reconstructed $\ee \to \pts \gisr \to \jpsi
\pp \gisr$ events and study the spectrum of the mass recoiling against
$\gisr$; a broad peak around the \pts\ mass is expected. A small
shift between MC and data is observed, from which a correction factor
for the photon energy is determined to be $0.9980\pm 0.0004$. We apply
this correction throughout the analysis.

\section{Data sample and event selection}
\label{select}

In the first Belle analysis~\cite{belle:ddst}, the strategy was
to select a clean sample of the studied process with minimal
background contribution to provide the most reliable result. It was
demonstrated there that all backgrounds were well under control
and were subtracted reliably using the data. The aim of the present
analysis is to improve the accuracy of the cross section
measurement. Therefore, we reoptimize the selection criteria and add
more $D^{(*)}$ decay modes.

All charged tracks are required to be consistent with origination from
the interaction point (IP): we require $dr<2 \, {\mathrm{cm}}$ and
$|dz|<4\,{\mathrm{cm}}$, where $dr$ and $|dz|$ are the impact
parameters perpendicular to and along the beam direction,
respectively, with respect to the IP. Information from the TOF, the
number of the photoelectrons from the ACC and the $dE/dx$ measurement
in the CDC are combined to form a likelihood $\mathcal L$ for hadron
identification. Charged kaon candidates are required to have a
kaon/pion likelihood ratio ${\mathcal{P}}_{K/\pi} = \mathcal{L}_K /
(\mathcal{L}_K + \mathcal{L}_\pi)>0.6$. The identification
efficiencies typically exceed 90\%, while misidentification
probabilities are less than 10\%~\cite{PID}. No identification
requirements are applied for pion candidates, as the pion multiplicity
is much higher than that of other hadrons.  \ks\ candidates are
reconstructed by combining \pp\ pairs with an invariant mass within
$\pm 15\mevc$ ($\approx\ \pm\ 5 \sigma$) of the \ks\ mass. The
distance between the two pion tracks at the \ks\ vertex must be less
than $1\,\mathrm{cm}$. The transverse flight distance from the
interaction point is required to be greater than $0.1\,\mathrm{cm}$,
and the angle between the \ks\ momentum direction and the line joining
the decay vertex and the IP in the transverse plane should be smaller
than $0.1\,\mathrm{rad}$. Photons are reconstructed in the
electromagnetic calorimeter as showers with energy greater than
$50\mev$ that are not associated with charged tracks. ISR photon
candidates are required to have a center-of-mass energy above
$3.0\gev$.  Combinations of two photons are treated as a $\pi^0$
candidates if their invariant mass lies within $15\mevc$ ($\approx
\pm\ 3 \sigma$) of the nominal $\pi^0$ mass. Such combinations are
then refitted with a $\pi^0$ mass constraint to improve the $\pi^0$
momentum resolution.

$D^0$ candidates are reconstructed in nine decay modes: $K^- \pi^+$,
$K^- K^+$, $K^- \pi^- \pi^+ \pi^+$, $\ks \pi^+ \pi^-$, $K^- \pi^+
\pi^0$, $\ks K^+ K^-$, $\ks \pi^0$, $K^- K^+ \pi^- \pi^+$, and $\ks
\pi^+ \pi^- \pi^0$. $D^+$ candidates are reconstructed using the
$K^+ \pi^+ \pi^-$, $\ks \pi^+$, and $\ks K^+$ channels. The mass
window for all modes without (with) $\pi^0$ is $\pm 15\mevc$
($\pm\ 20\mevc$) of the nominal $D^0$ mass ~\cite{pdg}, corresponding
to $\approx \pm\ 3 \sigma$ in each case. A mass-constrained vertex fit
is applied to the $D^0$ and $D^+$ candidates to improve their momentum
resolution.

\Dstp\ candidates are selected in the $D^0 \pi^+$ decay mode; the
$D^{*0}$ candidates are selected in the $D^0 \pi^0$ mode and are used
for background studies. In both cases, the signal window for the $D^*$
mass is chosen to be $\pm 3\mevc$ ($\approx 3 \sigma$) of the nominal
\Dstp\ mass ~\cite{pdg}.

To increase the efficiency, the $\mrec(\Dap \gisr)$ signal region is
defined to be $\pm 300\mevc$ of the \Dstm\ meson mass (see
Fig.~\ref{rmdg}). The process $\ee \to \DDa \pmis^0 \gisr$ has the
$\mrec(\Dap \gisr)$ spectrum shifted to higher values, but still
overlaps this signal window. We eschewed a further extension of the
signal window to the higher masses since this would lead to an to
increased contribution from the background process with an extra
$\pi^0$. The \dmrecf\ value is required to be within $\pm 3\mevc$ of
the $m_{\Dstp}-m_{D^0}$ value (see Fig.~\ref{rmd}).

The fraction of events with more than one candidate passing all
selection criteria is small (3\%). For these, we select the one with
the smallest value of
\begin{linenomath*}
\begin{equation}
\chi^2 = \chi^2_{M(D)}~ (+ \chi^2_{M(D^*)} )+ \chi^2_{\dmrecf} \, ,
\end{equation}
\end{linenomath*}
where $\chi^2_{i}$ is defined as a squared ratio of the difference
between the measured and expected observable $i$ to the corresponding
resolution. For the background studies, we use $M(\Dap)$ and
$\dmrecf(\Dap \gisr)$ sidebands. In the sideband regions, a
similar selection of a single candidate per event is performed based
on a $\chi^2$ calculated with respect to the center of the
corresponding sideband interval.

\section{\label{analy}Data analysis}

As the process $\ee \to \DstDst \gisr$ contributes to the $\ee \to
\DDst \gisr$ when one of the $D^{*\pm}$ mesons decays into $D^+ \pi^0$
or $D^+ \gamma$, we study the process $\ee \to \DstDst$ first. Then we
use the obtained result to correct the $\ee \to \DstDst \gisr$ Monte
Carlo and use the latter to extract the information that is needed to
subtract the \DstDst\ contribution from the $\ee \to \DDst \gisr$
process.

\subsection{Measurement of $\ee \to \DstDst \gisr$}
\label{sec:dstdst}

The \DstDst\ mass spectrum, after applying all the selection
criteria, is shown in Fig.~\ref{mdd_bg_b}. {Taking advantage of} the
fine \Mdst\ resolution at the threshold and higher statistics in
comparison with the first Belle paper~\cite{belle:ddst}, we study the
mass structure near the threshold with finer bins. The
\DstDst\ mass spectrum  is presented in the
inset of Fig.~\ref{mdd_bg_b}.

\begin{figure}[h!]
\centering 
\includegraphics[width=0.48\textwidth]{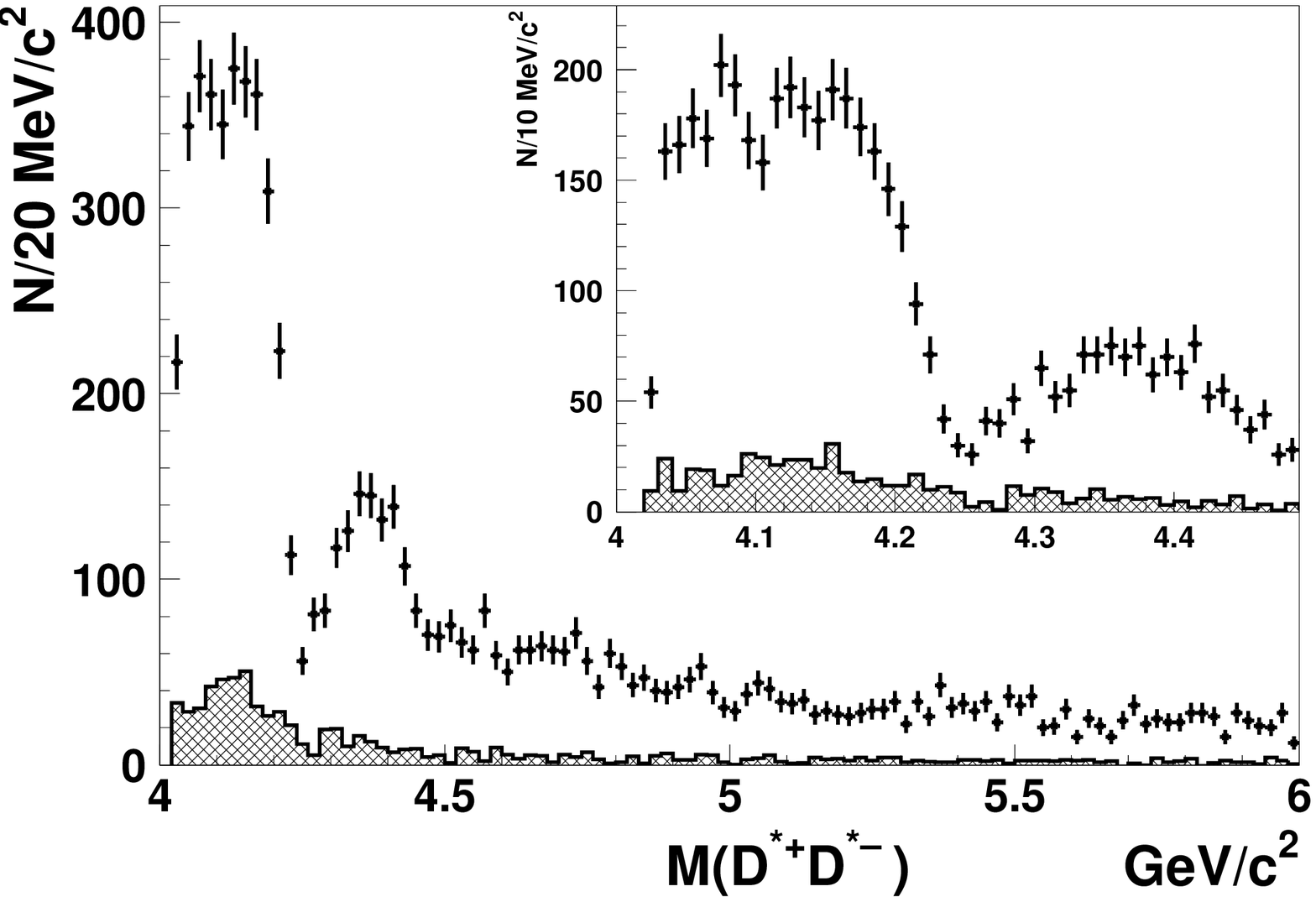}
\caption{The \Mdst\ spectrum in the data after applying all selection
  criteria (points with error bars). The sum of backgrounds (1)-(3) is
  shown as the hatched histogram. The inset shows the zoomed spectrum
  near the threshold with a finer (half-size) bin width of
  $10\mevc$.}
\label{mdd_bg_b}
\end{figure}

We consider the following background contributions:
\begin{enumerate}
\item combinatorial background under the reconstructed \Dstp\ peak,
  when the slow pion is truly a \Dstp\ daughter;
\item real \Dstp\ mesons combined with a combinatorial (random) slow
  pion;
\item both the \Dstp\ meson and slow pion are combinatorial;
\item reflections from the processes $\ee \to \DstDst \pi^0 \gisr$
  where the $\pi^0$ is lost;
\item $\ee \to \DstDst \pif^0$ where the hard $\pif^0$ is
  misidentified as $\gisr$. This can happen if the two photons from
  the $\pif^0$ decay merge into one ECL cluster or, in the case of an
  asymmetric $\pif^0$ decay, if one of the photons carries a large
  fraction of the $\pif^0$ energy.
\end{enumerate}

The contribution from the combinatorial backgrounds, \textit{i.e.,}
the first three background sources, is extracted using the $M(\Dstp)$
and \dmrecf\ sidebands. The $M(\Dstp)$ {\it vs} \dmrecf\ scatter plot
is shown in Fig.~\ref{md_rmd_b} for the data. The signal region
(indicated as box \bA) is defined by the requirement that $M(\Dstp)$
and \dmrecf\ lie within $\pm 3\mevc$ of the corresponding nominal
values. The contributions of combinatorial \Dstp\ candidates or random
slow pions --- backgrounds (1) and (2) --- are estimated from the
regions shown by boxes \bB\ and \bC, respectively. The sidebands of
double the width are shifted from the signal region to avoid signal
over-subtraction. Background (3) is present in both $M(\Dstp)$ and
\dmrecf\ sidebands and therefore is subtracted twice. To correct for
this over-subtraction, the double sideband region \bD\ is used.

\begin{figure}[h!]
\centering \includegraphics[width=0.48\textwidth]{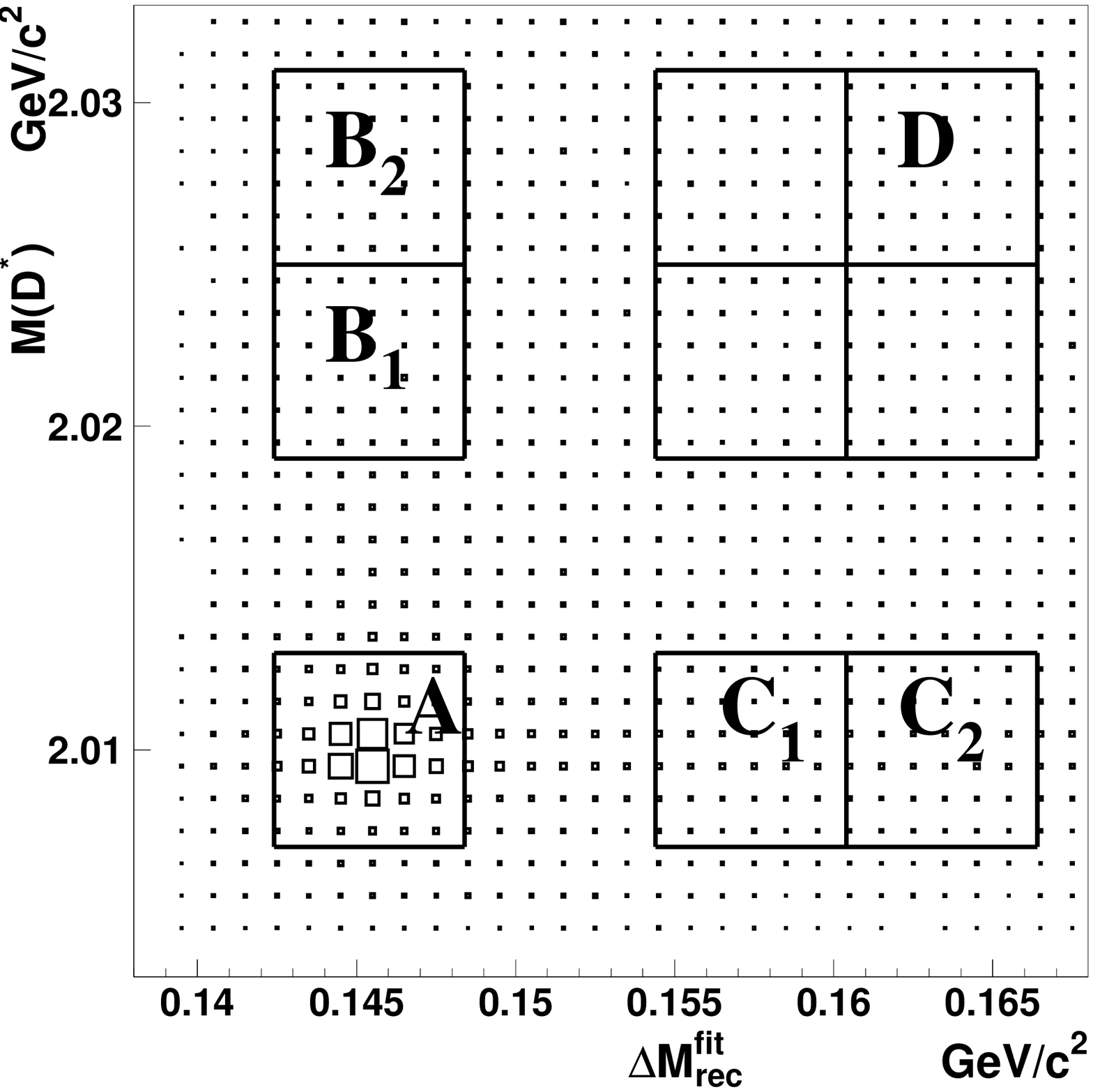}
\caption{The scatter plot of $M(\Dstp)$ {\it vs.} \dmrecf\ in the
  data.}
\label{md_rmd_b} 
\end{figure}

The total contribution to the \Mdst\ spectrum from the
combinatorial backgrounds (1)-(3) is calculated according to the
formula
\begin{linenomath*}
\begin{equation}
M_{\text{bg~(1)-(3)}}=0.58\cdot M_{\bB}+0.53\cdot
M_{\bC}-0.307\cdot M_{\bD},
\label{bg1}
\end{equation}
\end{linenomath*}
where $M_{\bB,~\bC,~\bD}$ are the \DstDst\ mass spectra from the \bB,
\bC, and \bD\ sidebands, respectively, and the scaling factors are
calculated to provide normalization of the corresponding background
contributions within the signal window. To obtain these scaling
factors, the distributions of $M(\Dstp)$ and \dmrecf\ are fit
using signal shapes fixed to those from the MC simulation and the
following background parameterization:
\begin{linenomath*}\begin{equation}
f_{\rm bg}=a\sqrt{x-x_{\rm thr}}\cdot (1+b x+c x^2) \, ,
\label{fbg1}
\end{equation}\end{linenomath*}
where $x$ is the $M(\Dstp)$ or \dmrecf, $x_{\rm thr}$ is a
corresponding threshold values, and $a,~b,$ and $~c$ are free parameters. The
fit results are shown in Fig.~\ref{fit_b}.

\begin{figure}[h!]
\centering
\includegraphics[width=0.48\textwidth]{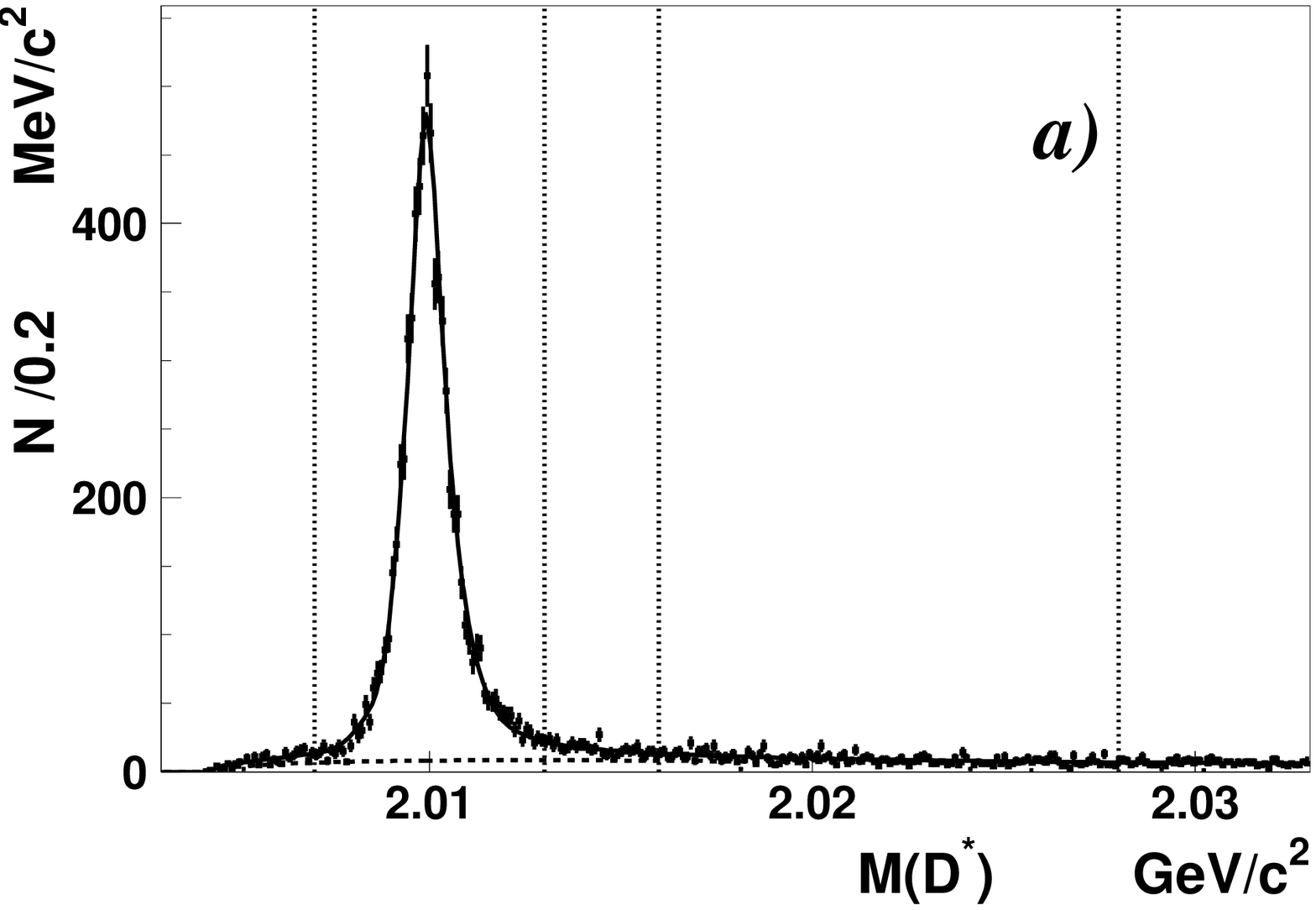} \\
\includegraphics[width=0.48\textwidth]{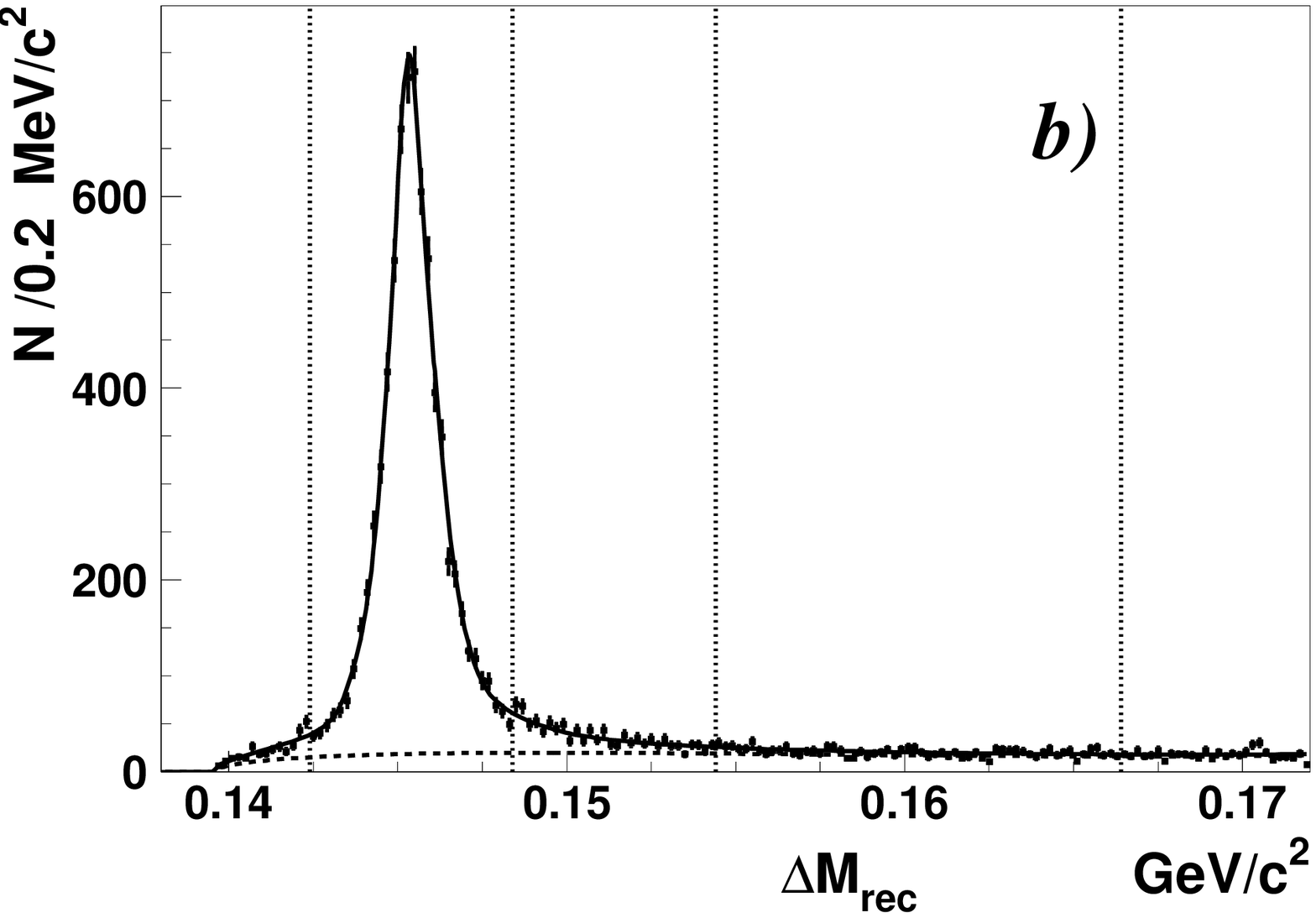}
\caption{Distributions of \fa\ $M(\Dstp)$; \fb\ \dmrecf. The
  solid line represents the result of fit described in the text. The
  background contributions are shown as dotted curves. The signal and
  sidebands regions are indicated by vertical lines.}
\label{fit_b}
\end{figure}

The total combinatorial background (1-3), calculated according to
Eq.~(\ref{bg1}), is shown in Fig.~\ref{mdd_bg_b} as the hatched
histogram.

To estimate the contribution from background (4), the process
$\ee \to \DstDst \pmis^0 \gisr$, we study the isospin-conjugated
process $\ee \to D^{*0} D^{*-} \pmis^+ \gisr$ using a similar partial
reconstruction method. Because of charge imbalance of the
reconstructed combination ($D^{*0} \pis^- \gisr$), only events with
a missing $\pmis^+$ can contribute to the recoil mass difference
peak (indicating the presence of a \Dstm\ meson). The $D^{*0}
D^{*-}$ mass spectrum is obtained by applying all requirements and
subtracting combinatorial backgrounds. To recalculate the background
(4) contribution from the isospin-conjugated $M(D^{*0} D^{*-})$
spectrum, one needs to correct the latter for the ratio of the
$D^{*+}$ and $D^{*0}$ reconstruction efficiencies (equal to 1.77,
according to the MC) and the isospin factor $1/2$. As in the
previous Belle analysis~\cite{belle:ddst}, this contribution is
found to be consistent with zero (Fig.~\ref{m_dst0dstm}). However, to
avoid additional systematic errors, the spectrum of background
(4) obtained from the data is subtracted bin-by-bin, and the
statistical errors of the subtraction of this contribution are
included in the final result.

The \DstDst\ mass spectrum after bin-by-bin combinatorial
background subtraction is shown in Fig.~\ref{mdd_d1_6}.

\begin{figure}[h!]
\centering
\includegraphics[width=0.48\textwidth]{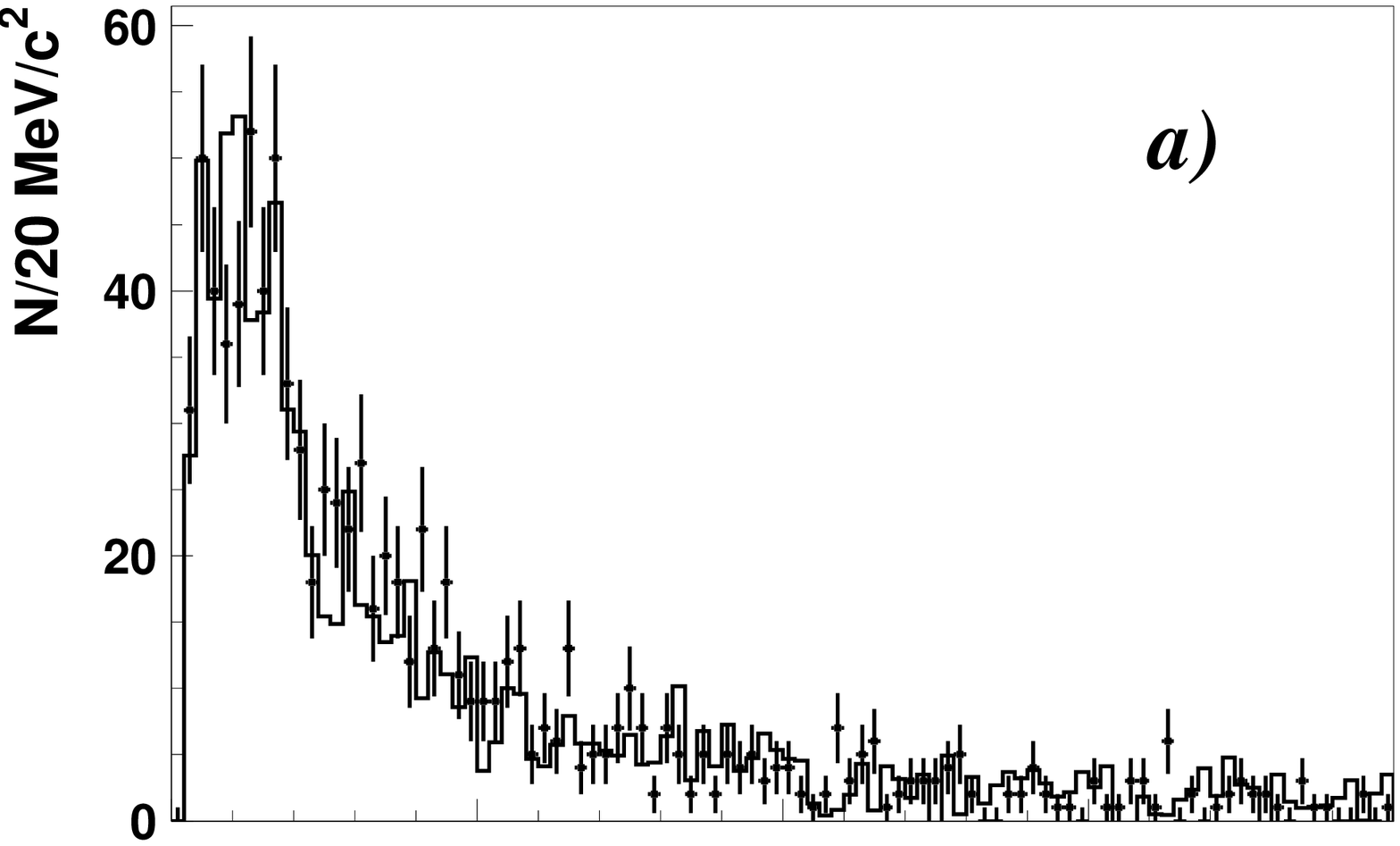} \\[-1.1cm]
\includegraphics[width=0.48\textwidth]{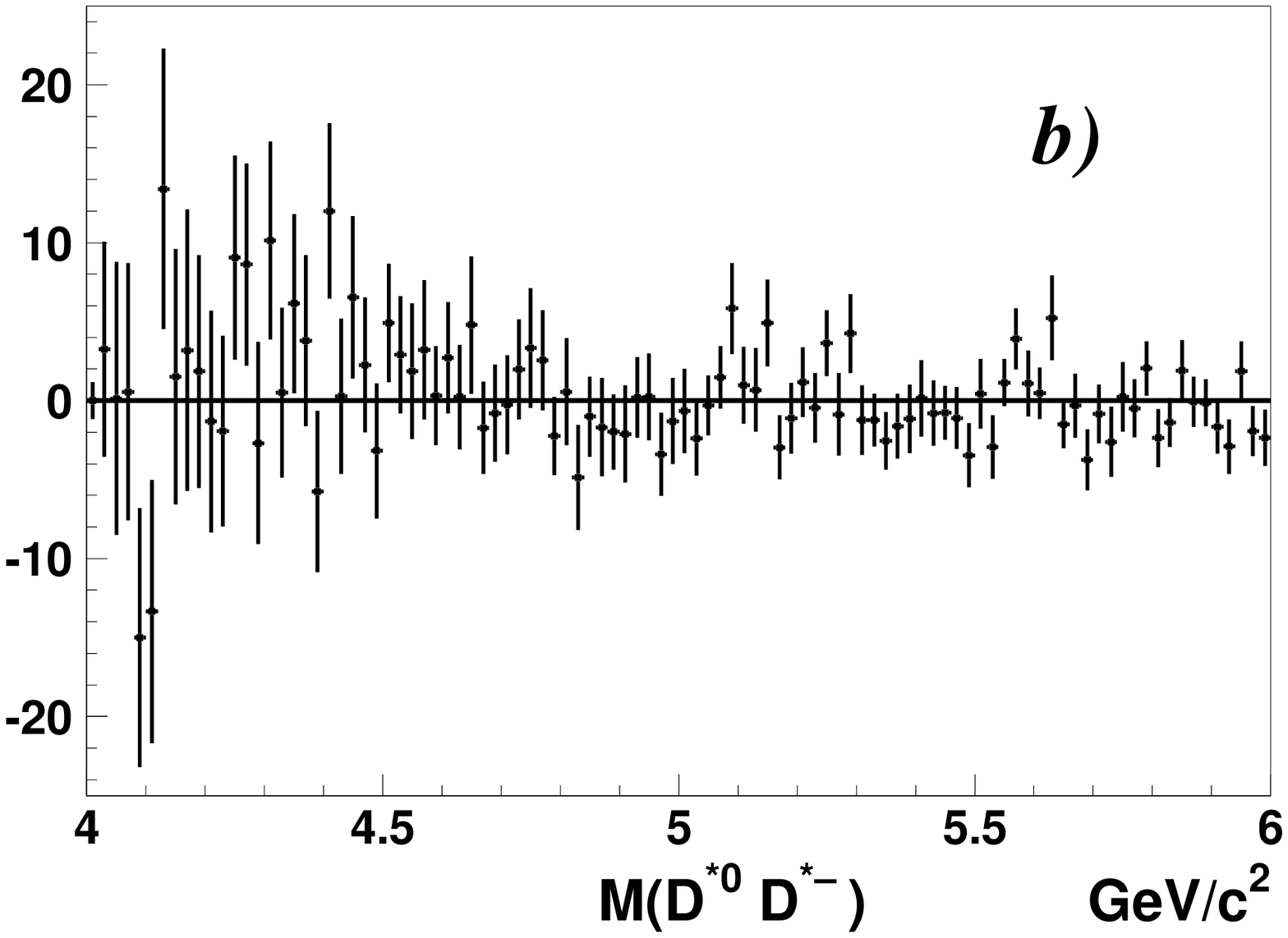}
\caption{The $M(D^{*0} D^{*-})$ spectrum after applying all
  requirements \fa\ before and \fb\ after combinatorial background
  subtraction. The solid histogram corresponds to the total
  contribution of the combinatorial background.}
\label{m_dst0dstm}
\end{figure}

\begin{figure}[h!]
\centering
\includegraphics[width=0.48\textwidth]{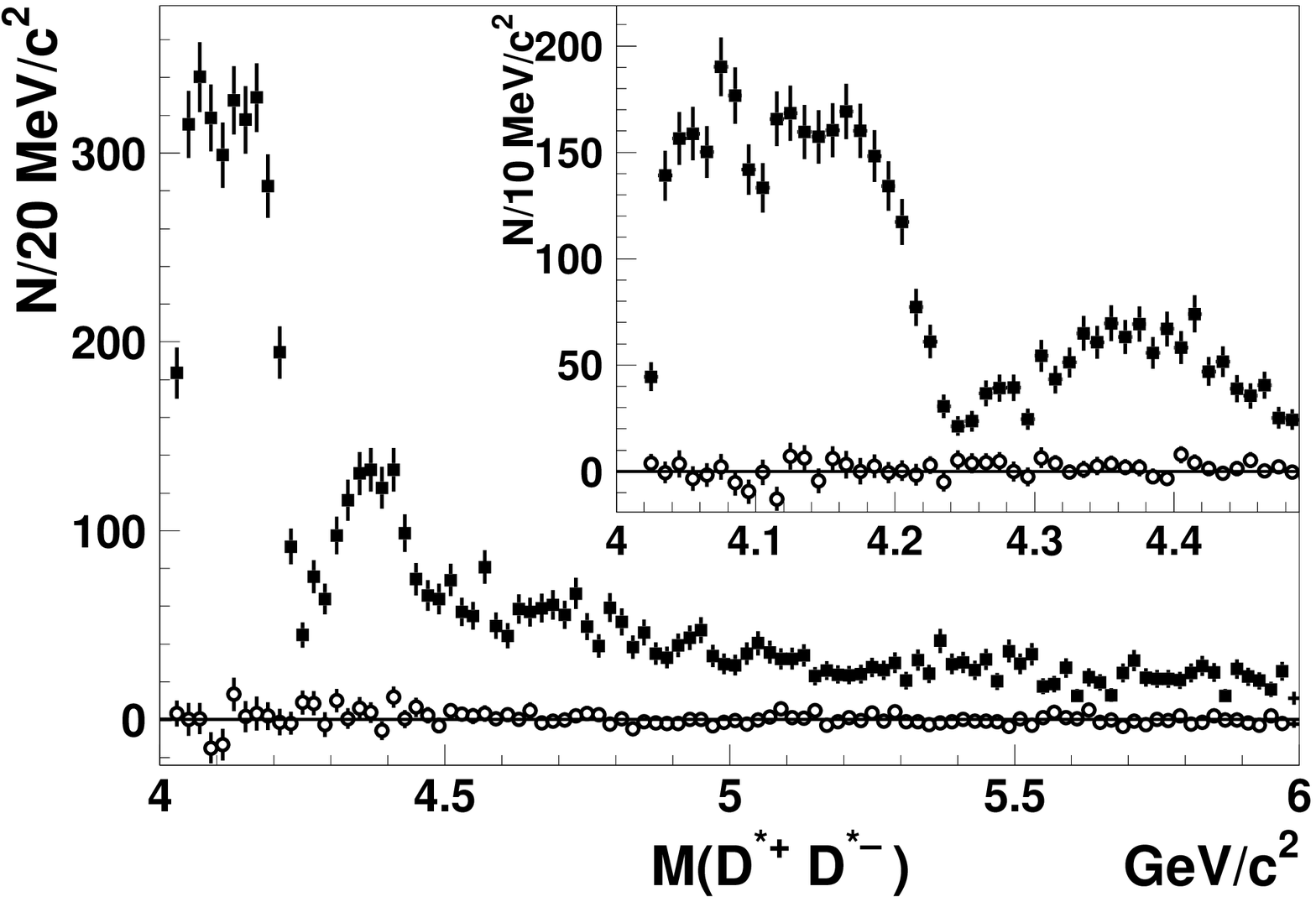}
\caption{The $\Mdst$ spectrum after the combinatorial background
  (1)-(3) subtraction. The contribution from the process $\ee \to
  \DstDst \pmis^0 \gisr$ is shown as open circles with error bars.
  The inset shows the spectrum near the threshold with finer bins.}
\label{mdd_d1_6}
\end{figure}

Background (5) is also estimated from the data using the same method
of partial reconstruction, substituting \gisr\ for a fully
reconstructed energetic $\pi^0$. From the fit to the $\pi^0$ mass
distribution, we find $56 \pm 12$ events corresponding to the process
$\ee \to \DstDst \pi^0$ with $\Mdst \leq 6 \gevc$. From the MC
simulation, we estimate the ratio of efficiencies for background
process (5) to be reconstructed as $D^{*+} D^{*-} \gisr$ and $D^{*+}
D^{*-} \pi^0$ to be equal to $0.39$. Thus, we conclude that only
$22 \pm 5$ events contribute to our reconstructed ISR spectrum in the
region $\Mdst \leq 6 \gevc$, {\it i.e.}, under $0.3\%$ of the
  signal. As the \Mdst\ spectrum from $\ee \to \DstDst \pi^0$ is
  distributed uniformly, far less than one event per bin from
  this background is expected. Therefore, we incorporate
  this background contribution into the systematic uncertainty.

\subsection{Measurement of $\ee \to \DDst \gisr$} 

The \DDst\ mass spectrum, after applying all the selection
criteria, is shown in Fig.~\ref{mdd_bg_a}.
\begin{figure}[h!]
\centering
\includegraphics[width=0.48\textwidth]{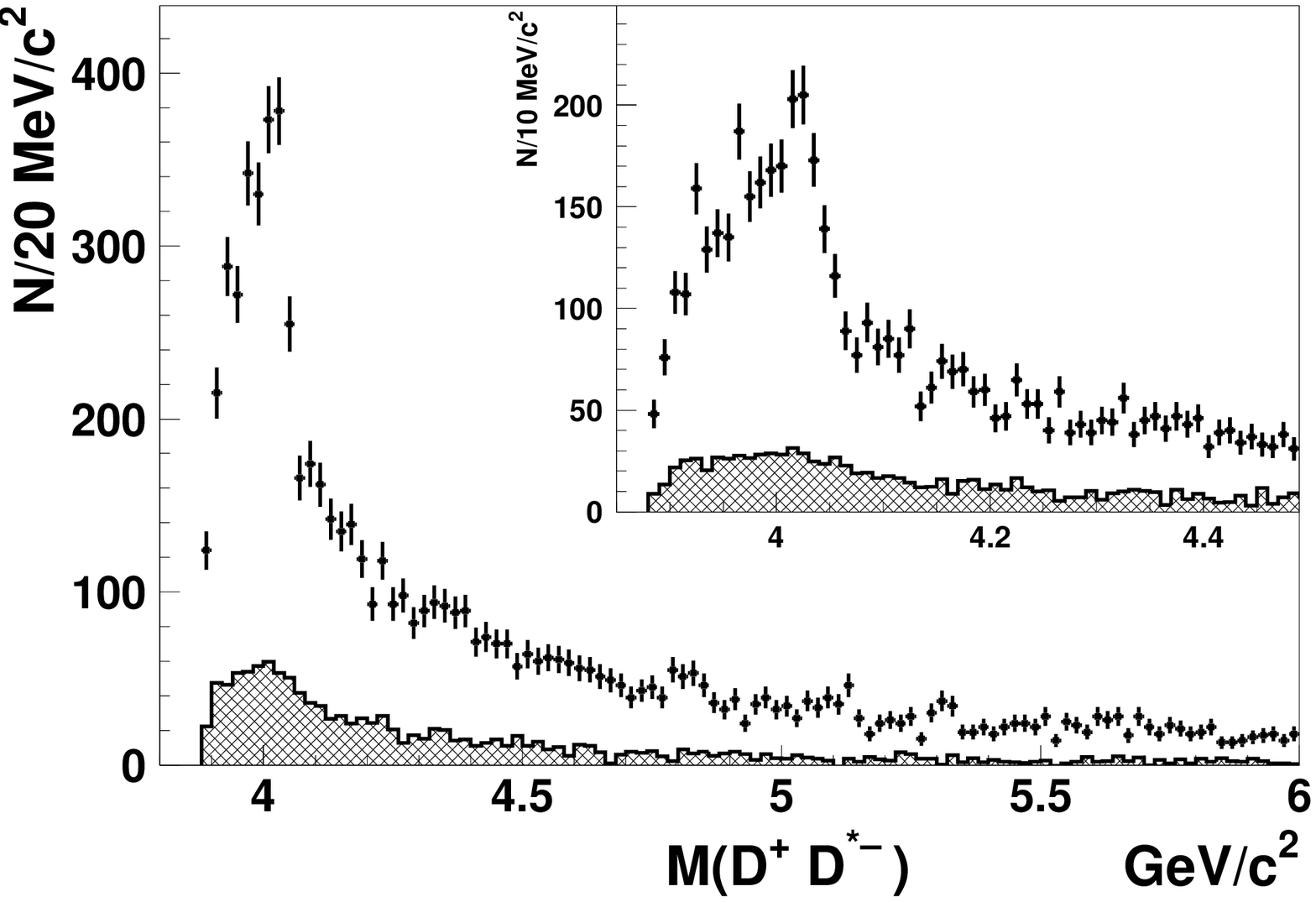}
\caption{The $M(\DDst)$ spectrum in the data after applying all
  selection criteria (points with error bars). The sum of backgrounds
  (1)-(3) is shown as the hatched histogram. The inset shows the
  spectrum near the threshold with finer bins}.
\label{mdd_bg_a}
\end{figure}

For $\ee \to \DDst \gisr$, we consider similar background sources
and use the same methods to subtract them:
\begin{enumerate}
\item[1.] combinatorial \Dp\ candidate combined with a true slow pion
  from $D^{*-}$ decay;
\item[2.] real \Dp\ mesons combined with a combinatorial slow pion;
\item[3.] both \Dp and $\pis^-$ are combinatorial;
\item[4a.] reflection from the processes $\ee \to \DDst \pmis^0 \gisr$
  with a lost $\pi^0$ in the final state, including $\ee \to
  \DstDst \gisr$ followed by $\Dstp \to D^+ \pi^0$;
\item[4b.] reflection from $\ee \to \DstDst \gisr$ followed by $\Dstp
  \to D^+\gamma$;
\item[5.] contribution from $\ee \to \DDst \pif^0$, where the fast
  $\pif^0$ is misidentified as \gisr.
\end{enumerate}

The contribution from the combinatorial backgrounds (1)-(3) is
estimated using two-dimensional sideband regions of the \Dp\ candidate
mass versus the recoil mass difference (Fig.~\ref{md_rmd_a}).
\begin{figure}[h!]
\centering
\includegraphics[width=0.48\textwidth]{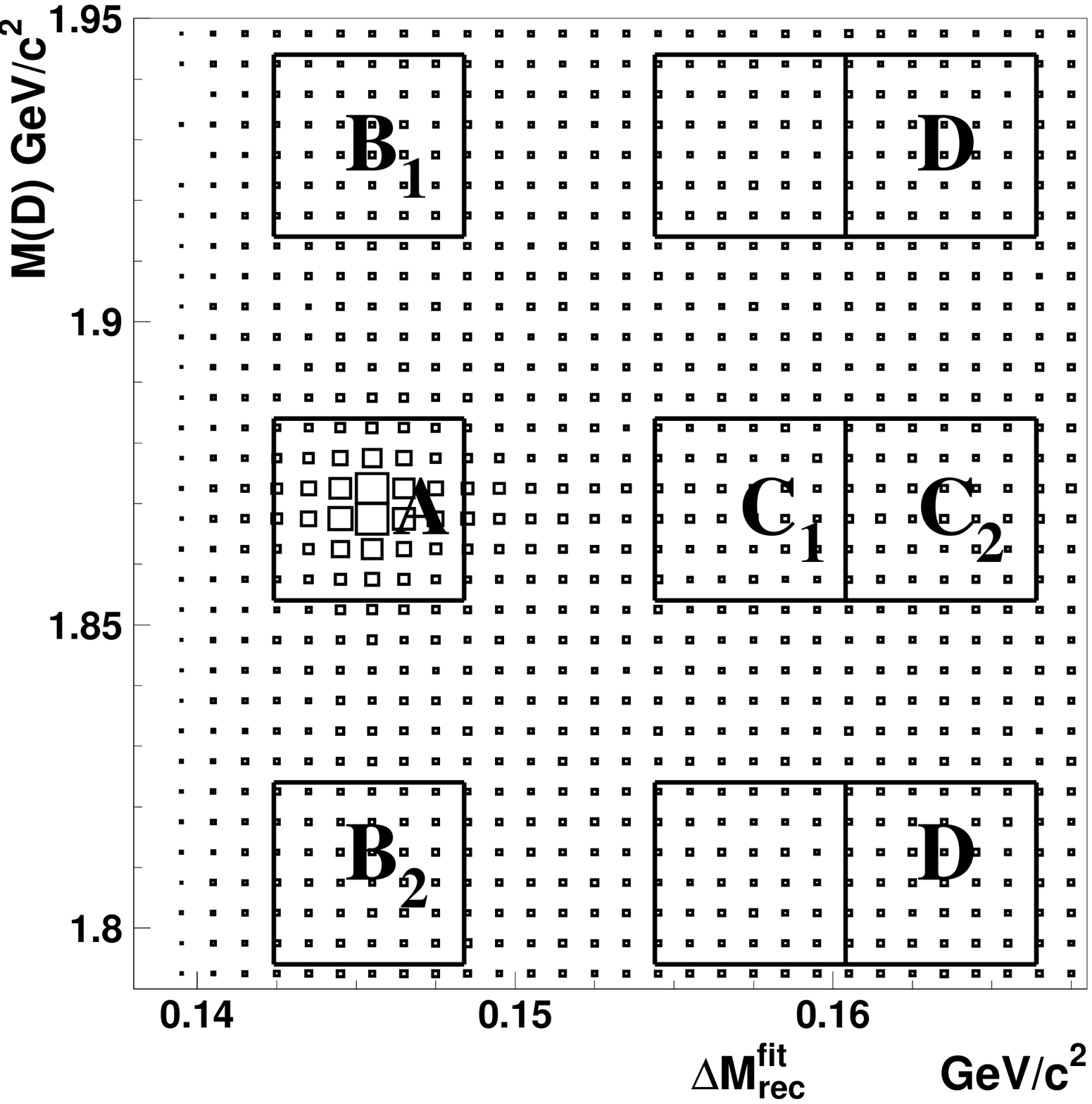} 
\caption{The $M(\Dp)$ {\it vs.} \dmrecf\ scatter plot in the data.}
\label{md_rmd_a} 
\end{figure}

The total contribution to the \Md\ spectra of the combinatorial
backgrounds (1)-(3) is calculated according to the formula
\begin{linenomath*}\begin{equation}
M_{\text{bg~(1)-(3)}}=0.5\cdot M_{\bB}+0.43\cdot M_{\bC}-0.215\cdot
M_{\bD},
\label{bg2}
\end{equation}\end{linenomath*}
where $M_{\bB,~\bC,~\bD}$ are the \DDst\ mass spectra from the \bB,
\bC, and \bD\ sidebands, respectively. The scaling factors are
determined from the fits to the $M(\Dp)$ and \dmrecf\ distributions
shown in Fig.~\ref{fit_a}. In the case of the $M(\Dp)$
sidebands, the scaling factor turns out to be exactly related to
the ratio of widths for the sideband and signal windows, as the
background is well described by a linear function. We note that
background (4) contributes to the \dmrecf\ spectrum peak. However,
this peak is wider than the signal's as the refitting procedure
of the recoil mass against $\Dap \gisr$ into the \Dstm\ mass for this
background process works improperly and does not improve the
\dmrecf\ resolution. We fix this contribution (shown in
Fig.~\ref{fit_a}~\ffb\ as the dashed-dotted line) from the MC.

\begin{figure}[h!]
\centering
\includegraphics[width=0.48\textwidth]{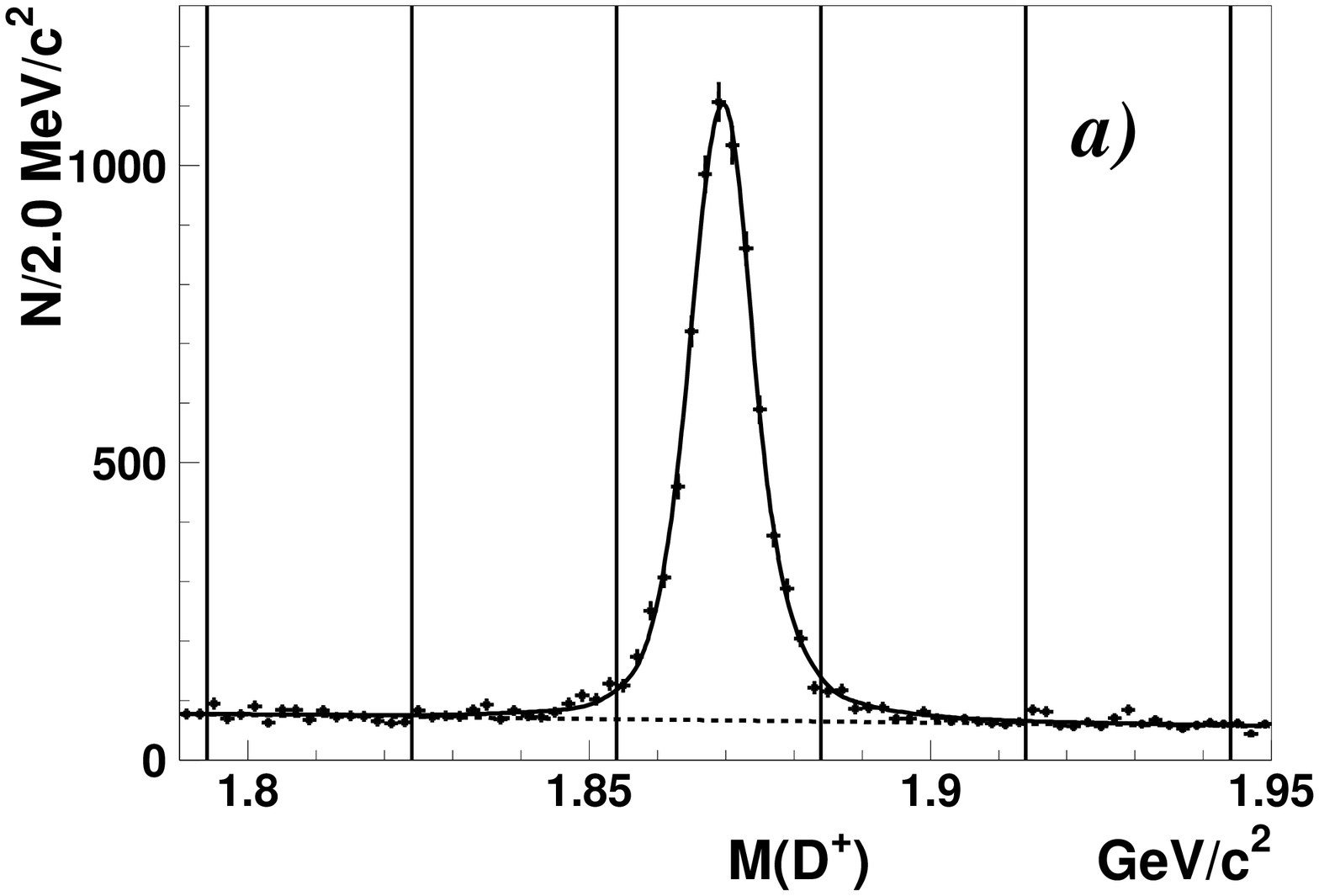} \\[-0.1cm]
\includegraphics[width=0.48\textwidth]{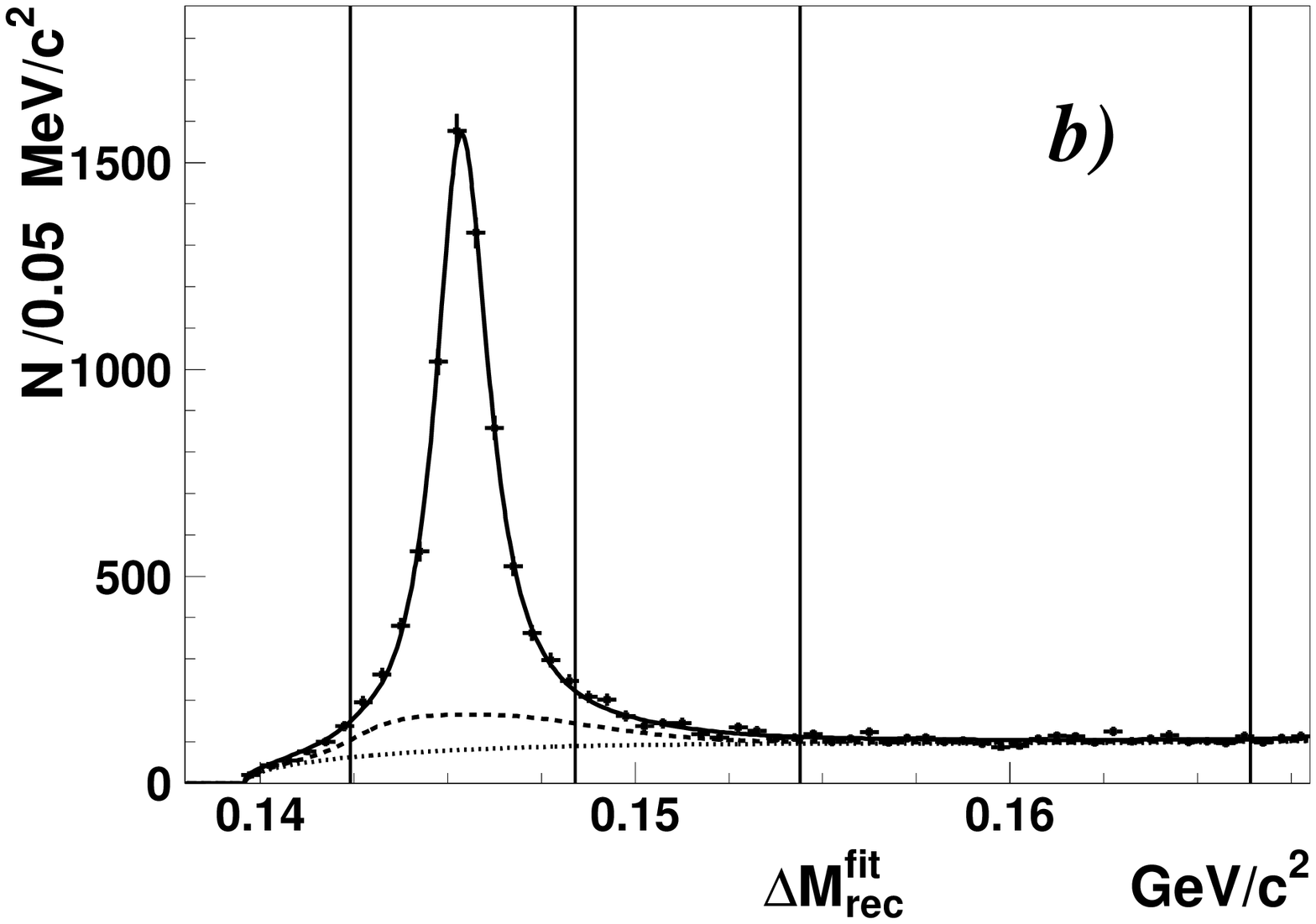}
\caption{Distributions of the \fa\ $M(\Dp)$; \fb\ \dmrecf. The solid
  line represents the result of the fit described in the text. The
  combinatorial background contribution is shown as the dotted
  curve. The dashed curve corresponds to background (4). The signal
  and sidebands regions are indicated by vertical lines.}
\label{fit_a}
\end{figure}

To estimate the contribution of the background process (4a), the
isospin-conjugated process $\ee \to D^{0} D^{*-} \pi^+ \gisr$ is
studied using the same method of partial reconstruction by replacing
$D^+$ with $D^0$. Using the MC simulation, we verify that this method
gives an accurate subtraction of the background without bias. The
measured $M(D^0 D^{*-})$ spectrum is shown in Fig.~\ref{md0d}\ffa\
  as the points with error bars. We repeat the procedure of the
subtraction of combinatorial background (shown by the hatched
histogram in Fig.~\ref{md0d}\ffa\ similarly to the studied process and
obtain the net $M(D^0 D^{*-})$ spectrum as shown in
Fig.~\ref{md0d}\ffb. The background process $\ee \to \DDst \pi^0 \gisr$
includes contributions from $\ee \to \DstDst \gisr$ with $D^{*+} \to
D^+ \pi^0$ as well as higher resonance $D^{**+} \to D^+ \pi^0_{\rm
  miss}$ or non-resonant $D^{+} \pi^0_{\rm miss}$ contributions. For
the two latter sources, the ratio $D^{+} \pi^0$ to $D^{0} \pi^+$ is
related by the isospin factor $1/2$. A similar factor is valid 
  for the ratio of $\Br(\Dstp \to \Dp \pi^0)/\Br (\Dstp \to D^0
\pi^+)=0.453\pm 0.011$~\cite{pdg}. We do not study all of these
contributions separately, but rather scale the measured $D^0
D^{*-}$ spectrum for the ratio of efficiencies for $D^+$ and $D^0$
($0.28$ according to MC) and the factor $0.453$, assuming the main
contribution has $\pmis^0$ coming from \Dstp. The difference
between $\Br(\Dstp \to \Dp \pi^0)/\Br (\Dstp \to D^0 \pi^+)$ and the
isospin factor is taken into account in the systematic error of the
result. The measured contribution from the process $\ee \to \DDst
\pi^0 \gisr$ to the $D^+ D^{*-}$ mass spectrum is shown in
Fig.~\ref{mdd_dst_a} as the open circles.
\begin{figure}[h!]
\centering 
\includegraphics[width=0.48\textwidth]{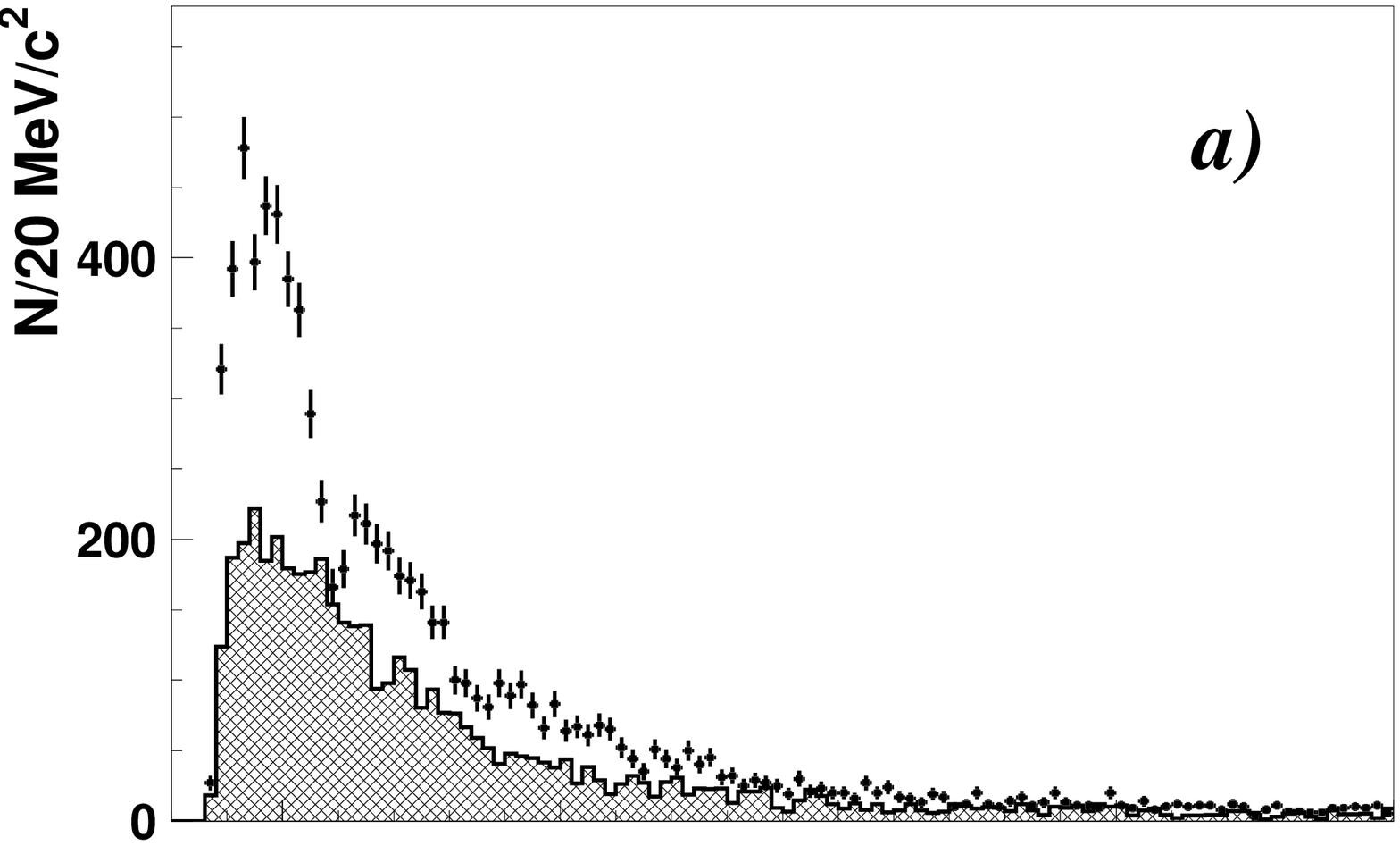} \\[-1.1cm]
\includegraphics[width=0.48\textwidth]{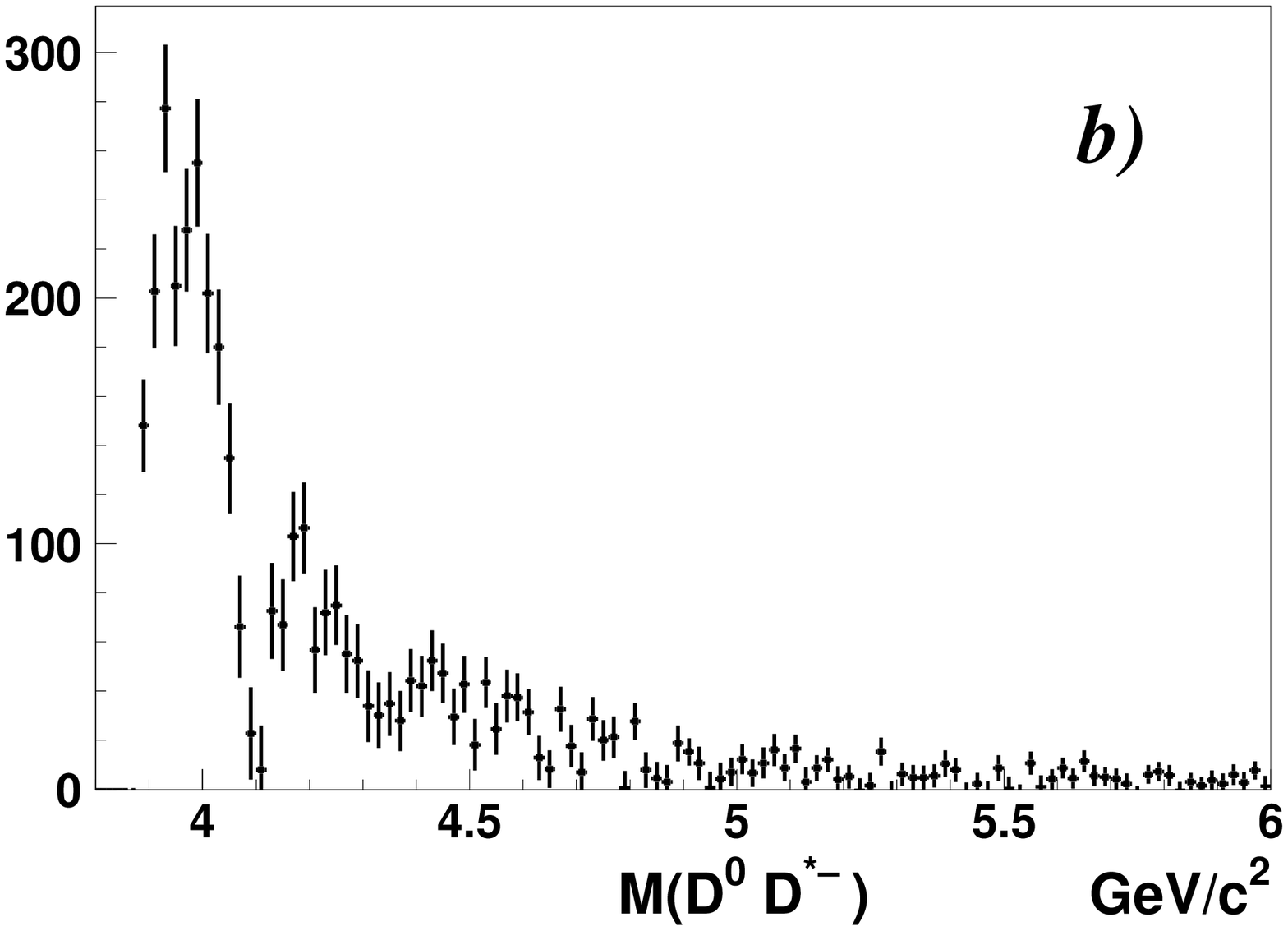}
\caption{The $M(D^0 D^{*-})$ spectrum \fa\ before and \fb\ after
  subtraction of combinatorial background.}
\label{md0d} 
\end{figure}

\begin{figure}[h!]
\centering \includegraphics[width=0.48\textwidth]{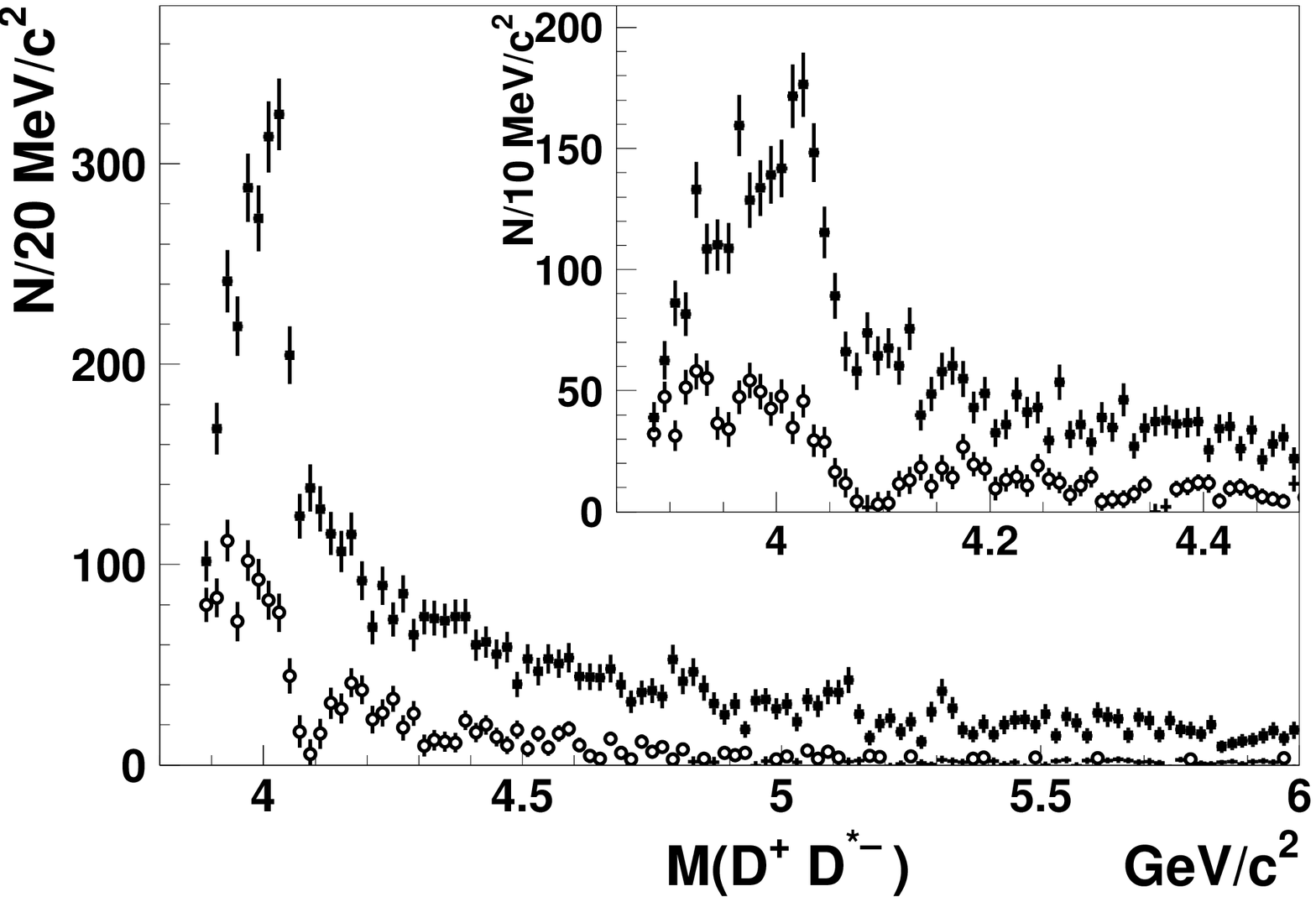}
\caption{The $M(\DDst)$ spectrum after subtraction of combinatorial
  background. Contribution from the $\ee \to \DstDst \gisr$ is shown
  as open circles with error bars.}
\label{mdd_dst_a} 
\end{figure}

The small contribution of background (4b), from $\ee \to \DstDst
\gisr$ with $D^{*+} \to D^+ \gamma$, is estimated using the MC
simulation. In the $\ee \to \DstDst \gisr$ MC sample, the
\DstDst\ spectrum is set according to our measurement. Background
  (4b) is tiny in comparison with (4a) because of the small
$\Br(D^{*+} \to D^+ \gamma)$ branching fraction and a greatly smeared
\dmrecf\ distribution in the case of a lost $\gamma$.

Background (5) is estimated similarly to the study of $\ee\to \DstDst$
and found to be negligibly small. Its contribution is incorporated
into the systematic uncertainty.

\section{Cross-checks}

We performed several checks to determine that the background
subtraction procedure does not bias the measured spectra.

To verify that the combinatorial background subtraction is performed
correctly, each of \bB\ and \bC\ sideband regions is divided into two
equal intervals: $\bB_1$-$\bB_2$ and $\bC_1$-$\bC_2$, respectively, as
shown in Fig.~\ref{md_rmd_a}. Then, the consistency of both the
$M(\DstDst)$ shapes and normalizations in the pairs of subintervals is
checked. The differences between these two pairs of control spectra
are consistent with zero. The same procedure, repeated for the
processes $\ee \to \DDst$, also demonstrates good agreement between
the $M(\DDst)$ spectra from different sidebands regions. We thus
conclude that the combinatorial background shapes and normalizations
are well understood.

We check that, after the energy correction of fast photons, the
spectra of masses recoiling against $\Dap \gisr$ combinations in the
data are consistent with the MC
expectations. Figures~\ref{rmy}\ffa\ and ~\ref{rmy}\ffb\ show the
$\mrec(D^{(*)+} \gisr)$ spectra in the data (points with errors) and
signal MC events (hatched histogram) for the processes $\ee \to \DDst
\gisr$ and $\ee \to \DstDst \gisr$, respectively. The good agreement
between MC and data spectra demonstrates not only the correctness of
the photon energy calibration but also the proper background
subtraction.

\begin{figure}[h!]
\centering 
\includegraphics[width=0.48\textwidth]{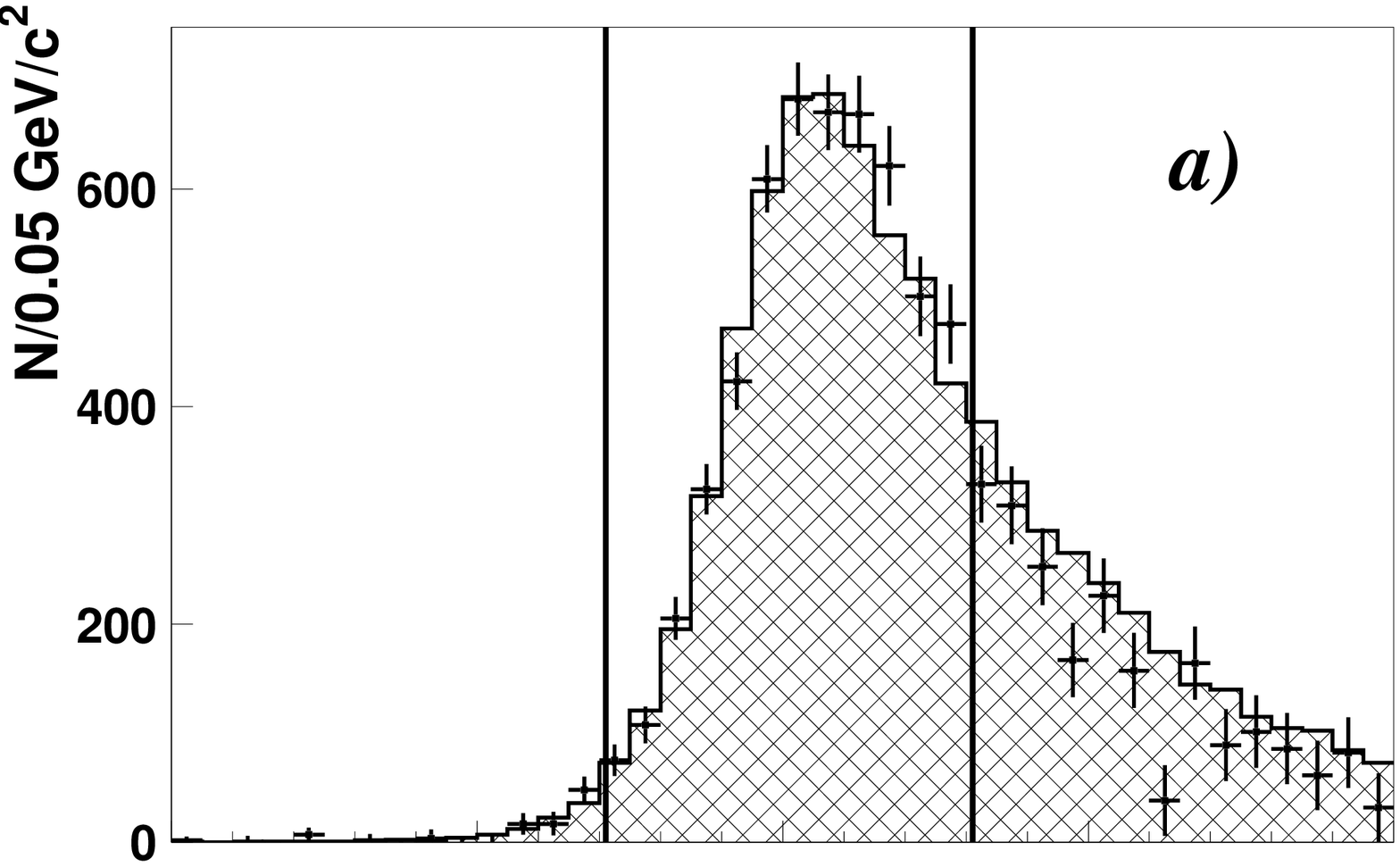} \\[-1.1cm]
\includegraphics[width=0.48\textwidth]{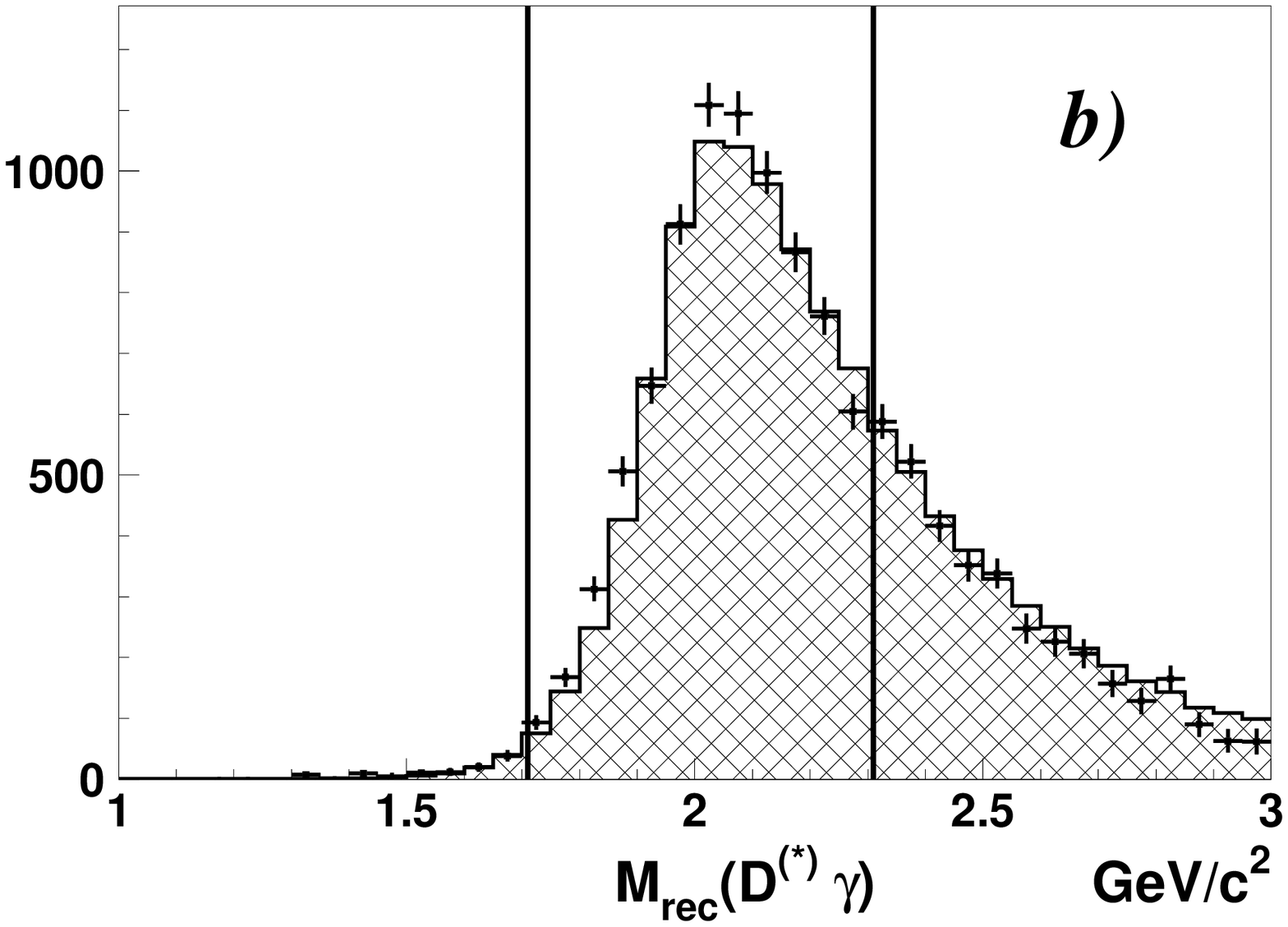}
\caption{\fa\ The $\mrec(\Dp \gisr)$ spectrum after subtraction of the
  combinatorial background and the contribution from the process $\ee
  \to \DstDst\ \gisr$ for $\ee \to \DDst\ \gisr$.\\ \fb\ The
  $\mrec(\Dstp \gisr)$ spectrum after combinatorial-background
  subtraction for $\ee \to \DstDst \gisr$. Points with error bars show
  the data after $\gisr$ energy correction; the hatched histogram
  corresponds to the MC simulation. The signal window lies between
    the vertical lines. }
\label{rmy}
\end{figure}

\section{Cross section calculation}

We calculate the exclusive cross sections of the processes $\ee \to
\DDst$ and $\ee \to \DstDst$ as a function of $\sqrt{s}$ according
to the formula
\begin{linenomath*}\begin{equation}
\sigma_{\ee\ \to\ \DDa}=\frac{dN/dM}{\eta_{\rm tot}(M)\cdot dL/dM},
\end{equation}\end{linenomath*}
where $M$ is the $\DDa$ mass, equivalent to $\sqrt{s}$, $dN/dM$ is the
measured mass spectrum, $\eta_{\rm tot}$ is a $M$-dependent total
efficiency, and $dL/dM$ is the differential luminosity.

The dependence of the efficiency on $M$ is calculated using the MC
simulation and is defined as the ratio of reconstructed mass spectrum
for true MC candidates after applying all requirements to the
generated spectrum (Fig.~\ref{eff}). Alternatively, we calculate the
efficiency without the requirement for the selected MC candidates to
be the true combination, but repeating in this case the procedure of
the background subtraction. The latter method checks for possible
oversubtraction of the signal by the applied procedure of the
combinatorial background subtraction due to tails in the $M(D^{*+})$
and recoil mass difference resolution functions. It is found that the
oversubtraction is negligibly small, and both methods result in the
same $\eta_{tot}(M)$ dependence.

\begin{figure}[h!]
\centering
\includegraphics[width=0.48\textwidth]{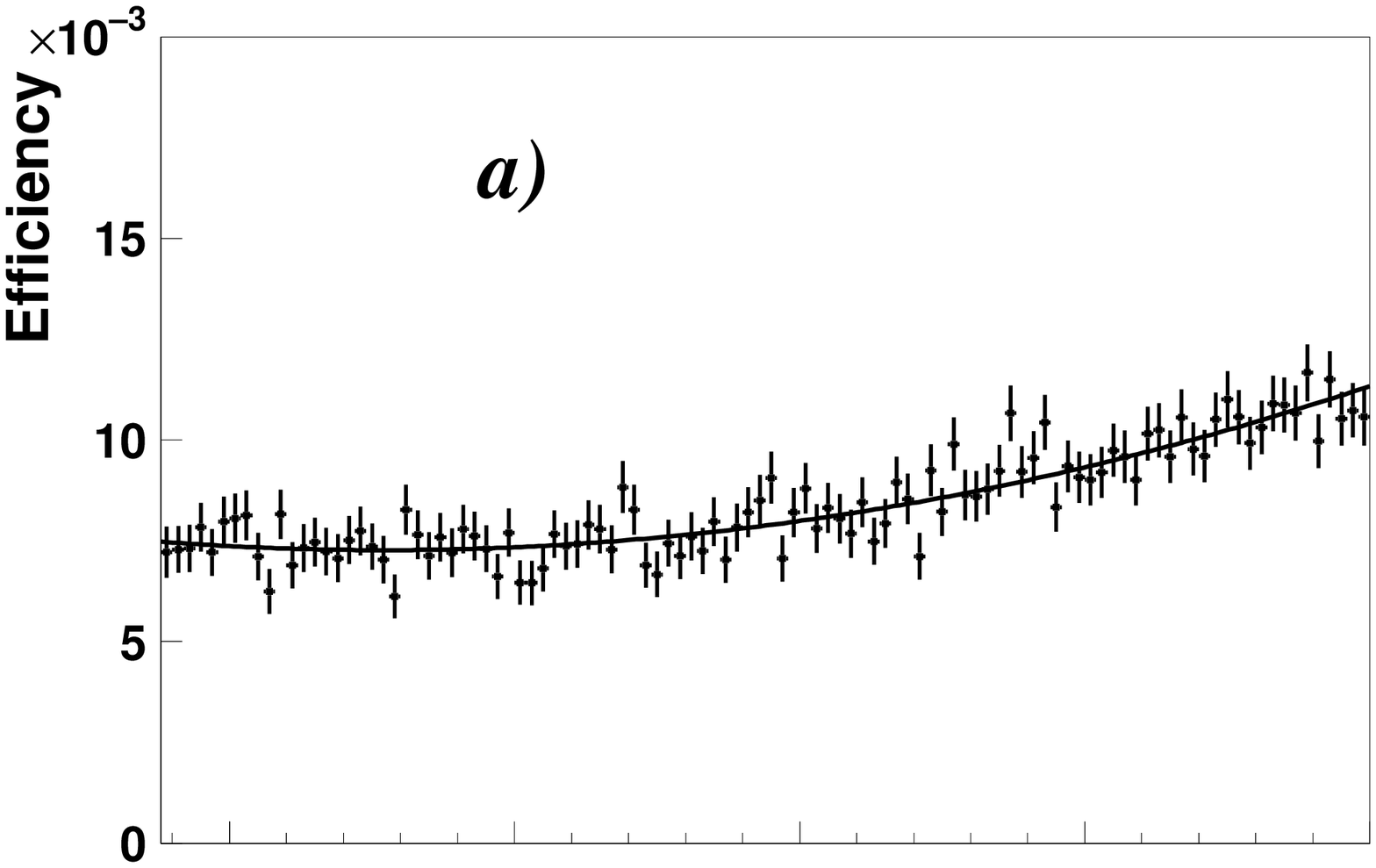} \\[-1.1cm]
\includegraphics[width=0.48\textwidth]{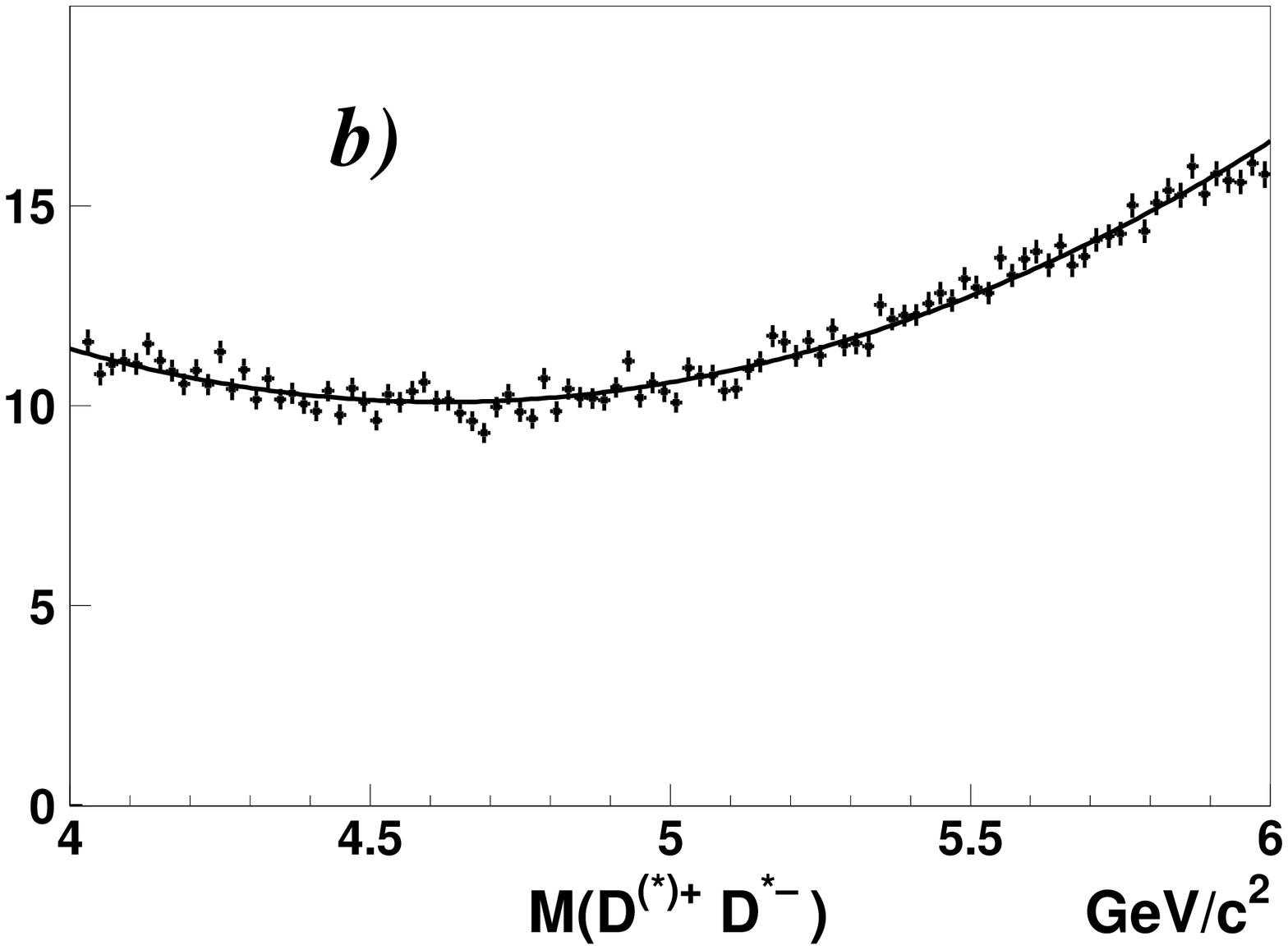}  
\caption{The total efficiency as a function of \Ma\ for \fa\ $\ee \to
  \DDst \gisr$ and \fb\ $\ee \to \DstDst \gisr$.}
\label{eff} 
\end{figure}

The differential luminosity is calculated as a sum over all energy
points --- \Ufo, \Ufi, and continuum --- using the known luminosities of
each data sub-sample. We use the $dL/dM$ formula that includes the
second-order QED corrections~\cite{kuraev}. The latter varies from 2.5
to 3.5\% of the leading contribution in the studied $\sqrt{s}$
interval. In the previous Belle paper, this was treated as a
systematic uncertainty.

Finally, the obtained exclusive $\ee \to \DDst$ and $\ee \to \DstDst$
cross sections are shown in Fig.~\ref{cross}.

\begin{figure}[h!]
\centering
\includegraphics[width=0.48\textwidth]{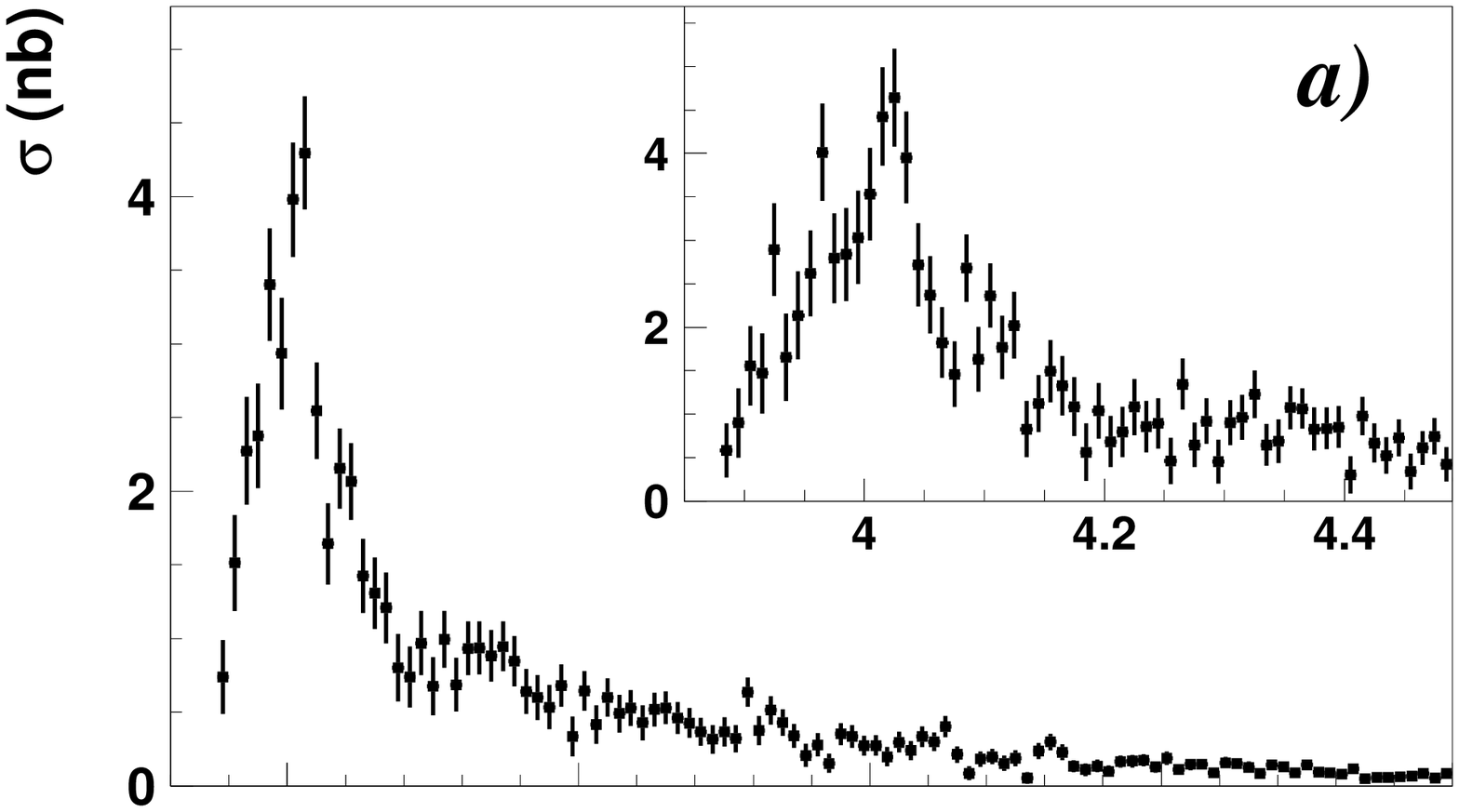} \\[-1.1cm]
\includegraphics[width=0.48\textwidth]{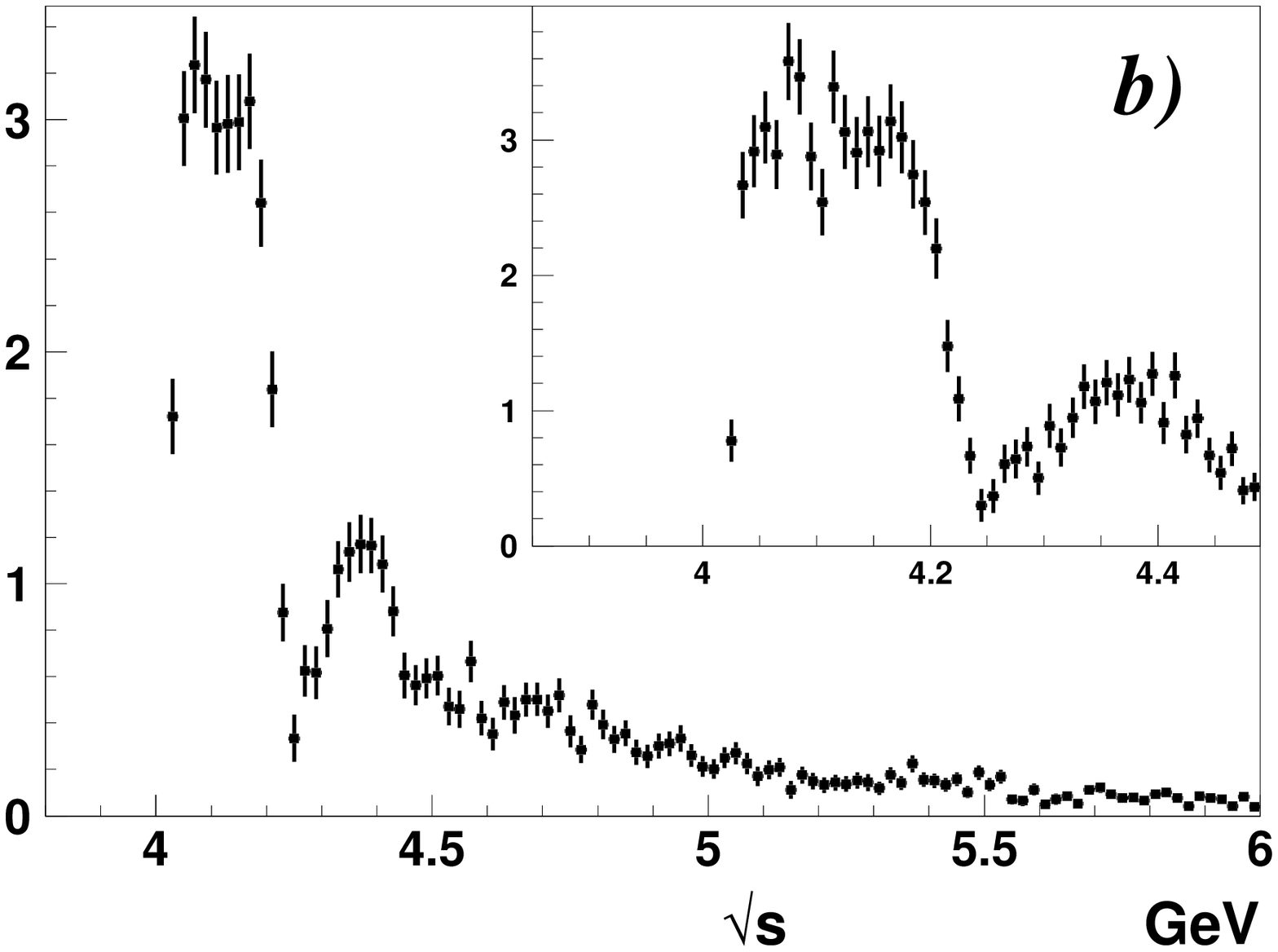} 
\caption{The exclusive cross sections as functions of $\sqrt{s}$ for
  \fa\ $\ee \to \DDst$ and \fb\ $\ee \to \DstDst$.}
\label{cross} 
\end{figure}

\section{Angular analysis}

We study the $D^{*\pm}$ helicities from both processes. The $D^{*\pm}$
helicity angle, $\theta$, is defined as the angle between the
$\pis^\pm$ from $D^{*\pm}$ decay and the \DDa\ system, seen in the
$D^{*\pm}$ rest frame. The angular distributions of the \Dstm\ decays
can be studied even in the case of partial \Dstm\ reconstruction. The
refit procedure of $\mrec(\Dstp \gisr)$ to $m_{\Dstm}$ provides
sufficient accuracy in the unreconstructed \Dstm\ momentum
determination. The helicity angle resolution of the partially
reconstructed $D^{*\pm}$ mesons ($\sigma_\theta\sim 0.06$\,rad) is
slightly worse than for fully reconstructed $D^{*\pm}$ mesons
($\sigma_\theta\sim 0.05$\,rad).

For the $\ee \to \DDst$ process, the helicity of the \Dstm\ meson is
uniquely defined by the angular momentum and parity conservation: the
\Dstm\ meson polarization should be transverse. Thus, we perform
the \Dstm\ angular analysis for this process to verify the method
only.

The \Dstm\ helicity angle distribution is analyzed in each bin of
$M(\DDst)$. Figure~\ref{h_i} illustrates the obtained $\cos{\theta}$
distributions in different $M(\DDst)$ regions by combining several
$M(\DDst)$ bins.
\begin{figure}[h!]
\centering
\begin{tabular}{cc}
\includegraphics[width=0.49\linewidth]{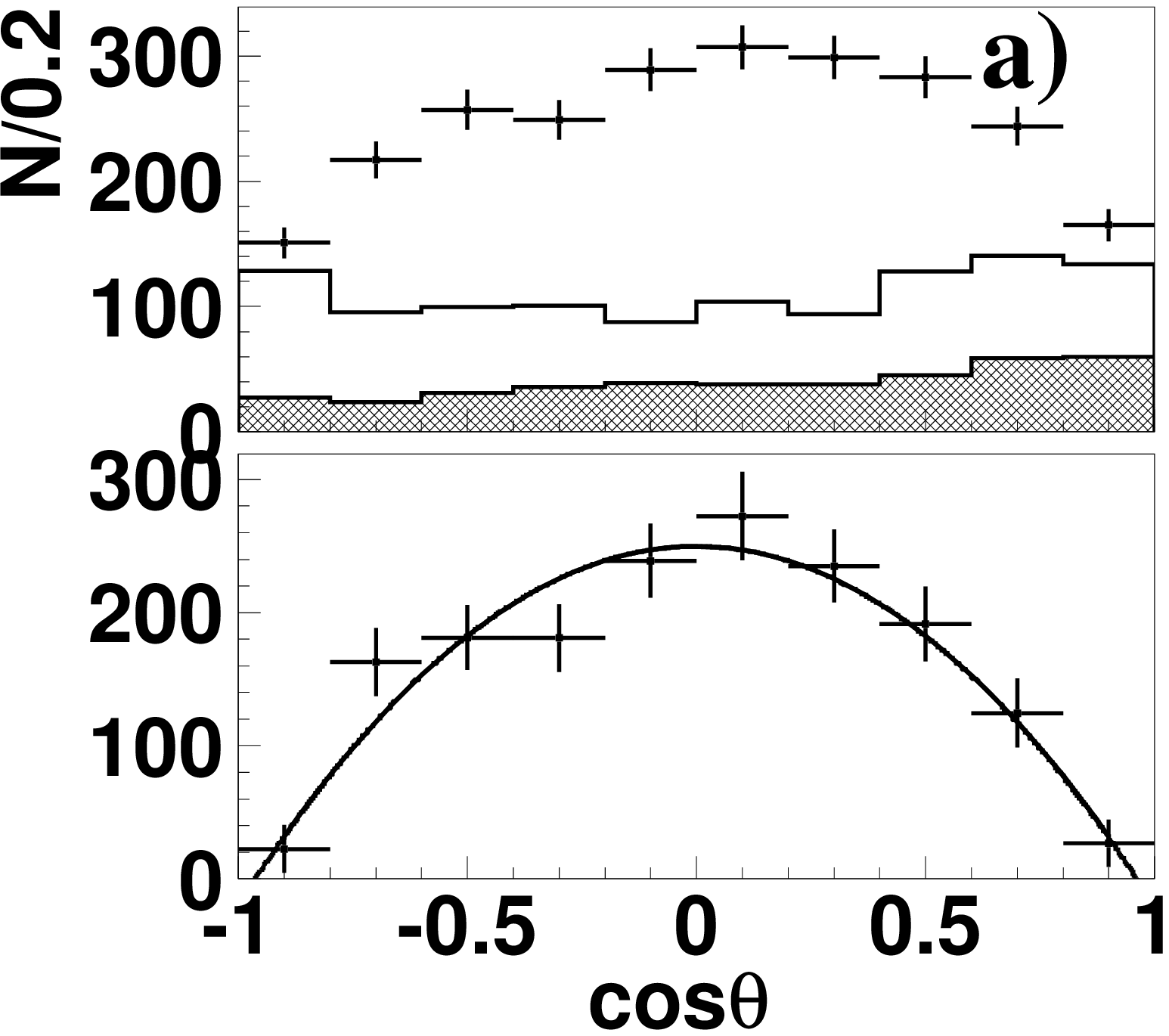} &
\includegraphics[width=0.49\linewidth]{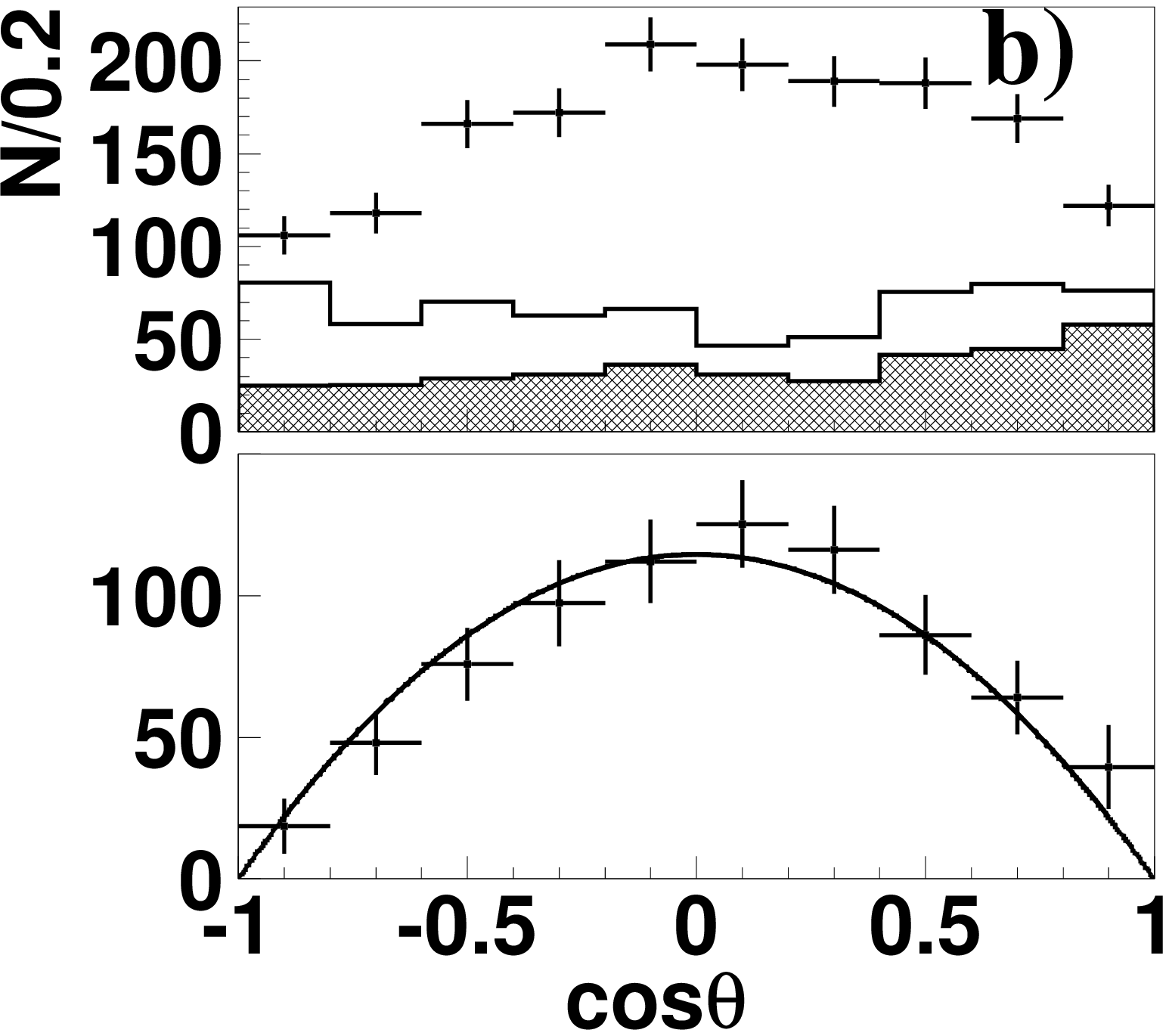} \\ [-0.5cm] 
\includegraphics[width=0.49\linewidth]{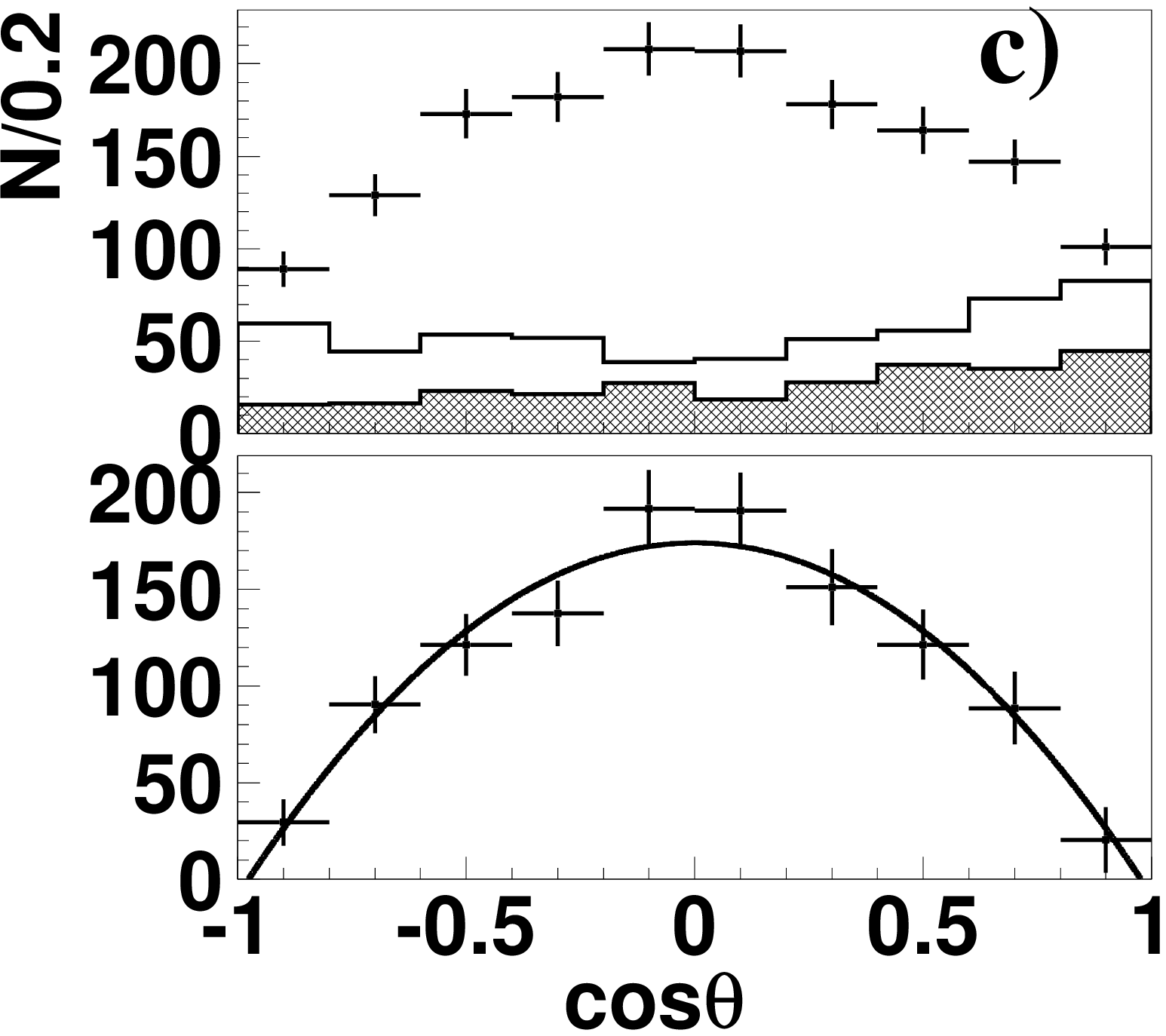} &
\includegraphics[width=0.49\linewidth]{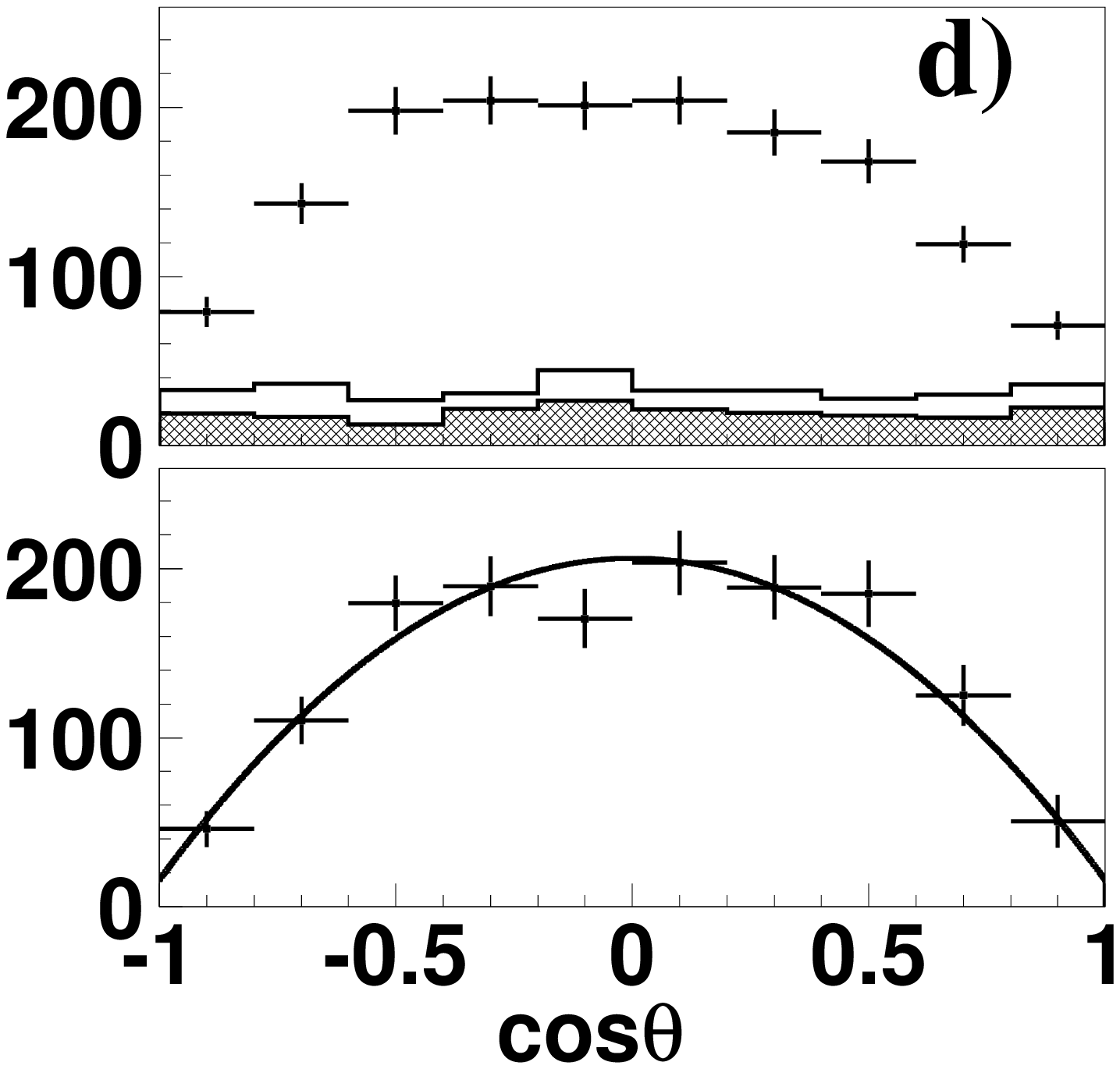} 
\end{tabular}
\caption{The distribution of $\cos{\theta}$ before (top panels)
  and after (bottom panels) background subtraction for different
  mass range: \fa\ $M(\DDst)<4.05 \gevc$, \fb\ $4.05 \gevc
  <M(\DDst)<4.3 \gevc$, \fc\ $4.3 \gevc <M(\DDst)<4.8 \gevc$,
  \fd\ $M(\DDst)>4.8 \gevc$. The contribution from the combinatorial
  background only is shown as the hatched histogram. The total
  contribution from the backgrounds sources is shown as the open
  histogram. }
\label{h_i} 
\end{figure}
Here, the combinatorial background contribution, taken from the $M(\Dp)$
and \dmrecf\ sidebands, is shown as the hatched histogram, and the sum
of the combinatorial background and $\Dstp\Dstm$ feed-down is shown as
the open histogram; the latter contribution is obtained from the $D^0
D^-$ data, corrected for the $D^0$ and $D^-$ efficiency ratio and the
ratio of $\Br(\Dstp \to \Dp \pi^0)/\Br (\Dstp \to D^0 \pi^+)$. In each
bin of $M(\DDst)$, the $\cos{\theta}$ distribution after subtraction of
all background contributions is fitted with the function
\begin{linenomath*}\begin{equation}
F(\cos{\theta})=\eta(\cos{\theta}) \cdot dM/dL \cdot (f_{L}+f_{T}),
\end{equation}\end{linenomath*}
where $\eta(\cos{\theta})$ is the efficiency depending on the helicity
of the \Dstm, $dL/dM$ is calculated in each $M$ bin, and
$f_{L}=\sigma_L \cdot \cos^2{\theta}$ and $f_{T}=\sigma_T \cdot
(1-\cos^2{\theta})$ are the contributions from distinct
\Dstm\ helicity states. The subscripts $L$ and $T$ refer to
longitudinally and transversely polarized $D^{*\pm}$ mesons,
correspondingly. The fitting procedure scans over $M$ bins and the
fits return the cross sections $\sigma_{T,L}(\sqrt{s})$ for $T$ and
$L$ components. Figure~\ref{h1} shows the resulting cross sections for
$\ee \to \DDst_T$ and $\DDst_L$. The latter is consistent with zero,
as expected.

\begin{figure}[h!]
\centering
\includegraphics[width=0.48\textwidth]{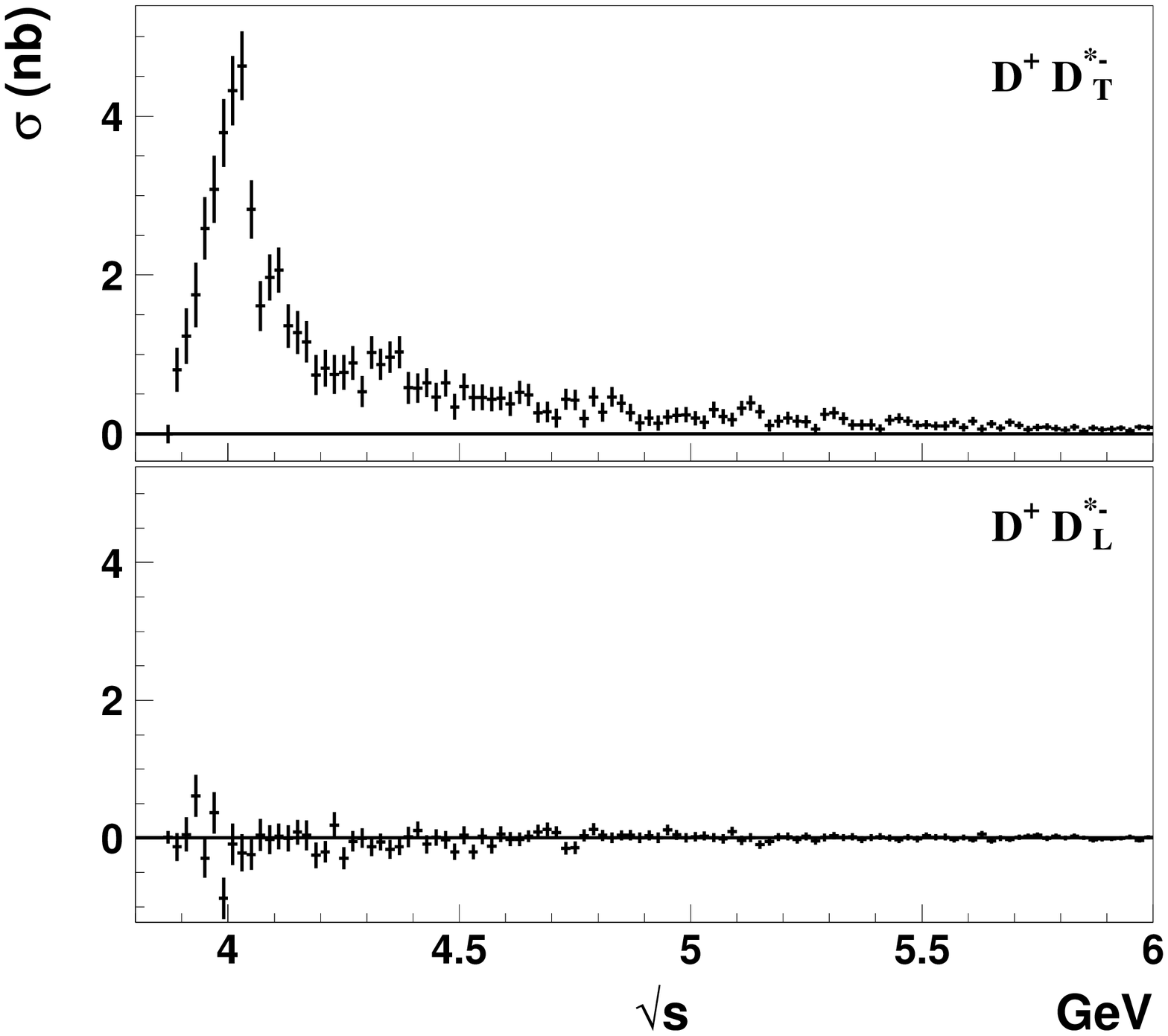} 
\caption{The components of the $\ee \to \DDst \gisr$ cross section
  corresponding to the different $D^{*-}$ helicity: top is transverse;
  bottom is longitudinal.}
\label{h1}
\end{figure}

For the process $\ee \to \DstDst$, according to theoretical predictions
for the processes $\ee\ \to \DstDst$~\cite{hel_form}, the helicity
composition of the $\DstDst$ final state is a mixture of $\Dst_T
\Dst_T$, $\Dst_T \Dst_L$ and $D^{*}_L D^{*}_L$. We perform a study of
the $D^{*\pm}$ helicity angle distribution in each bin of \Mdst.

\begin{figure}[h!]
\centering
\begin{tabular}{cc}
\includegraphics[width=0.22\textwidth]{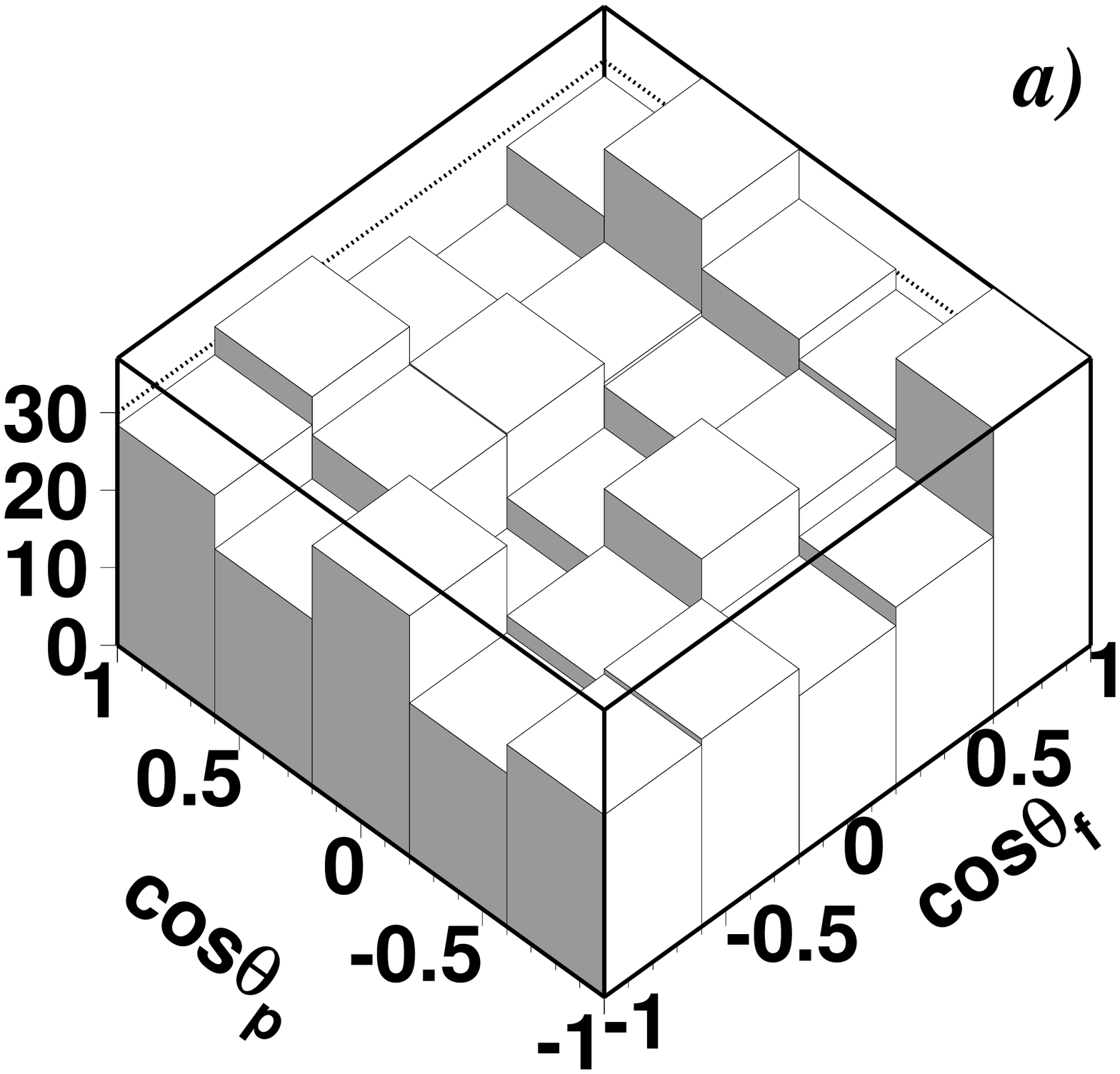} &
\includegraphics[width=0.22\textwidth]{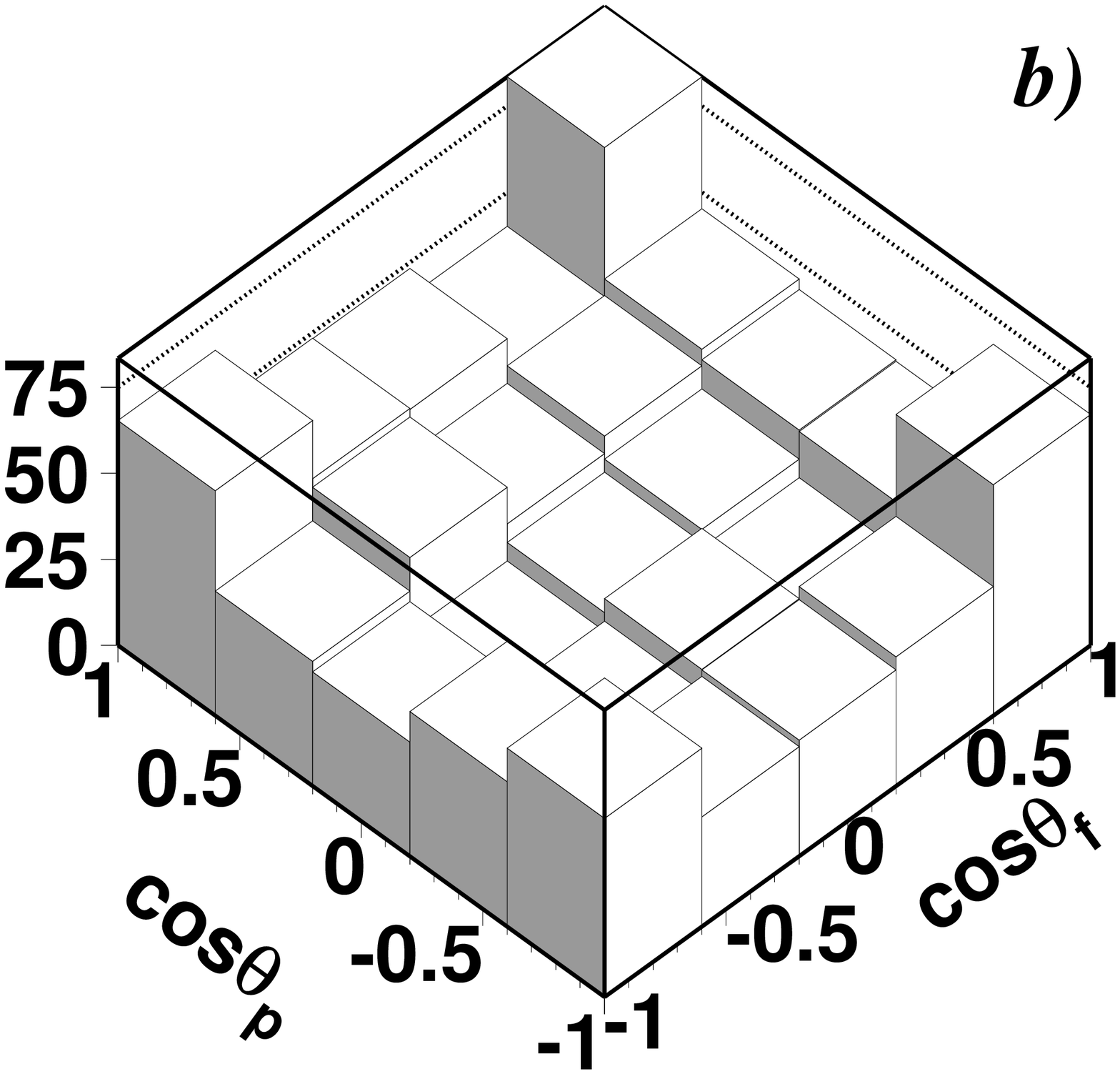} \\[-0.4cm]
\includegraphics[width=0.22\textwidth]{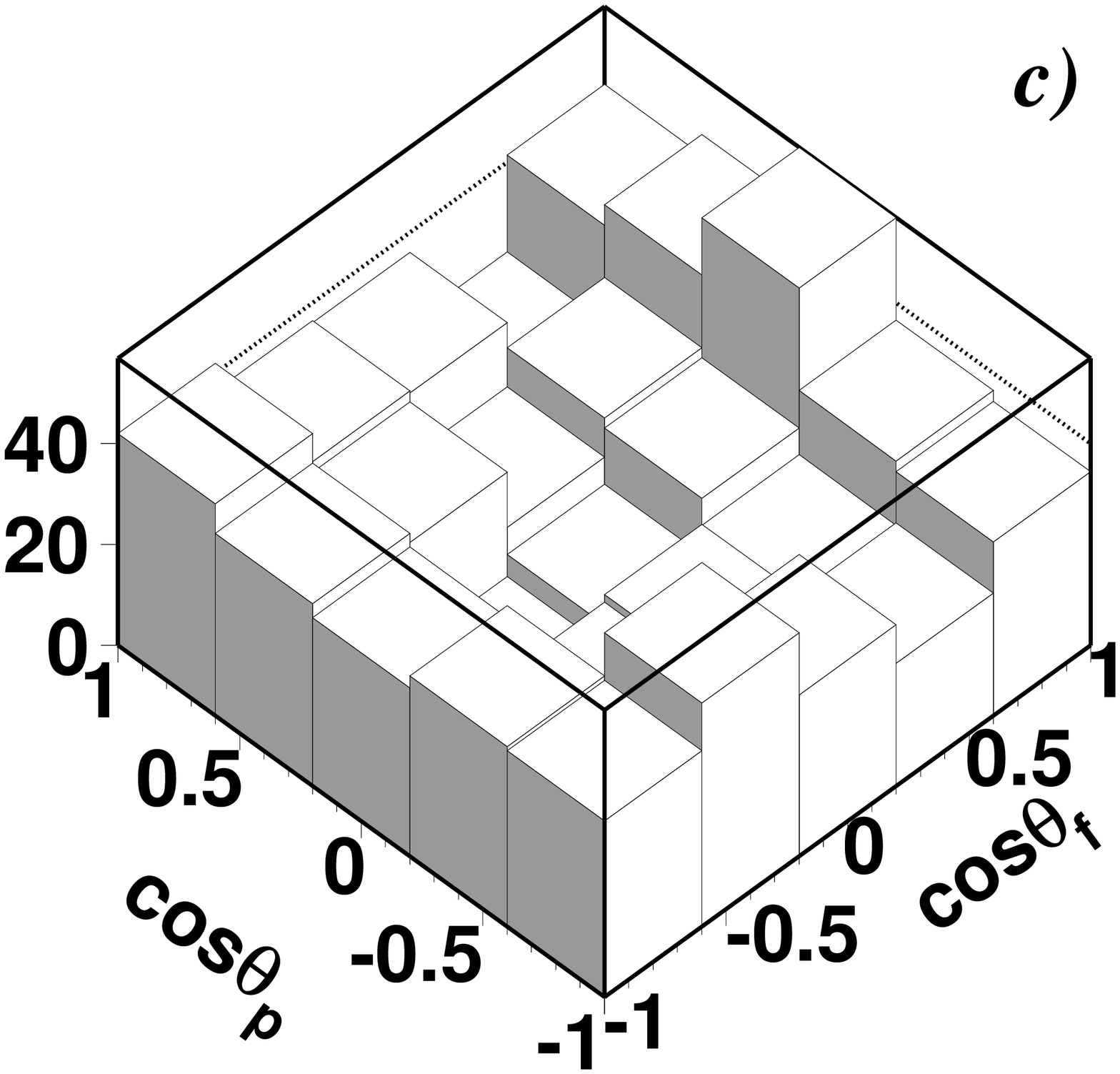} &
\includegraphics[width=0.22\textwidth]{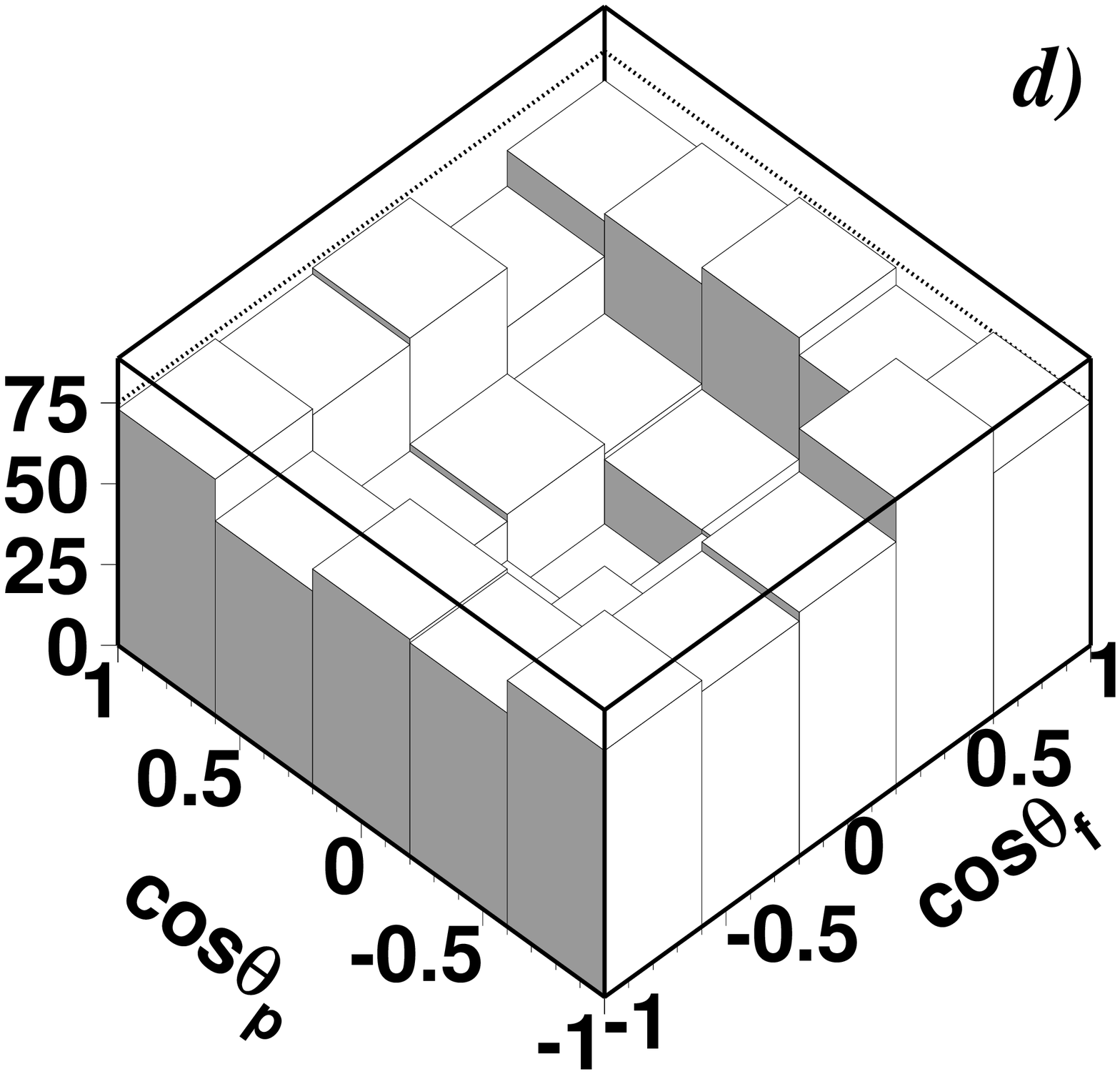} \\
\end{tabular}
\caption{Two-dimensional distribution of $\cos{\theta_f}$ {\it vs.}
  $\cos{\theta_p}$ for different mass ranges: \fa\ $4.0 \gevc
  <M(\DstDst)<4.1 \gevc$, \fb\ $4.1 \gevc <M(\DstDst)<4.25 \gevc$,
  \fc\ $4.25 \gevc <M(\DstDst)<4.6 \gevc$, \fd\ $M(\DstDst)>4.6
  \gevc$.}
\label{c1c2} 
\end{figure}

We analyze the two-dimensional distribution of $c_1 \equiv
\cos{\theta_f}$ \textit{vs.}  $c_2 \equiv \cos{\theta_p}$, where the
first helicity angle, $\theta_f$, corresponds to the fully
reconstructed \Dstp, and the second, $\theta_p$, is calculated for the
partially reconstructed \Dstm. As an illustration, the two-dimensional
distributions of $c_1$ versus $c_2$ for four mass ranges are
shown in Fig.~\ref{c1c2}.

We perform binned maximum likelihood fits to these distributions in
bins of \Mdst. The fitting function is an incoherent sum of the $LL$ and
$TL$ and $TT$ contributions and the background component:
\begin{linenomath*}
\begin{equation}
f=\eta(c_1,c_2)\cdot\ dL/dM \cdot(f_{LL}+f_{TL}+f_{TT}) +f_{\rm bg}.
\end{equation}\end{linenomath*}
Here, $\eta(c_1,c_2)$ is the efficiency, depending on the two $D^*$
helicity angles, and $f_{LL}$, $f_{TL}$, $f_{TT}$, and $f_{\rm bg}$
are the contributions from the three mutually orthogonal signal
components and from the background, respectively. In this study, we
take into account the combinatorial background only, and ignore the
contributions of backgrounds (4) and (5) because they are very
small. We define:
\begin{eqnarray}
\begin{aligned}
f_{TT}= & ~ \sigma_{TT} \cdot (1-c_1^2) \cdot (1-c_2^2), \nonumber \\
f_{TL}= & ~ \sigma_{TL} \cdot \big( (1-c_1^2) \cdot c_2^2 +
c_1^2 \cdot (1-c_2^2) \big),  \nonumber \\
f_{LL}= & ~ \sigma_{LL} \cdot c_1^2 \cdot c_2^2,  \\
f_{\rm bg}= & ~ N_{\rm bg} \big( 0.58 \cdot f_{\bB}(c_1,c_2) + \nonumber \\ 
& \qquad \quad  0.53 \cdot
f_{\bC}(c_1,c_2) - 0.307 \cdot f_{\bD}(c_1,c_2) \big) \, ,  \nonumber
\label{helic}
\end{aligned}
\end{eqnarray}
where $\sigma_{LL}$, $\sigma_{TL}$, and $\sigma_{TT}$ are free
parameters in the fit that represent the cross sections of each
component in $M$ bins. The $f_{\bB, ~\bC,~\bD}(c_1,c_2)$ are elements
of the matrices that are constructed from the binned $c_1$ \textit{vs.}
$c_2$ histograms for the corresponding sideband regions; the
background normalization, $N_{\rm bg}$, is fixed in each \Mdst\ bin
within the error.

The efficiency map is shown in Fig.~\ref{eff_map}\,\ffa. The efficiency
is almost symmetric with respect to $c_1$ {\it vs.}\ $c_2$, as it
depends mainly on the momenta of the two slow pions. The combinatorial
background $f_{\rm bg}$ is presented in Fig.~\ref{eff_map}\,\ffb.

\begin{figure}[h!]
\centering
  \includegraphics[width=0.48\textwidth]{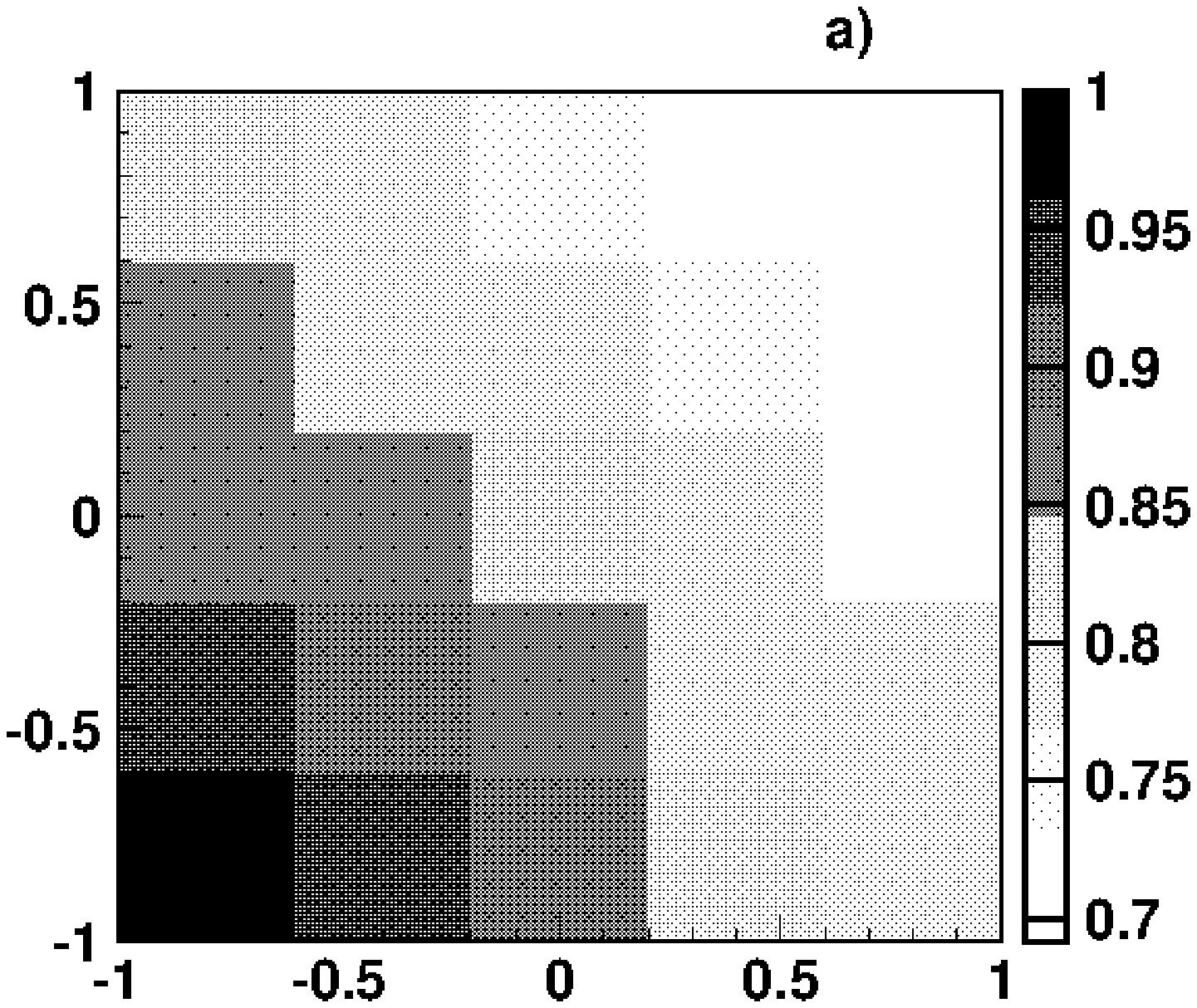}\\
  \includegraphics[width=0.48\textwidth]{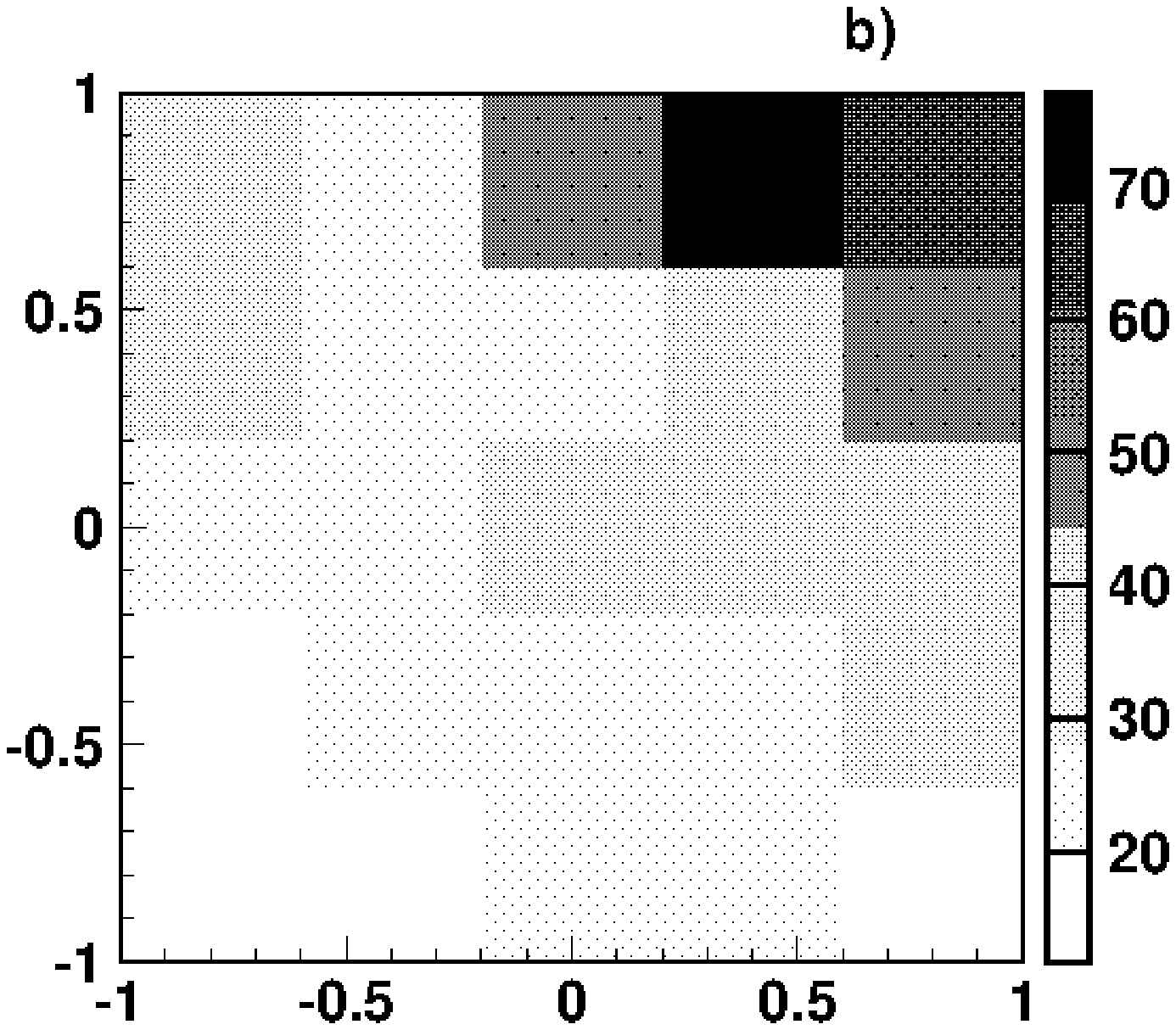}
\caption{The $\ee \to \DstDst \gisr$ process: \fa\ reconstruction
  efficiency as a function $c_1$ and $c_2$; \fb\ total combinatorial
  background distribution.}
\label{eff_map} 
\end{figure}

\begin{figure}[h!]
\centering
\includegraphics[width=0.99\linewidth]{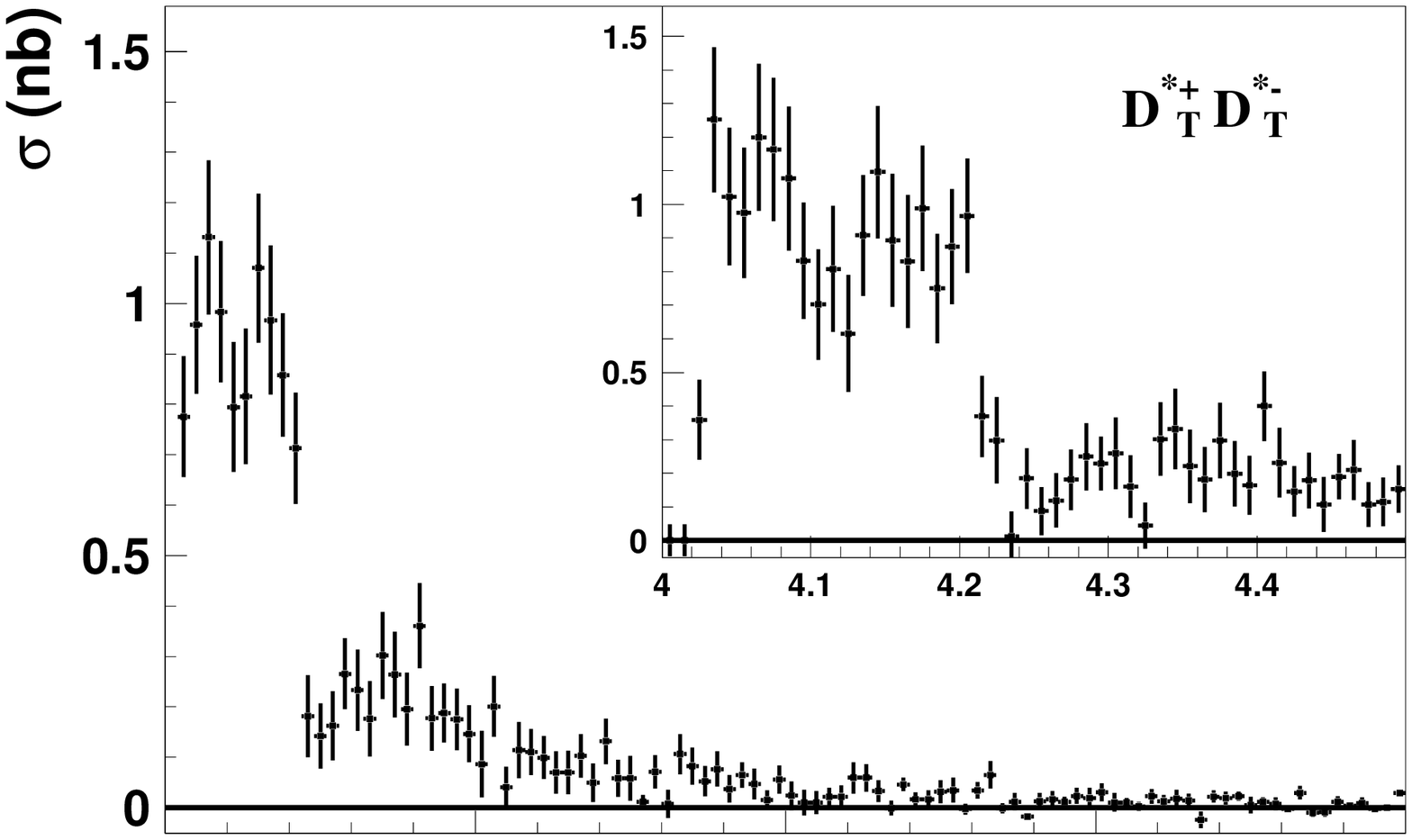} \\[-1.11cm]
\includegraphics[width=0.99\linewidth]{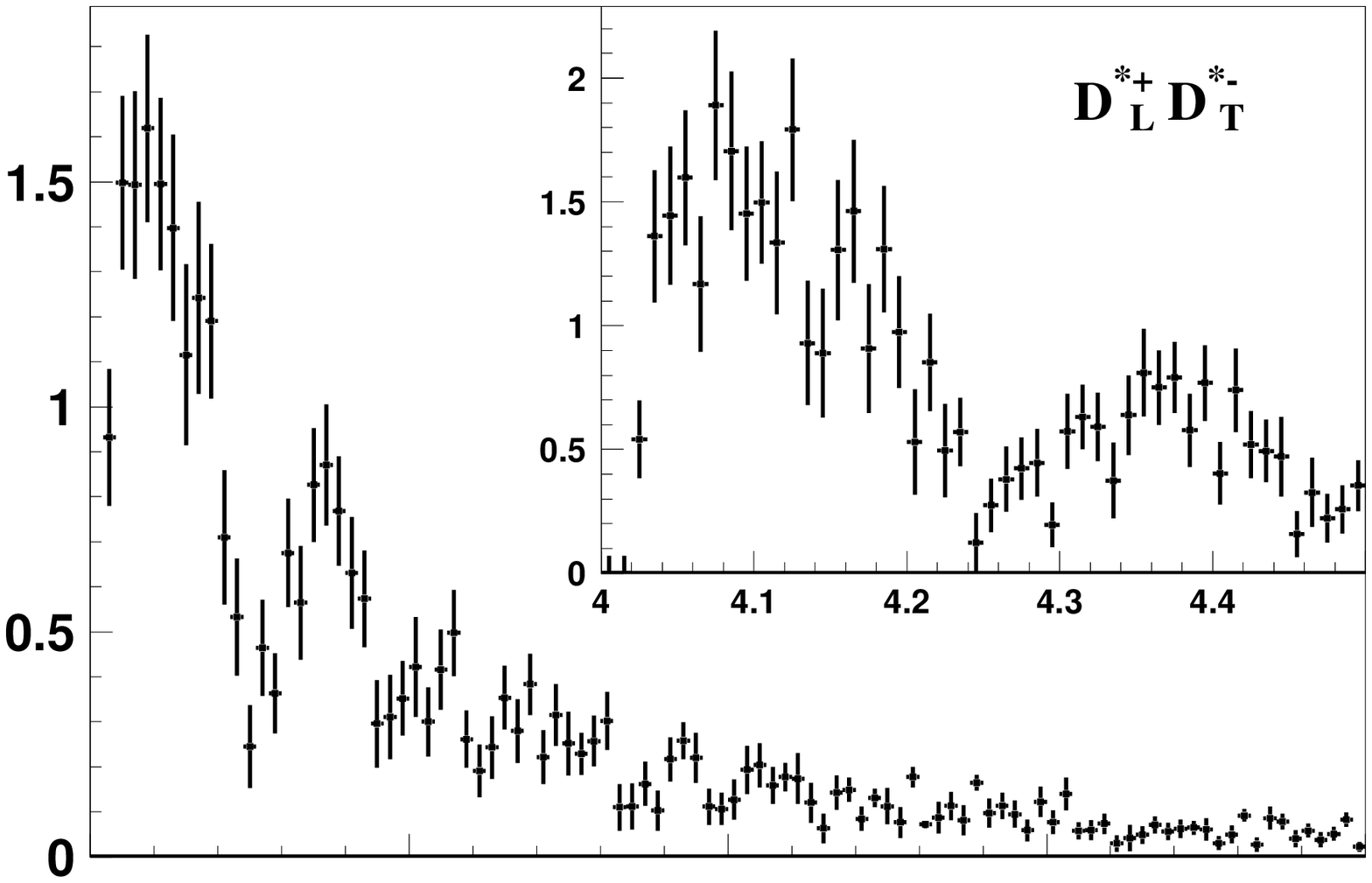} \\[-1.11cm]
\includegraphics[width=0.99\linewidth]{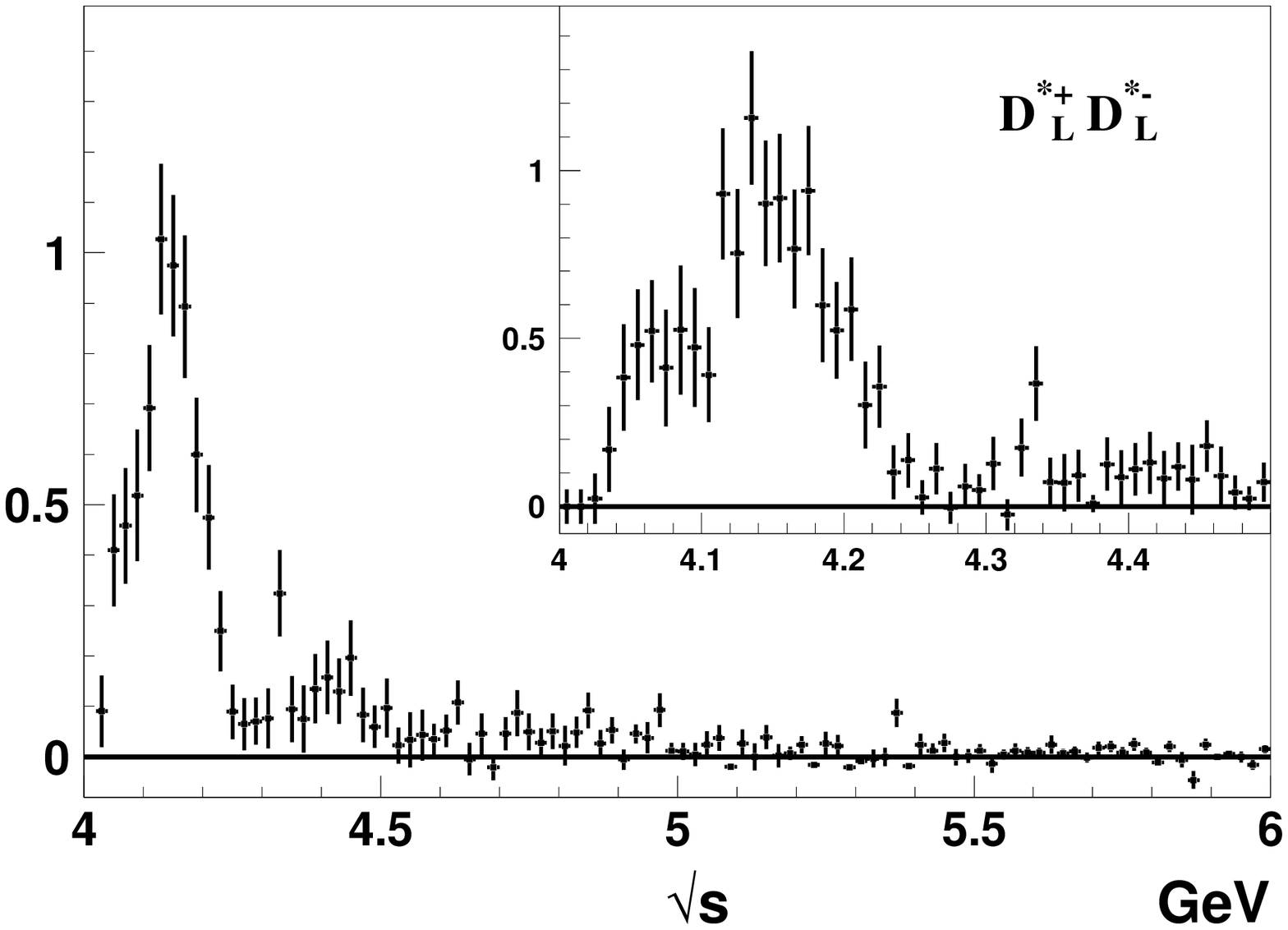} 
\caption{The components of the $\ee \to \DstDst \gisr$ cross section
  corresponding to the different $D^{*\pm}$'s helicities.}
\label{cross_hel}
\end{figure}

The angular-fit results are plotted in Fig.~\ref{cross_hel}. We
observe that the cross sections corresponding to the different
$\DstDst$ helicities have distinct dependencies on $\sqrt{s}$. Near
the threshold, the $TT$ and $TL$ components have a similar sharp rise,
while the $LL$ component rises slowly. This can be explained by the
high centrifugal barrier for the $LL$ component, which originates from
the $F$ wave (chapter 48 of Ref.~\cite{pdg}). All three components
reach the same value of $\sim 1\,\mathrm{nb}$ at $\sqrt{s}\sim
4.15\gev$ and fall into a common dip at $\sqrt{s}\sim 4.25\gev$. The
$LL$ and $TT$ components are attenuated and only the $TL$ component
survives in the region of high $\sqrt{s} \gtrsim 4.5\gev$, which is in
good agreement with theoretical predictions~\cite{grozin}.

\section{Systematic errors}

The systematic errors in the cross section calculation for the studied
processes are summarized in Table~\ref{sys}.


\begin{table}[h]
\begin{ruledtabular}
\caption{The summary of the systematic errors in the cross section
  calculation.}\label{sys}
\begin{tabular}{l|c|c}
Source & $\DDst$ & $\DstDst$ \\ \hline Background subtraction & $ 2
\%$ & $ 2 \%$ \\ Reconstruction & $ 3 \%$ & $ 4 \%$ \\ Selection & $ 1
\%$ & $ 1 \%$ \\ Angular distribution & $-$ & $ 2 \%$ \\ Cross section
calculation~~ & $ 1.5 \%$ & $1.5 \%$ \\ $\Br(D^{(*)})$ & $ 2 \%$ & $ 3
\%$ \\ MC statistics & $ 1 \%$ & $ 2 \%$ \\ \hline Total & $ 5 \%$ & $
7 \%$ \\ 
\end{tabular}
\end{ruledtabular}
\end{table}

The systematic errors arising from the background subtractions include
the uncertainty in the calculation of the scaling factors for the
sideband distributions in Eqs.~\ref{bg1} and ~\ref{bg2}, systematic
errors in the determination of background (4) with missing neutrals
and unsubtracted background (5). To estimate the uncertainty in the
calculation of the scaling factors, we perform fits to
Figs.~\ref{fit_b} and ~\ref{fit_a} with different parameterizations
and in different $M(\DstDst)$ intervals. As a result, the scaling
factor extracted from the integral under the signal and sideband
regions varies within $\pm15\%$. In spite of large uncertainties in
this scaling factor, the final systematic error due to the subtraction
of combinatorial backgrounds (1)--(3) is estimated to be only 2\% as
these backgrounds are small (only 15\% of the signal). The systematic
error associated with subtraction of background (4) includes the
uncertainty of the isospin-conjugated spectra and efficiency ratio for
the isospin-conjugated final states. The upper limit on
background (5) is considered as an extra contribution to the
systematic error.

The main contribution to the total systematic error comes from the
uncertainties in track and photon reconstruction, estimated as 0.35\%
per track and 1.5\% per photon. An extra uncertainty of 1\% is
ascribed to slow pion(s), and 2\% to \ks\ reconstruction. Uncertainty
of the kaon identification efficiency is 1\%, evaluated by the study
of inclusive $D^*$ mesons.

The systematic error of the selection efficiency comes from the
possible difference of resolution and calibration in the MC and data.
Particularly, the calibration of the fast photon can influence the
efficiency of the requirement $|\mrec(\Dap \gisr)-m_{\Dstm}|<300
\mevc$. From the error in the photon energy correction factor obtained
in our study, we estimate this uncertainty as $\pm 1\%$. Other
contributions are estimated by comparison of $M(D^{(*)})$ and
\dmrecf\ distributions in the MC and data.

The uncertainty due to the $D^*$ meson helicity distributions are
reduced in comparison to the first Belle analysis~\cite{belle:ddst},
because the angular distributions are analyzed. We set the measured
$D^*$ helicity in the MC simulation to reduce the uncertainty of the
slow pion reconstruction due to angular distribution in $D^*$ decays.

The systematic error ascribed to the cross section calculation is
estimated from a study of the $\cos{\theta_0}$ dependence ($\theta_0$
defines the polar angle range for \gisr\ in the \ee\ c.m. frame:
$\theta_0 <\gisr< 180^\circ-\theta_0$) of the final result and
includes a 1.4\% error on the total luminosity. We add uncertainties
of the same origin linearly, while independent uncertainties for
different types of particles are summed quadratically. Different $D$
modes have slightly different reconstruction uncertainties; the
average error is calculated according to the weight of each mode in
the final sample.

Other contributions come from the uncertainty in absolute the
$D^{(*)}$ branching fractions~\cite{pdg} and MC statistics.

We divide the total systematic errors in two parts: correlated and
uncorrelated.

The $\sqrt{s}$-independent correlated errors come from track
reconstruction, selection efficiency, cross section calculation and
uncertanty from the $D^{(*)}$ branching fractions. These errors
influence the normalization of the measured cross section as a whole.
Therefore, to take into account the correlated errors, the measured
cross section should be multiplied by a factor of $1 \pm\ 0.05$ for
$\ee\ \to\ \DDst$ and $1 \pm\ 0.06$ for $\ee \to \DstDst$.

On the contrary, the uncorrelated errors vary in each $\sqrt{s}$
bin. These errors come from the background subtraction (for $\ee \to
\DDst$) and from calculating efficiency dependence on the helicity of
the \DstDst\ final state (for $\ee \to \DstDst$). Uncorrelated
systematic errors are given in Tables~\ref{tab1} and ~\ref{tab2}.

\section{Summary}

In summary, we report measurements of the exclusive $\ee \to \DDst$
and $\ee \to \DstDst$ cross sections at $\sqrt{s}$ near $\DDst$ and
$\DstDst$ thresholds. These results supersede those of
Ref.~\cite{belle:ddst}. Due to the increased size of the data sample,
an improved track reconstruction efficiency, and additional modes for
the charmed meson reconstruction, the accuracy of the cross section
measurements is increased by a factor of two compared to
Ref.~\cite{belle:ddst}. The systematic uncertainties are also
significantly improved. We also extend the studied region up to
$\sqrt{s}=6\gev$ and, taking advantage of the improved resolution and
high statistics, we halve the size of the $\sqrt{s}$ steps close
to threshold.

The complex shape of the $\ee \to \DstDst$ cross sections can be
explained by the fact that its components can interfere
  constructively or destructively. The fit of this cross section is
not trivial, because it must take into account the threshold and
coupled-channels effects~\cite{Uglov:2016orr}.

The first angular analysis of the $\ee \to \DstDst$ process allows
  us to decompose the corresponding exclusive cross section into
three possible components for the longitudinally, and
transversely-polarized $D^{*\pm}$ mesons. The obtained components have
distinct behavior near the $\DstDst$ threshold. The only
non-vanishing component at higher energy is the $TL$ helicity of the
$\DstDst$ final state. The measured decomposition allows the
  future measurement of the couplings of vector charmonium states
into different helicity components, useful in identifying their
nature and in testing the heavy-quark symmetry~\cite{hqs}.

We thank the KEKB group for the excellent operation of the
accelerator; the KEK cryogenics group for the efficient operation of
the solenoid; and the KEK computer group, the National Institute of
Informatics, and the PNNL/EMSL computing group for valuable computing
and SINET5 network support.  We acknowledge support from the Ministry
of Education, Culture, Sports, Science, and Technology (MEXT) of
Japan, the Japan Society for the Promotion of Science (JSPS), and the
Tau-Lepton Physics Research Center of Nagoya University; the
Australian Research Council; Austrian Science Fund under Grant No.~P
26794-N20; the National Natural Science Foundation of China under
Contracts No.~10575109, No.~10775142, No.~10875115, No.~11175187,
No.~11475187, No.~11521505 and No.~11575017; the Chinese Academy of
Science Center for Excellence in Particle Physics; the Ministry of
Education, Youth and Sports of the Czech Republic under Contract
No.~LG14034; the Carl Zeiss Foundation, the Deutsche
Forschungsgemeinschaft, the Excellence Cluster Universe, and the
VolkswagenStiftung; the Department of Science and Technology of India;
the Istituto Nazionale di Fisica Nucleare of Italy; the WCU program of
the Ministry of Education, National Research Foundation (NRF) of Korea
Grants No. 2011-0029457, No. 2012-0008143, No.2014R1A2A2A01005286,
No.2014R1A2A2A01002734,No.2015R1A2A2A01003280,
No.2015H1A2A1033649,No.2016R1D1A1B01010135,
No.2016K1A3A7A09005603,No.2016K1A3A7A09005604,
No.2016R1D1A1B02012900,No.2016K1A3A7A09005606,
No. NRF-2013K1A3A7A06056592; the Brain Korea 21-Plus program and
Radiation Science Research Institute; the Polish Ministry of Science
and Higher Education and the National Science Center; the Ministry of
Education and Science of the Russian Federation under contracts
No.~3.2989.2017 and the Russian Foundation for Basic Research
No.~17-02-00485; the Slovenian Research Agency; Ikerbasque, Basque
Foundation for Science and the Euskal Herriko Unibertsitatea (UPV/EHU)
under program UFI 11/55 (Spain); the Swiss National Science
Foundation; the Ministry of Education and the Ministry of Science and
Technology of Taiwan; and the U.S.\ Department of Energy and the
National Science Foundation.


%
\section{Appendix}

\begin{widetext} 
\begin{tabularx}{\linewidth}{l|c|c|}
\caption{The values of the $\ee\ \to\ \DDst$ and $\ee \to \DstDst$
  cross sections and their statistical and uncorrelated systematic
  errors. Correlated systematic errors equal to $5\%$ and $6\%$ for
  the $\ee\ \to\ \DDst$ and $\ee \to \DstDst$ process,
  respectively. The first tabulated error is statistical and the
  second is systematic.}\label{tab1}\\ $\sqrt{s}$ &
$\sigma_{\ee\ \to\ \DDst}$ & $\sigma_{\ee\ \to\ \DstDst}$ \\

3.875	&	0.018	$\pm$	0.028	 $\pm$	0.000	 &	 \\
3.885	&	0.579	$\pm$	0.312	 $\pm$	0.022	 &	 \\
3.895	&	0.898	$\pm$	0.396	 $\pm$	0.036	 &	 \\
3.905	&	1.556	$\pm$	0.457	 $\pm$	0.045	 &  \\
3.915	&	1.470	$\pm$	0.461	 $\pm$	0.046	 &	 \\
3.925	&	2.893	$\pm$	0.533	 $\pm$	0.054	 &	 \\
3.935	&	1.654	$\pm$	0.500	 $\pm$	0.057	 &  \\
3.945	&	2.135	$\pm$	0.507	 $\pm$	0.053	 &	 \\
3.955	&	2.617	$\pm$	0.494	 $\pm$	0.042	 &	 \\
3.965	&	4.014	$\pm$	0.561	 $\pm$	0.049	 &	 \\
3.975	&	2.796	$\pm$	0.518	 $\pm$	0.051	 &	 \\
3.985	&	2.837	$\pm$	0.535	 $\pm$	0.055	 &	 \\
3.995	&	3.030	$\pm$	0.536	 $\pm$	0.055	 &	 \\
4.005	&	3.532	$\pm$	0.529	 $\pm$	0.046	 &	 \\
4.015	&	4.424	$\pm$	0.566	 $\pm$	0.050	 &  \\
4.025	&	4.642	$\pm$	0.561	 $\pm$	0.047	 &	0.788	$\pm$	0.156	 $\pm$	0.025	  \\
4.035	&	3.953	$\pm$	0.526	 $\pm$	0.038	 &	2.702	$\pm$	0.249	 $\pm$	0.085	 \\
4.045	&	2.719	$\pm$	0.479	 $\pm$	0.040	 &	2.958	$\pm$	0.272	 $\pm$	0.094	 \\
4.055	&	2.372	$\pm$	0.445	 $\pm$	0.031	 &	3.137	$\pm$	0.268	 $\pm$	0.101	 \\
4.065	&	1.825	$\pm$	0.404	 $\pm$	0.023	 &	2.932	$\pm$	0.260	 $\pm$	0.096	 \\
4.075	&	1.460	$\pm$	0.379	 $\pm$	0.022	 &	3.629	$\pm$	0.290	 $\pm$	0.120	 \\
4.085	&	2.681	$\pm$	0.391	 $\pm$	0.009	 &	3.514	$\pm$	0.281	 $\pm$	0.117	 \\
4.095	&	1.629	$\pm$	0.374	 $\pm$	0.021	 &	2.919	$\pm$	0.254	 $\pm$	0.098	 \\
4.105	&	2.364	$\pm$	0.369	 $\pm$	0.009	 &	2.574	$\pm$	0.251	 $\pm$	0.087	 \\
4.115	&	1.768	$\pm$	0.365	 $\pm$	0.016	 &	3.436	$\pm$	0.271	 $\pm$	0.118	 \\
4.125	&	2.022	$\pm$	0.383	 $\pm$	0.019	 &	3.100	$\pm$	0.278	 $\pm$	0.107	 \\
4.135	&	0.828	$\pm$	0.324	 $\pm$	0.018	 &	2.943	$\pm$	0.270	 $\pm$	0.103	 \\
4.145	&	1.122	$\pm$	0.329	 $\pm$	0.018	 &	3.102	$\pm$	0.264	 $\pm$	0.109	 \\
4.155	&	1.492	$\pm$	0.357	 $\pm$	0.018	 &	2.956	$\pm$	0.266	 $\pm$	0.105	 \\
4.165	&	1.327	$\pm$	0.341	 $\pm$	0.018	 &	3.178	$\pm$	0.277	 $\pm$	0.114	 \\
4.175	&	1.087	$\pm$	0.338	 $\pm$	0.024	 &	3.060	$\pm$	0.269	 $\pm$	0.110	 \\
4.185	&	0.562	$\pm$	0.329	 $\pm$	0.027	 &	2.780	$\pm$	0.255	 $\pm$	0.101	 \\
4.195	&	1.038	$\pm$	0.321	 $\pm$	0.018	 &	2.570	$\pm$	0.240	 $\pm$	0.094	 \\
4.205	&	0.683	$\pm$	0.294	 $\pm$	0.016	 &	2.226	$\pm$	0.226	 $\pm$	0.082	 \\
4.215	&	0.793	$\pm$	0.290	 $\pm$	0.014	 &	1.495	$\pm$	0.193	 $\pm$	0.056	 \\
4.225	&	1.081	$\pm$	0.321	 $\pm$	0.020	 &	1.100	$\pm$	0.168	 $\pm$	0.041	 \\
4.235	&	0.856	$\pm$	0.297	 $\pm$	0.016	 &	0.674	$\pm$	0.133	 $\pm$	0.025	 \\
4.245	&	0.894	$\pm$	0.290	 $\pm$	0.016	 &	0.303	$\pm$	0.122	 $\pm$	0.011	 \\
4.255	&	0.460	$\pm$	0.267	 $\pm$	0.016	 &	0.374	$\pm$	0.126	 $\pm$	0.014	 \\
4.265	&	1.344	$\pm$	0.293	 $\pm$	0.010	 &	0.614	$\pm$	0.141	 $\pm$	0.024	 \\
4.275	&	0.647	$\pm$	0.254	 $\pm$	0.011	 &	0.650	$\pm$	0.146	 $\pm$	0.025	 \\
4.285	&	0.920	$\pm$	0.261	 $\pm$	0.008	 &	0.741	$\pm$	0.147	 $\pm$	0.029	 \\
4.295	&	0.452	$\pm$	0.252	 $\pm$	0.015	 &	0.507	$\pm$	0.124	 $\pm$	0.020	 \\
4.305	&	0.903	$\pm$	0.258	 $\pm$	0.010	 &	0.897	$\pm$	0.165	 $\pm$	0.035	 \\
4.315	&	0.961	$\pm$	0.256	 $\pm$	0.008	 &	0.735	$\pm$	0.143	 $\pm$	0.029	 \\
4.325	&	1.228	$\pm$	0.273	 $\pm$	0.010	 &	0.959	$\pm$	0.149	 $\pm$	0.038	 \\
4.335	&	0.646	$\pm$	0.237	 $\pm$	0.010	 &	1.191	$\pm$	0.167	 $\pm$	0.047	 \\
4.345	&	0.688	$\pm$	0.250	 $\pm$	0.014	 &	1.078	$\pm$	0.165	 $\pm$	0.043	 \\
4.355	&	1.074	$\pm$	0.249	 $\pm$	0.007	 &	1.221	$\pm$	0.169	 $\pm$	0.049	 \\
4.365	&	1.064	$\pm$	0.233	 $\pm$	0.003	 &	1.127	$\pm$	0.162	 $\pm$	0.045	 \\
4.375	&	0.829	$\pm$	0.245	 $\pm$	0.012	 &	1.241	$\pm$	0.169	 $\pm$	0.050	 \\
4.385	&	0.837	$\pm$	0.241	 $\pm$	0.009	 &	1.069	$\pm$	0.153	 $\pm$	0.043	 \\
4.395	&	0.852	$\pm$	0.242	 $\pm$	0.010	 &	1.285	$\pm$	0.163	 $\pm$	0.052	 \\
4.405	&	0.299	$\pm$	0.213	 $\pm$	0.013	 &	0.919	$\pm$	0.157	 $\pm$	0.037	 \\
4.415	&	0.979	$\pm$	0.220	 $\pm$	0.003	 &	1.272	$\pm$	0.172	 $\pm$	0.051	 \\
4.425	&	0.670	$\pm$	0.221	 $\pm$	0.010	 &	0.831	$\pm$	0.141	 $\pm$	0.034	 \\
4.435	&	0.525	$\pm$	0.211	 $\pm$	0.009	 &	0.951	$\pm$	0.143	 $\pm$	0.039	 \\
4.445	&	0.728	$\pm$	0.213	 $\pm$	0.007	 &	0.677	$\pm$	0.129	 $\pm$	0.028	 \\
4.455	&	0.340	$\pm$	0.207	 $\pm$	0.012	 &	0.545	$\pm$	0.127	 $\pm$	0.022	 \\
4.465	&	0.612	$\pm$	0.195	 $\pm$	0.006	 &	0.725	$\pm$	0.128	 $\pm$	0.030	 \\
4.475	&	0.745	$\pm$	0.211	 $\pm$	0.007	 &	0.412	$\pm$	0.102	 $\pm$	0.017	 \\
4.485	&	0.423	$\pm$	0.197	 $\pm$	0.009	 &	0.440	$\pm$	0.107	 $\pm$	0.018	 \\
4.495	&	0.248	$\pm$	0.185	 $\pm$	0.010	 &	0.754	$\pm$	0.121	 $\pm$	0.031	 \\
4.505	&	0.725	$\pm$	0.200	 $\pm$	0.006	 &	0.635	$\pm$	0.115	 $\pm$	0.026	 \\
4.515	&	0.564	$\pm$	0.179	 $\pm$	0.004	 &	0.584	$\pm$	0.119	 $\pm$	0.024	 \\
4.525	&	0.295	$\pm$	0.176	 $\pm$	0.009	 &	0.573	$\pm$	0.104	 $\pm$	0.023	 \\
4.535	&	0.535	$\pm$	0.192	 $\pm$	0.008	 &	0.378	$\pm$	0.105	 $\pm$	0.015	 \\
4.545	&	0.757	$\pm$	0.194	 $\pm$	0.005	 &	0.652	$\pm$	0.119	 $\pm$	0.027	 \\
4.555	&	0.439	$\pm$	0.174	 $\pm$	0.006	 &	0.275	$\pm$	0.091	 $\pm$	0.011	 \\
4.565	&	0.519	$\pm$	0.185	 $\pm$	0.007	 &	0.566	$\pm$	0.121	 $\pm$	0.023	 \\
4.575	&	0.460	$\pm$	0.179	 $\pm$	0.007	 &	0.776	$\pm$	0.124	 $\pm$	0.031	 \\
4.585	&	0.464	$\pm$	0.173	 $\pm$	0.007	 &	0.452	$\pm$	0.096	 $\pm$	0.018	 \\
4.595	&	0.593	$\pm$	0.170	 $\pm$	0.004	 &	0.396	$\pm$	0.091	 $\pm$	0.016	 \\
4.605	&	0.506	$\pm$	0.178	 $\pm$	0.007	 &	0.323	$\pm$	0.088	 $\pm$	0.013	 \\
4.615	&	0.351	$\pm$	0.157	 $\pm$	0.006	 &	0.387	$\pm$	0.094	 $\pm$	0.016	 \\
4.625	&	0.474	$\pm$	0.158	 $\pm$	0.004	 &	0.507	$\pm$	0.101	 $\pm$	0.020	 \\
4.635	&	0.558	$\pm$	0.171	 $\pm$	0.005	 &	0.479	$\pm$	0.098	 $\pm$	0.019	 \\
4.645	&	0.333	$\pm$	0.145	 $\pm$	0.004	 &	0.438	$\pm$	0.103	 $\pm$	0.017	 \\
4.655	&	0.724	$\pm$	0.162	 $\pm$	0.001	 &	0.434	$\pm$	0.103	 $\pm$	0.017	 \\
4.665	&	0.460	$\pm$	0.152	 $\pm$	0.003	 &	0.456	$\pm$	0.097	 $\pm$	0.018	 \\
4.675	&	0.460	$\pm$	0.147	 $\pm$	0.004	 &	0.551	$\pm$	0.096	 $\pm$	0.022	 \\
4.685	&	0.485	$\pm$	0.152	 $\pm$	0.004	 &	0.357	$\pm$	0.088	 $\pm$	0.014	 \\
4.695	&	0.365	$\pm$	0.136	 $\pm$	0.003	 &	0.653	$\pm$	0.106	 $\pm$	0.025	 \\
4.705	&	0.312	$\pm$	0.132	 $\pm$	0.003	 &	0.391	$\pm$	0.093	 $\pm$	0.015	 \\
4.715	&	0.419	$\pm$	0.137	 $\pm$	0.002	 &	0.517	$\pm$	0.093	 $\pm$	0.020	 \\
4.725	&	0.303	$\pm$	0.137	 $\pm$	0.004	 &	0.481	$\pm$	0.098	 $\pm$	0.018	 \\
4.735	&	0.327	$\pm$	0.138	 $\pm$	0.005	 &	0.565	$\pm$	0.102	 $\pm$	0.022	 \\
4.745	&	0.301	$\pm$	0.129	 $\pm$	0.003	 &	0.481	$\pm$	0.101	 $\pm$	0.018	 \\
4.755	&	0.430	$\pm$	0.148	 $\pm$	0.004	 &	0.254	$\pm$	0.078	 $\pm$	0.010	 \\
4.765	&	0.155	$\pm$	0.112	 $\pm$	0.003	 &	0.255	$\pm$	0.081	 $\pm$	0.010	 \\
4.775	&	0.482	$\pm$	0.144	 $\pm$	0.003	 &	0.320	$\pm$	0.076	 $\pm$	0.012	 \\
4.785	&	0.846	$\pm$	0.159	 $\pm$	0.001	 &	0.498	$\pm$	0.091	 $\pm$	0.018	 \\
4.795	&	0.427	$\pm$	0.119	 $\pm$	0.000	 &	0.465	$\pm$	0.088	 $\pm$	0.017	 \\
4.805	&	0.439	$\pm$	0.145	 $\pm$	0.005	 &	0.392	$\pm$	0.090	 $\pm$	0.014	 \\
4.815	&	0.310	$\pm$	0.129	 $\pm$	0.004	 &	0.399	$\pm$	0.084	 $\pm$	0.014	 \\
4.825	&	0.397	$\pm$	0.127	 $\pm$	0.003	 &	0.237	$\pm$	0.073	 $\pm$	0.008	 \\
4.835	&	0.628	$\pm$	0.146	 $\pm$	0.002	 &	0.426	$\pm$	0.079	 $\pm$	0.015	 \\
4.845	&	0.416	$\pm$	0.128	 $\pm$	0.003	 &	0.374	$\pm$	0.082	 $\pm$	0.013	 \\
4.855	&	0.443	$\pm$	0.125	 $\pm$	0.001	 &	0.341	$\pm$	0.074	 $\pm$	0.012	 \\
4.865	&	0.368	$\pm$	0.114	 $\pm$	0.001	 &	0.229	$\pm$	0.073	 $\pm$	0.008	 \\
4.875	&	0.311	$\pm$	0.114	 $\pm$	0.002	 &	0.319	$\pm$	0.070	 $\pm$	0.011	 \\
4.885	&	0.226	$\pm$	0.106	 $\pm$	0.002	 &	0.158	$\pm$	0.054	 $\pm$	0.005	 \\
4.895	&	0.187	$\pm$	0.110	 $\pm$	0.004	 &	0.357	$\pm$	0.074	 $\pm$	0.012	 \\
4.905	&	0.225	$\pm$	0.104	 $\pm$	0.003	 &	0.384	$\pm$	0.073	 $\pm$	0.013	 \\
4.915	&	0.330	$\pm$	0.118	 $\pm$	0.003	 &	0.222	$\pm$	0.072	 $\pm$	0.007	 \\
4.925	&	0.169	$\pm$	0.083	 $\pm$	0.001	 &	0.273	$\pm$	0.068	 $\pm$	0.009	 \\
4.935	&	0.135	$\pm$	0.104	 $\pm$	0.004	 &	0.351	$\pm$	0.076	 $\pm$	0.011	 \\
4.945	&	0.341	$\pm$	0.110	 $\pm$	0.002	 &	0.328	$\pm$	0.069	 $\pm$	0.010	 \\
4.955	&	0.359	$\pm$	0.098	 $\pm$	0.000	 &	0.343	$\pm$	0.081	 $\pm$	0.011	 \\
4.965	&	0.334	$\pm$	0.118	 $\pm$	0.003	 &	0.293	$\pm$	0.066	 $\pm$	0.009	 \\
4.975	&	0.338	$\pm$	0.098	 $\pm$	0.000	 &	0.229	$\pm$	0.063	 $\pm$	0.007	 \\
4.985	&	0.347	$\pm$	0.104	 $\pm$	0.001	 &	0.257	$\pm$	0.061	 $\pm$	0.008	 \\
4.995	&	0.197	$\pm$	0.090	 $\pm$	0.002	 &	0.170	$\pm$	0.058	 $\pm$	0.005	 \\
5.005	&	0.215	$\pm$	0.093	 $\pm$	0.002	 &	0.233	$\pm$	0.061	 $\pm$	0.007	 \\
5.015	&	0.330	$\pm$	0.101	 $\pm$	0.001	 &	0.172	$\pm$	0.055	 $\pm$	0.005	 \\
5.025	&	0.113	$\pm$	0.078	 $\pm$	0.002	 &	0.260	$\pm$	0.060	 $\pm$	0.007	 \\
5.035	&	0.281	$\pm$	0.099	 $\pm$	0.002	 &	0.244	$\pm$	0.061	 $\pm$	0.007	 \\
5.045	&	0.300	$\pm$	0.095	 $\pm$	0.001	 &	0.282	$\pm$	0.065	 $\pm$	0.008	 \\
5.055	&	0.289	$\pm$	0.103	 $\pm$	0.002	 &	0.261	$\pm$	0.062	 $\pm$	0.007	 \\
5.065	&	0.264	$\pm$	0.093	 $\pm$	0.002	 &	0.185	$\pm$	0.054	 $\pm$	0.005	 \\
5.075	&	0.212	$\pm$	0.093	 $\pm$	0.002	 &	0.265	$\pm$	0.062	 $\pm$	0.007	 \\
5.085	&	0.334	$\pm$	0.095	 $\pm$	0.001	 &	0.160	$\pm$	0.053	 $\pm$	0.004	 \\
5.095	&	0.330	$\pm$	0.098	 $\pm$	0.001	 &	0.182	$\pm$	0.063	 $\pm$	0.004	 \\
5.105	&	0.311	$\pm$	0.097	 $\pm$	0.002	 &	0.178	$\pm$	0.055	 $\pm$	0.004	 \\
5.115	&	0.283	$\pm$	0.082	 $\pm$	0.000	 &	0.221	$\pm$	0.057	 $\pm$	0.005	 \\
5.125	&	0.438	$\pm$	0.104	 $\pm$	0.001	 &	0.250	$\pm$	0.061	 $\pm$	0.006	 \\
5.135	&	0.364	$\pm$	0.098	 $\pm$	0.001	 &	0.170	$\pm$	0.053	 $\pm$	0.004	 \\
5.145	&	0.262	$\pm$	0.086	 $\pm$	0.001	 &	0.183	$\pm$	0.055	 $\pm$	0.004	 \\
5.155	&	0.163	$\pm$	0.073	 $\pm$	0.001	 &	0.042	$\pm$	0.042	 $\pm$	0.001	 \\
5.165	&	0.142	$\pm$	0.073	 $\pm$	0.001	 &	0.108	$\pm$	0.043	 $\pm$	0.002	 \\
5.175	&	0.028	$\pm$	0.063	 $\pm$	0.002	 &	0.245	$\pm$	0.053	 $\pm$	0.005	 \\
5.185	&	0.204	$\pm$	0.071	 $\pm$	0.000	 &	0.220	$\pm$	0.054	 $\pm$	0.004	 \\
5.195	&	0.158	$\pm$	0.076	 $\pm$	0.002	 &	0.080	$\pm$	0.036	 $\pm$	0.002	 \\
5.205	&	0.240	$\pm$	0.083	 $\pm$	0.001	 &	0.196	$\pm$	0.050	 $\pm$	0.004	 \\
5.215	&	0.155	$\pm$	0.064	 $\pm$	0.000	 &	0.070	$\pm$	0.038	 $\pm$	0.001	 \\
5.225	&	0.081	$\pm$	0.066	 $\pm$	0.002	 &	0.121	$\pm$	0.046	 $\pm$	0.002	 \\
5.235	&	0.223	$\pm$	0.078	 $\pm$	0.001	 &	0.166	$\pm$	0.044	 $\pm$	0.003	 \\
5.245	&	0.152	$\pm$	0.075	 $\pm$	0.002	 &	0.149	$\pm$	0.051	 $\pm$	0.002	 \\
5.255	&	0.225	$\pm$	0.077	 $\pm$	0.001	 &	0.126	$\pm$	0.042	 $\pm$	0.002	 \\
5.265	&	0.038	$\pm$	0.063	 $\pm$	0.002	 &	0.108	$\pm$	0.042	 $\pm$	0.002	 \\
5.275	&	0.064	$\pm$	0.054	 $\pm$	0.001	 &	0.197	$\pm$	0.049	 $\pm$	0.003	 \\
5.285	&	0.286	$\pm$	0.080	 $\pm$	0.001	 &	0.142	$\pm$	0.048	 $\pm$	0.002	 \\
5.295	&	0.184	$\pm$	0.068	 $\pm$	0.000	 &	0.145	$\pm$	0.047	 $\pm$	0.002	 \\
5.305	&	0.342	$\pm$	0.080	 $\pm$	0.000	 &	0.070	$\pm$	0.032	 $\pm$	0.001	 \\
5.315	&	0.251	$\pm$	0.076	 $\pm$	0.001	 &	0.169	$\pm$	0.045	 $\pm$	0.002	 \\
5.325	&	0.185	$\pm$	0.077	 $\pm$	0.002	 &	0.153	$\pm$	0.043	 $\pm$	0.002	 \\
5.335	&	0.265	$\pm$	0.078	 $\pm$	0.001	 &	0.199	$\pm$	0.047	 $\pm$	0.002	 \\
5.345	&	0.178	$\pm$	0.058	 $\pm$	0.000	 &	0.179	$\pm$	0.044	 $\pm$	0.002	 \\
5.355	&	0.084	$\pm$	0.058	 $\pm$	0.001	 &	0.104	$\pm$	0.036	 $\pm$	0.001	 \\
5.365	&	0.143	$\pm$	0.056	 $\pm$	0.000	 &	0.245	$\pm$	0.051	 $\pm$	0.002	 \\
5.375	&	0.074	$\pm$	0.060	 $\pm$	0.002	 &	0.205	$\pm$	0.050	 $\pm$	0.002	 \\
5.385	&	0.135	$\pm$	0.064	 $\pm$	0.001	 &	0.201	$\pm$	0.047	 $\pm$	0.002	 \\
5.395	&	0.133	$\pm$	0.057	 $\pm$	0.001	 &	0.112	$\pm$	0.037	 $\pm$	0.001	 \\
5.405	&	0.087	$\pm$	0.052	 $\pm$	0.001	 &	0.156	$\pm$	0.041	 $\pm$	0.001	 \\
5.415	&	0.109	$\pm$	0.058	 $\pm$	0.001	 &	0.147	$\pm$	0.044	 $\pm$	0.001	 \\
5.425	&	0.166	$\pm$	0.054	 $\pm$	0.000	 &	0.134	$\pm$	0.039	 $\pm$	0.001	 \\
5.435	&	0.160	$\pm$	0.062	 $\pm$	0.001	 &	0.133	$\pm$	0.038	 $\pm$	0.001	 \\
5.445	&	0.169	$\pm$	0.057	 $\pm$	0.000	 &	0.152	$\pm$	0.041	 $\pm$	0.001	 \\
5.455	&	0.165	$\pm$	0.060	 $\pm$	0.001	 &	0.165	$\pm$	0.041	 $\pm$	0.001	 \\
5.465	&	0.180	$\pm$	0.057	 $\pm$	0.000	 &	0.066	$\pm$	0.029	 $\pm$	0.000	 \\
5.475	&	0.170	$\pm$	0.057	 $\pm$	0.000	 &	0.138	$\pm$	0.038	 $\pm$	0.000	 \\
5.485	&	0.168	$\pm$	0.054	 $\pm$	0.000	 &	0.192	$\pm$	0.042	 $\pm$	0.000	 \\
5.495	&	0.086	$\pm$	0.057	 $\pm$	0.002	 &	0.181	$\pm$	0.043	 $\pm$	0.000	 \\
5.505	&	0.158	$\pm$	0.055	 $\pm$	0.000	 &	0.158	$\pm$	0.039	 $\pm$	0.000	 \\
5.515	&	0.215	$\pm$	0.067	 $\pm$	0.001	 &	0.110	$\pm$	0.037	 $\pm$	0.000	 \\
5.525	&	0.100	$\pm$	0.041	 $\pm$	0.000	 &	0.176	$\pm$	0.042	 $\pm$	0.000	 \\
5.535	&	0.120	$\pm$	0.042	 $\pm$	0.000	 &	0.163	$\pm$	0.039	 $\pm$	0.000	 \\
5.545	&	0.148	$\pm$	0.058	 $\pm$	0.001	 &	0.072	$\pm$	0.028	 $\pm$	0.000	 \\
5.555	&	0.140	$\pm$	0.051	 $\pm$	0.000	 &	0.074	$\pm$	0.030	 $\pm$	0.000	 \\
5.565	&	0.173	$\pm$	0.055	 $\pm$	0.000	 &	0.042	$\pm$	0.028	 $\pm$	0.000	 \\
5.575	&	0.117	$\pm$	0.049	 $\pm$	0.000	 &	0.089	$\pm$	0.032	 $\pm$	0.000	 \\
5.585	&	0.103	$\pm$	0.052	 $\pm$	0.001	 &	0.069	$\pm$	0.030	 $\pm$	0.000	 \\
5.595	&	0.075	$\pm$	0.045	 $\pm$	0.001	 &	0.156	$\pm$	0.038	 $\pm$	0.001	 \\
5.605	&	0.104	$\pm$	0.050	 $\pm$	0.001	 &	0.039	$\pm$	0.021	 $\pm$	0.000	 \\
5.615	&	0.206	$\pm$	0.059	 $\pm$	0.000	 &	0.063	$\pm$	0.026	 $\pm$	0.000	 \\
5.625	&	0.100	$\pm$	0.047	 $\pm$	0.001	 &	0.035	$\pm$	0.025	 $\pm$	0.000	 \\
5.635	&	0.199	$\pm$	0.056	 $\pm$	0.000	 &	0.106	$\pm$	0.037	 $\pm$	0.001	 \\
5.645	&	0.173	$\pm$	0.058	 $\pm$	0.001	 &	0.077	$\pm$	0.026	 $\pm$	0.001	 \\
5.655	&	0.078	$\pm$	0.050	 $\pm$	0.001	 &	0.094	$\pm$	0.029	 $\pm$	0.001	 \\
5.665	&	0.075	$\pm$	0.040	 $\pm$	0.000	 &	0.051	$\pm$	0.021	 $\pm$	0.000	 \\
5.675	&	0.092	$\pm$	0.046	 $\pm$	0.001	 &	0.055	$\pm$	0.026	 $\pm$	0.001	 \\
5.685	&	0.104	$\pm$	0.049	 $\pm$	0.001	 &	0.123	$\pm$	0.032	 $\pm$	0.001	 \\
5.695	&	0.177	$\pm$	0.054	 $\pm$	0.001	 &	0.098	$\pm$	0.028	 $\pm$	0.001	 \\
5.705	&	0.130	$\pm$	0.047	 $\pm$	0.000	 &	0.142	$\pm$	0.034	 $\pm$	0.002	 \\
5.715	&	0.127	$\pm$	0.043	 $\pm$	0.000	 &	0.102	$\pm$	0.030	 $\pm$	0.001	 \\
5.725	&	0.157	$\pm$	0.049	 $\pm$	0.000	 &	0.085	$\pm$	0.029	 $\pm$	0.001	 \\
5.735	&	0.020	$\pm$	0.030	 $\pm$	0.001	 &	0.102	$\pm$	0.028	 $\pm$	0.001	 \\
5.745	&	0.097	$\pm$	0.040	 $\pm$	0.000	 &	0.059	$\pm$	0.025	 $\pm$	0.001	 \\
5.755	&	0.182	$\pm$	0.047	 $\pm$	0.000	 &	0.097	$\pm$	0.028	 $\pm$	0.001	 \\
5.765	&	0.111	$\pm$	0.045	 $\pm$	0.001	 &	0.088	$\pm$	0.029	 $\pm$	0.001	 \\
5.775	&	0.076	$\pm$	0.041	 $\pm$	0.001	 &	0.071	$\pm$	0.023	 $\pm$	0.001	 \\
5.785	&	0.124	$\pm$	0.044	 $\pm$	0.000	 &	0.055	$\pm$	0.025	 $\pm$	0.001	 \\
5.795	&	0.049	$\pm$	0.032	 $\pm$	0.000	 &	0.078	$\pm$	0.025	 $\pm$	0.001	 \\
5.805	&	0.043	$\pm$	0.036	 $\pm$	0.001	 &	0.103	$\pm$	0.028	 $\pm$	0.002	 \\
5.815	&	0.111	$\pm$	0.043	 $\pm$	0.001	 &	0.085	$\pm$	0.025	 $\pm$	0.002	 \\
5.825	&	0.064	$\pm$	0.034	 $\pm$	0.000	 &	0.088	$\pm$	0.026	 $\pm$	0.002	 \\
5.835	&	0.167	$\pm$	0.045	 $\pm$	0.000	 &	0.114	$\pm$	0.028	 $\pm$	0.002	 \\
5.845	&	0.048	$\pm$	0.032	 $\pm$	0.000	 &	0.058	$\pm$	0.021	 $\pm$	0.001	 \\
5.855	&	0.042	$\pm$	0.034	 $\pm$	0.001	 &	0.095	$\pm$	0.029	 $\pm$	0.002	 \\
5.865	&	0.072	$\pm$	0.032	 $\pm$	0.000	 &	0.041	$\pm$	0.019	 $\pm$	0.001	 \\
5.875	&	0.040	$\pm$	0.032	 $\pm$	0.001	 &	0.041	$\pm$	0.018	 $\pm$	0.001	 \\
5.885	&	0.046	$\pm$	0.033	 $\pm$	0.001	 &	0.103	$\pm$	0.027	 $\pm$	0.002	 \\
5.895	&	0.070	$\pm$	0.031	 $\pm$	0.000	 &	0.070	$\pm$	0.022	 $\pm$	0.002	 \\
5.905	&	-0.013	$\pm$	0.025	 $\pm$	0.001	 &	0.080	$\pm$	0.023	 $\pm$	0.002	 \\
5.915	&	0.135	$\pm$	0.042	 $\pm$	0.000	 &	0.073	$\pm$	0.022	 $\pm$	0.002	 \\
5.925	&	0.078	$\pm$	0.036	 $\pm$	0.001	 &	0.079	$\pm$	0.023	 $\pm$	0.002	 \\
5.935	&	0.053	$\pm$	0.032	 $\pm$	0.000	 &	0.065	$\pm$	0.020	 $\pm$	0.002	 \\
5.945	&	0.106	$\pm$	0.037	 $\pm$	0.000	 &	0.030	$\pm$	0.016	 $\pm$	0.001	 \\
5.955	&	0.059	$\pm$	0.030	 $\pm$	0.000	 &	0.056	$\pm$	0.023	 $\pm$	0.001	 \\
5.965	&	0.046	$\pm$	0.028	 $\pm$	0.000	 &	0.107	$\pm$	0.025	 $\pm$	0.003	 \\
5.975	&	0.060	$\pm$	0.032	 $\pm$	0.000	 &	0.056	$\pm$	0.019	 $\pm$	0.002	 \\
5.985	&	0.070	$\pm$	0.033	 $\pm$	0.000	 &	0.042	$\pm$	0.017	 $\pm$	0.001	 \\
5.995	&	0.098	$\pm$	0.032	 $\pm$	0.000	 &	0.038	$\pm$	0.015	 $\pm$	0.001	 \\

\end{tabularx}
\end{widetext}

\begin{widetext}
\begin{tabularx}{\linewidth}{l|c|c|c|}
\caption{The values of the $\ee \to \DstDst$ cross sections components
  and their statistical and uncorrelated systematic errors. Correlated
  systematic error is equal to $6 \%$. The first tabulated error is
  statistical and the second is systematic.}
\label{tab2}\\
$\sqrt{s}$ & $\sigma_{TT}$ & $\sigma_{TL}$ & $\sigma_{LL}$ \\

4.025	&	0.359	$\pm$	0.130	$\pm$	0.041	&	0.517	$\pm$	0.171	$\pm$	0.009	&	0.023	$\pm$	0.080	$\pm$	0.010	\\
4.035	&	1.252	$\pm$	0.220	$\pm$	0.131	&	1.293	$\pm$	0.262	$\pm$	0.030	&	0.170	$\pm$	0.129	$\pm$	0.031	\\
4.045	&	1.023	$\pm$	0.211	$\pm$	0.119	&	1.37	$\pm$	0.277	$\pm$	0.037	&	0.384	$\pm$	0.160	$\pm$	0.051	\\
4.055	&	0.975	$\pm$	0.204	$\pm$	0.105	&	1.536	$\pm$	0.285	$\pm$	0.016	&	0.481	$\pm$	0.173	$\pm$	0.049	\\
4.065	&	1.199	$\pm$	0.227	$\pm$	0.126	&	1.103	$\pm$	0.273	$\pm$	0.033	&	0.522	$\pm$	0.153	$\pm$	0.046	\\
4.075	&	1.163	$\pm$	0.221	$\pm$	0.128	&	1.812	$\pm$	0.300	$\pm$	0.025	&	0.413	$\pm$	0.176	$\pm$	0.055	\\
4.085	&	1.077	$\pm$	0.213	$\pm$	0.121	&	1.627	$\pm$	0.295	$\pm$	0.032	&	0.525	$\pm$	0.184	$\pm$	0.063	\\
4.095	&	0.831	$\pm$	0.173	$\pm$	0.093	&	1.452	$\pm$	0.269	$\pm$	0.021	&	0.473	$\pm$	0.175	$\pm$	0.059	\\
4.105	&	0.702	$\pm$	0.163	$\pm$	0.080	&	1.498	$\pm$	0.241	$\pm$	0.001	&	0.392	$\pm$	0.140	$\pm$	0.039	\\
4.115	&	0.808	$\pm$	0.185	$\pm$	0.099	&	1.335	$\pm$	0.282	$\pm$	0.030	&	0.930	$\pm$	0.193	$\pm$	0.079	\\
4.125	&	0.616	$\pm$	0.173	$\pm$	0.089	&	1.791	$\pm$	0.289	$\pm$	0.022	&	0.753	$\pm$	0.192	$\pm$	0.076	\\
4.135	&	0.908	$\pm$	0.180	$\pm$	0.101	&	0.930	$\pm$	0.250	$\pm$	0.039	&	1.157	$\pm$	0.198	$\pm$	0.089	\\
4.145	&	1.096	$\pm$	0.190	$\pm$	0.115	&	0.888	$\pm$	0.246	$\pm$	0.041	&	0.902	$\pm$	0.180	$\pm$	0.074	\\
4.155	&	0.892	$\pm$	0.207	$\pm$	0.119	&	1.305	$\pm$	0.303	$\pm$	0.052	&	0.918	$\pm$	0.198	$\pm$	0.083	\\
4.165	&	0.830	$\pm$	0.190	$\pm$	0.109	&	1.461	$\pm$	0.272	$\pm$	0.031	&	0.766	$\pm$	0.171	$\pm$	0.068	\\
4.175	&	0.987	$\pm$	0.187	$\pm$	0.111	&	0.908	$\pm$	0.260	$\pm$	0.049	&	0.940	$\pm$	0.193	$\pm$	0.083	\\
4.185	&	0.750	$\pm$	0.163	$\pm$	0.080	&	1.309	$\pm$	0.256	$\pm$	0.008	&	0.599	$\pm$	0.169	$\pm$	0.055	\\
4.195	&	0.874	$\pm$	0.171	$\pm$	0.089	&	0.974	$\pm$	0.228	$\pm$	0.015	&	0.524	$\pm$	0.147	$\pm$	0.045	\\
4.205	&	0.966	$\pm$	0.165	$\pm$	0.092	&	0.530	$\pm$	0.199	$\pm$	0.030	&	0.587	$\pm$	0.147	$\pm$	0.050	\\
4.215	&	0.369	$\pm$	0.120	$\pm$	0.045	&	0.852	$\pm$	0.193	$\pm$	0.008	&	0.302	$\pm$	0.127	$\pm$	0.032	\\
4.225	&	0.298	$\pm$	0.132	$\pm$	0.045	&	0.496	$\pm$	0.197	$\pm$	0.022	&	0.357	$\pm$	0.123	$\pm$	0.033	\\
4.235	&	0.012	$\pm$	0.035	$\pm$	0.019	&	0.570	$\pm$	0.126	$\pm$	0.007	&	0.101	$\pm$	0.079	$\pm$	0.014	\\
4.245	&	0.185	$\pm$	0.091	$\pm$	0.016	&	0.123	$\pm$	0.121	$\pm$	0.011	&	0.137	$\pm$	0.081	$\pm$	0.031	\\
4.255	&	0.088	$\pm$	0.071	$\pm$	0.007	&	0.274	$\pm$	0.108	$\pm$	0.036	&	0.027	$\pm$	0.051	$\pm$	0.008	\\
4.265	&	0.119	$\pm$	0.076	$\pm$	0.010	&	0.379	$\pm$	0.118	$\pm$	0.046	&	0.112	$\pm$	0.071	$\pm$	0.026	\\
4.275	&	0.181	$\pm$	0.086	$\pm$	0.014	&	0.423	$\pm$	0.117	$\pm$	0.058	&	-0.005	$\pm$	0.045	$\pm$	0.003	\\
4.285	&	0.249	$\pm$	0.099	$\pm$	0.022	&	0.446	$\pm$	0.135	$\pm$	0.058	&	0.060	$\pm$	0.064	$\pm$	0.016	\\
4.295	&	0.229	$\pm$	0.083	$\pm$	0.022	&	0.195	$\pm$	0.094	$\pm$	0.026	&	0.049	$\pm$	0.046	$\pm$	0.011	\\
4.305	&	0.260	$\pm$	0.106	$\pm$	0.022	&	0.572	$\pm$	0.151	$\pm$	0.074	&	0.128	$\pm$	0.080	$\pm$	0.029	\\
4.315	&	0.160	$\pm$	0.093	$\pm$	0.010	&	0.632	$\pm$	0.13	$\pm$	0.085	&	-0.025	$\pm$	0.047	$\pm$	0.002	\\
4.325	&	0.045	$\pm$	0.071	$\pm$	0.002	&	0.591	$\pm$	0.146	$\pm$	0.070	&	0.175	$\pm$	0.090	$\pm$	0.040	\\
4.335	&	0.302	$\pm$	0.117	$\pm$	0.027	&	0.375	$\pm$	0.166	$\pm$	0.042	&	0.365	$\pm$	0.115	$\pm$	0.075	\\
4.345	&	0.332	$\pm$	0.119	$\pm$	0.028	&	0.639	$\pm$	0.158	$\pm$	0.085	&	0.072	$\pm$	0.066	$\pm$	0.019	\\
4.355	&	0.221	$\pm$	0.110	$\pm$	0.017	&	0.810	$\pm$	0.176	$\pm$	0.101	&	0.070	$\pm$	0.085	$\pm$	0.025	\\
4.365	&	0.181	$\pm$	0.100	$\pm$	0.013	&	0.750	$\pm$	0.164	$\pm$	0.096	&	0.092	$\pm$	0.078	$\pm$	0.025	\\
4.375	&	0.298	$\pm$	0.112	$\pm$	0.025	&	0.790	$\pm$	0.143	$\pm$	0.103	&	0.009	$\pm$	0.028	$\pm$	0.012	\\
4.385	&	0.199	$\pm$	0.097	$\pm$	0.016	&	0.577	$\pm$	0.148	$\pm$	0.072	&	0.126	$\pm$	0.080	$\pm$	0.031	\\
4.395	&	0.164	$\pm$	0.088	$\pm$	0.013	&	0.769	$\pm$	0.154	$\pm$	0.096	&	0.086	$\pm$	0.080	$\pm$	0.025	\\
4.405	&	0.400	$\pm$	0.107	$\pm$	0.039	&	0.404	$\pm$	0.138	$\pm$	0.048	&	0.110	$\pm$	0.083	$\pm$	0.029	\\
4.415	&	0.231	$\pm$	0.103	$\pm$	0.019	&	0.740	$\pm$	0.168	$\pm$	0.090	&	0.130	$\pm$	0.094	$\pm$	0.036	\\
4.425	&	0.146	$\pm$	0.075	$\pm$	0.013	&	0.519	$\pm$	0.133	$\pm$	0.061	&	0.084	$\pm$	0.079	$\pm$	0.025	\\
4.435	&	0.179	$\pm$	0.087	$\pm$	0.015	&	0.494	$\pm$	0.135	$\pm$	0.062	&	0.118	$\pm$	0.076	$\pm$	0.028	\\
4.445	&	0.107	$\pm$	0.076	$\pm$	0.010	&	0.471	$\pm$	0.147	$\pm$	0.044	&	0.080	$\pm$	0.096	$\pm$	0.034	\\
4.455	&	0.189	$\pm$	0.068	$\pm$	0.019	&	0.159	$\pm$	0.090	$\pm$	0.016	&	0.179	$\pm$	0.075	$\pm$	0.036	\\
4.465	&	0.210	$\pm$	0.088	$\pm$	0.021	&	0.326	$\pm$	0.136	$\pm$	0.035	&	0.091	$\pm$	0.084	$\pm$	0.027	\\
4.475	&	0.107	$\pm$	0.061	$\pm$	0.009	&	0.221	$\pm$	0.087	$\pm$	0.029	&	0.041	$\pm$	0.040	$\pm$	0.010	\\
4.485	&	0.114	$\pm$	0.069	$\pm$	0.009	&	0.258	$\pm$	0.091	$\pm$	0.036	&	0.024	$\pm$	0.040	$\pm$	0.006	\\
4.495	&	0.153	$\pm$	0.071	$\pm$	0.014	&	0.354	$\pm$	0.105	$\pm$	0.044	&	0.072	$\pm$	0.060	$\pm$	0.018	\\
4.505	&	0.136	$\pm$	0.087	$\pm$	0.011	&	0.310	$\pm$	0.131	$\pm$	0.038	&	0.076	$\pm$	0.071	$\pm$	0.020	\\
4.515	&	-0.005	$\pm$	0.051	$\pm$	0.006	&	0.468	$\pm$	0.107	$\pm$	0.061	&	0.070	$\pm$	0.061	$\pm$	0.018	\\
4.525	&	0.241	$\pm$	0.087	$\pm$	0.023	&	0.192	$\pm$	0.106	$\pm$	0.025	&	0.042	$\pm$	0.055	$\pm$	0.011	\\
4.535	&	0.121	$\pm$	0.072	$\pm$	0.010	&	0.328	$\pm$	0.101	$\pm$	0.044	&	-0.004	$\pm$	0.035	$\pm$	0.002	\\
4.545	&	0.014	$\pm$	0.031	$\pm$	0.001	&	0.453	$\pm$	0.118	$\pm$	0.048	&	0.086	$\pm$	0.082	$\pm$	0.027	\\
4.555	&	0.067	$\pm$	0.052	$\pm$	0.005	&	0.271	$\pm$	0.062	$\pm$	0.036	&	-0.034	$\pm$	0.020	$\pm$	0.004	\\
4.565	&	0.089	$\pm$	0.071	$\pm$	0.007	&	0.425	$\pm$	0.128	$\pm$	0.049	&	0.038	$\pm$	0.071	$\pm$	0.016	\\
4.575	&	0.119	$\pm$	0.071	$\pm$	0.010	&	0.447	$\pm$	0.111	$\pm$	0.058	&	0.028	$\pm$	0.051	$\pm$	0.010	\\
4.585	&	0.056	$\pm$	0.051	$\pm$	0.004	&	0.260	$\pm$	0.086	$\pm$	0.033	&	0.067	$\pm$	0.049	$\pm$	0.014	\\
4.595	&	0.173	$\pm$	0.080	$\pm$	0.014	&	0.300	$\pm$	0.103	$\pm$	0.036	&	0.002	$\pm$	0.037	$\pm$	0.002	\\
4.605	&	0.113	$\pm$	0.059	$\pm$	0.010	&	0.115	$\pm$	0.077	$\pm$	0.012	&	0.125	$\pm$	0.061	$\pm$	0.020	\\
4.615	&	0.087	$\pm$	0.066	$\pm$	0.006	&	0.333	$\pm$	0.093	$\pm$	0.040	&	-0.009	$\pm$	0.031	$\pm$	0.000	\\
4.625	&	0.078	$\pm$	0.056	$\pm$	0.006	&	0.239	$\pm$	0.090	$\pm$	0.026	&	0.142	$\pm$	0.066	$\pm$	0.024	\\
4.635	&	0.058	$\pm$	0.068	$\pm$	0.002	&	0.323	$\pm$	0.123	$\pm$	0.036	&	0.105	$\pm$	0.076	$\pm$	0.020	\\
4.645	&	0.058	$\pm$	0.080	$\pm$	0.002	&	0.379	$\pm$	0.132	$\pm$	0.043	&	0.031	$\pm$	0.066	$\pm$	0.009	\\
4.655	&	0.080	$\pm$	0.060	$\pm$	0.006	&	0.435	$\pm$	0.078	$\pm$	0.050	&	-0.051	$\pm$	0.024	$\pm$	0.005	\\
4.665	&	0.141	$\pm$	0.074	$\pm$	0.012	&	0.337	$\pm$	0.093	$\pm$	0.031	&	0.000	$\pm$	0.009	$\pm$	0.009	\\
4.675	&	0.081	$\pm$	0.056	$\pm$	0.007	&	0.315	$\pm$	0.096	$\pm$	0.034	&	0.082	$\pm$	0.060	$\pm$	0.015	\\
4.685	&	0.154	$\pm$	0.071	$\pm$	0.013	&	0.104	$\pm$	0.087	$\pm$	0.012	&	0.084	$\pm$	0.055	$\pm$	0.015	\\
4.695	&	-0.018	$\pm$	0.062	$\pm$	0.004	&	0.738	$\pm$	0.140	$\pm$	0.078	&	-0.162	$\pm$	0.080	$\pm$	0.011	\\
4.705	&	0.189	$\pm$	0.076	$\pm$	0.016	&	0.187	$\pm$	0.093	$\pm$	0.021	&	0.050	$\pm$	0.054	$\pm$	0.010	\\
4.715	&	0.102	$\pm$	0.065	$\pm$	0.008	&	0.304	$\pm$	0.106	$\pm$	0.034	&	0.062	$\pm$	0.057	$\pm$	0.012	\\
4.725	&	0.153	$\pm$	0.070	$\pm$	0.013	&	0.215	$\pm$	0.096	$\pm$	0.022	&	0.113	$\pm$	0.068	$\pm$	0.021	\\
4.735	&	-0.057	$\pm$	0.025	$\pm$	0.006	&	0.513	$\pm$	0.089	$\pm$	0.051	&	0.094	$\pm$	0.075	$\pm$	0.022	\\
4.745	&	0.024	$\pm$	0.057	$\pm$	0.000	&	0.421	$\pm$	0.118	$\pm$	0.046	&	0.075	$\pm$	0.069	$\pm$	0.015	\\
4.755	&	0.101	$\pm$	0.064	$\pm$	0.008	&	0.139	$\pm$	0.088	$\pm$	0.016	&	0.047	$\pm$	0.049	$\pm$	0.009	\\
4.765	&	0.028	$\pm$	0.030	$\pm$	0.002	&	0.235	$\pm$	0.080	$\pm$	0.024	&	0.052	$\pm$	0.055	$\pm$	0.011	\\
4.775	&	-0.066	$\pm$	0.033	$\pm$	0.007	&	0.332	$\pm$	0.067	$\pm$	0.038	&	0.010	$\pm$	0.039	$\pm$	0.004	\\
4.785	&	-0.171	$\pm$	0.084	$\pm$	0.024	&	0.585	$\pm$	0.134	$\pm$	0.070	&	-0.021	$\pm$	0.046	$\pm$	0.000	\\
4.795	&	0.213	$\pm$	0.072	$\pm$	0.002	&	0.080	$\pm$	0.076	$\pm$	0.005	&	0.158	$\pm$	0.069	$\pm$	0.011	\\
4.805	&	0.023	$\pm$	0.044	$\pm$	0.001	&	0.335	$\pm$	0.095	$\pm$	0.037	&	0.035	$\pm$	0.052	$\pm$	0.008	\\
4.815	&	-0.006	$\pm$	0.024	$\pm$	0.001	&	0.333	$\pm$	0.084	$\pm$	0.035	&	0.024	$\pm$	0.052	$\pm$	0.008	\\
4.825	&	0.053	$\pm$	0.054	$\pm$	0.018	&	0.163	$\pm$	0.077	$\pm$	0.008	&	0.040	$\pm$	0.051	$\pm$	0.022	\\
4.835	&	0.161	$\pm$	0.064	$\pm$	0.015	&	0.106	$\pm$	0.074	$\pm$	0.012	&	0.076	$\pm$	0.049	$\pm$	0.012	\\
4.845	&	0.123	$\pm$	0.065	$\pm$	0.010	&	0.100	$\pm$	0.081	$\pm$	0.012	&	0.134	$\pm$	0.060	$\pm$	0.021	\\
4.855	&	0.064	$\pm$	0.045	$\pm$	0.005	&	0.139	$\pm$	0.067	$\pm$	0.014	&	0.085	$\pm$	0.053	$\pm$	0.015	\\
4.865	&	0.084	$\pm$	0.055	$\pm$	0.001	&	0.105	$\pm$	0.074	$\pm$	0.013	&	0.048	$\pm$	0.054	$\pm$	0.003	\\
4.875	&	0.023	$\pm$	0.037	$\pm$	0.002	&	0.275	$\pm$	0.079	$\pm$	0.028	&	-0.016	$\pm$	0.045	$\pm$	0.003	\\
4.885	&	0.064	$\pm$	0.045	$\pm$	0.004	&	0.077	$\pm$	0.038	$\pm$	0.010	&	-0.011	$\pm$	0.006	$\pm$	0.003	\\
4.895	&	0.082	$\pm$	0.055	$\pm$	0.006	&	0.167	$\pm$	0.076	$\pm$	0.020	&	0.097	$\pm$	0.050	$\pm$	0.016	\\
4.905	&	0.000	$\pm$	0.029	$\pm$	0.001	&	0.329	$\pm$	0.056	$\pm$	0.037	&	-0.039	$\pm$	0.017	$\pm$	0.004	\\
4.915	&	0.068	$\pm$	0.047	$\pm$	0.006	&	0.185	$\pm$	0.073	$\pm$	0.021	&	0.011	$\pm$	0.035	$\pm$	0.003	\\
4.925	&	0.107	$\pm$	0.055	$\pm$	0.019	&	0.122	$\pm$	0.071	$\pm$	0.008	&	0.106	$\pm$	0.052	$\pm$	0.002	\\
4.935	&	-0.018	$\pm$	0.037	$\pm$	0.027	&	0.386	$\pm$	0.079	$\pm$	0.009	&	-0.001	$\pm$	0.032	$\pm$	0.009	\\
4.945	&	0.103	$\pm$	0.057	$\pm$	0.008	&	0.149	$\pm$	0.078	$\pm$	0.017	&	0.057	$\pm$	0.048	$\pm$	0.011	\\
4.955	&	-0.025	$\pm$	0.011	$\pm$	0.044	&	0.409	$\pm$	0.013	$\pm$	0.015	&	0.029	$\pm$	0.013	$\pm$	0.000	\\
4.965	&	0.026	$\pm$	0.033	$\pm$	0.002	&	0.097	$\pm$	0.069	$\pm$	0.008	&	0.147	$\pm$	0.062	$\pm$	0.024	\\
4.975	&	0.010	$\pm$	0.028	$\pm$	0.001	&	0.142	$\pm$	0.066	$\pm$	0.014	&	0.077	$\pm$	0.051	$\pm$	0.014	\\
4.985	&	0.091	$\pm$	0.047	$\pm$	0.004	&	0.125	$\pm$	0.079	$\pm$	0.002	&	0.030	$\pm$	0.036	$\pm$	0.003	\\
4.995	&	0.032	$\pm$	0.046	$\pm$	0.001	&	0.154	$\pm$	0.065	$\pm$	0.019	&	0.002	$\pm$	0.021	$\pm$	0.000	\\
5.005	&	0.047	$\pm$	0.044	$\pm$	0.003	&	0.172	$\pm$	0.049	$\pm$	0.021	&	-0.021	$\pm$	0.015	$\pm$	0.003	\\
5.015	&	-0.017	$\pm$	0.017	$\pm$	0.004	&	0.155	$\pm$	0.044	$\pm$	0.020	&	0.022	$\pm$	0.028	$\pm$	0.004	\\
5.025	&	-0.030	$\pm$	0.039	$\pm$	0.026	&	0.330	$\pm$	0.059	$\pm$	0.020	&	-0.028	$\pm$	0.019	$\pm$	0.007	\\
5.035	&	0.043	$\pm$	0.043	$\pm$	0.004	&	0.156	$\pm$	0.082	$\pm$	0.014	&	0.044	$\pm$	0.056	$\pm$	0.011	\\
5.045	&	0.023	$\pm$	0.042	$\pm$	0.001	&	0.209	$\pm$	0.079	$\pm$	0.023	&	0.036	$\pm$	0.045	$\pm$	0.008	\\
5.055	&	-0.011	$\pm$	0.025	$\pm$	0.006	&	0.251	$\pm$	0.070	$\pm$	0.032	&	-0.017	$\pm$	0.046	$\pm$	0.015	\\
5.065	&	-0.01	$\pm$	0.027	$\pm$	0.003	&	0.227	$\pm$	0.044	$\pm$	0.027	&	-0.027	$\pm$	0.014	$\pm$	0.003	\\
5.075	&	0.039	$\pm$	0.033	$\pm$	0.003	&	0.164	$\pm$	0.067	$\pm$	0.017	&	0.094	$\pm$	0.052	$\pm$	0.016	\\
5.085	&	-0.021	$\pm$	0.016	$\pm$	0.009	&	0.196	$\pm$	0.048	$\pm$	0.015	&	-0.010	$\pm$	0.035	$\pm$	0.003	\\
5.095	&	0.058	$\pm$	0.041	$\pm$	0.004	&	0.199	$\pm$	0.050	$\pm$	0.024	&	-0.024	$\pm$	0.011	$\pm$	0.003	\\
5.105	&	0.056	$\pm$	0.044	$\pm$	0.006	&	0.307	$\pm$	0.048	$\pm$	0.025	&	-0.057	$\pm$	0.017	$\pm$	0.011	\\
5.115	&	0.073	$\pm$	0.039	$\pm$	0.013	&	0.086	$\pm$	0.053	$\pm$	0.013	&	0.104	$\pm$	0.044	$\pm$	0.014	\\
5.125	&	0.087	$\pm$	0.031	$\pm$	0.012	&	0.296	$\pm$	0.053	$\pm$	0.005	&	-0.014	$\pm$	0.021	$\pm$	0.005	\\
5.135	&	0.100	$\pm$	0.044	$\pm$	0.010	&	0.034	$\pm$	0.051	$\pm$	0.002	&	0.036	$\pm$	0.036	$\pm$	0.008	\\
5.145	&	0.058	$\pm$	0.036	$\pm$	0.001	&	0.16	$\pm$	0.048	$\pm$	0.010	&	-0.016	$\pm$	0.016	$\pm$	0.005	\\
5.155	&	0.012	$\pm$	0.008	$\pm$	0.000	&	-0.008	$\pm$	0.003	$\pm$	0.001	&	0.119	$\pm$	0.033	$\pm$	0.018	\\
5.165	&	-0.003	$\pm$	0.002	$\pm$	0.000	&	0.117	$\pm$	0.041	$\pm$	0.011	&	0.020	$\pm$	0.034	$\pm$	0.006	\\
5.175	&	0.000	$\pm$	0.020	$\pm$	0.001	&	0.195	$\pm$	0.060	$\pm$	0.019	&	-0.009	$\pm$	0.037	$\pm$	0.001	\\
5.185	&	0.047	$\pm$	0.023	$\pm$	0.013	&	0.182	$\pm$	0.053	$\pm$	0.002	&	0.027	$\pm$	0.040	$\pm$	0.002	\\
5.195	&	-0.004	$\pm$	0.030	$\pm$	0.018	&	0.154	$\pm$	0.033	$\pm$	0.015	&	-0.044	$\pm$	0.016	$\pm$	0.007	\\
5.205	&	0.038	$\pm$	0.033	$\pm$	0.014	&	0.073	$\pm$	0.054	$\pm$	0.012	&	0.007	$\pm$	0.025	$\pm$	0.001	\\
5.215	&	-0.042	$\pm$	0.005	$\pm$	0.005	&	0.120	$\pm$	0.015	$\pm$	0.013	&	0.020	$\pm$	0.027	$\pm$	0.004	\\
5.225	&	-0.006	$\pm$	0.027	$\pm$	0.012	&	0.175	$\pm$	0.047	$\pm$	0.007	&	0.032	$\pm$	0.025	$\pm$	0.001	\\
5.235	&	0.046	$\pm$	0.037	$\pm$	0.002	&	0.183	$\pm$	0.039	$\pm$	0.018	&	-0.148	$\pm$	0.045	$\pm$	0.027	\\
5.245	&	-0.022	$\pm$	0.030	$\pm$	0.007	&	0.154	$\pm$	0.063	$\pm$	0.005	&	0.007	$\pm$	0.031	$\pm$	0.005	\\
5.255	&	0.069	$\pm$	0.031	$\pm$	0.007	&	0.130	$\pm$	0.036	$\pm$	0.014	&	0.032	$\pm$	0.029	$\pm$	0.000	\\
5.265	&	0.051	$\pm$	0.030	$\pm$	0.005	&	0.008	$\pm$	0.037	$\pm$	0.000	&	0.054	$\pm$	0.033	$\pm$	0.009	\\
5.275	&	0.008	$\pm$	0.043	$\pm$	0.003	&	0.143	$\pm$	0.065	$\pm$	0.017	&	0.004	$\pm$	0.028	$\pm$	0.002	\\
5.285	&	-0.052	$\pm$	0.037	$\pm$	0.009	&	0.244	$\pm$	0.044	$\pm$	0.030	&	-0.028	$\pm$	0.011	$\pm$	0.003	\\
5.295	&	0.021	$\pm$	0.029	$\pm$	0.002	&	0.154	$\pm$	0.055	$\pm$	0.015	&	-0.004	$\pm$	0.033	$\pm$	0.003	\\
5.305	&	0.004	$\pm$	0.026	$\pm$	0.008	&	0.175	$\pm$	0.060	$\pm$	0.006	&	-0.152	$\pm$	0.114	$\pm$	0.017	\\
5.315	&	0.066	$\pm$	0.035	$\pm$	0.006	&	0.061	$\pm$	0.049	$\pm$	0.005	&	0.021	$\pm$	0.035	$\pm$	0.005	\\
5.325	&	0.031	$\pm$	0.049	$\pm$	0.002	&	0.116	$\pm$	0.078	$\pm$	0.013	&	0.010	$\pm$	0.038	$\pm$	0.003	\\
5.335	&	0.104	$\pm$	0.043	$\pm$	0.004	&	0.085	$\pm$	0.039	$\pm$	0.005	&	-0.011	$\pm$	0.013	$\pm$	0.001	\\
5.345	&	0.038	$\pm$	0.019	$\pm$	0.004	&	0.185	$\pm$	0.045	$\pm$	0.024	&	0.022	$\pm$	0.032	$\pm$	0.001	\\
5.355	&	-0.010	$\pm$	0.019	$\pm$	0.003	&	0.062	$\pm$	0.041	$\pm$	0.003	&	-0.016	$\pm$	0.019	$\pm$	0.006	\\
5.365	&	0.039	$\pm$	0.035	$\pm$	0.014	&	0.087	$\pm$	0.058	$\pm$	0.006	&	0.129	$\pm$	0.052	$\pm$	0.009	\\
5.375	&	-0.006	$\pm$	0.021	$\pm$	0.002	&	0.100	$\pm$	0.047	$\pm$	0.010	&	0.087	$\pm$	0.042	$\pm$	0.015	\\
5.385	&	-0.028	$\pm$	0.014	$\pm$	0.006	&	0.246	$\pm$	0.044	$\pm$	0.024	&	-0.041	$\pm$	0.031	$\pm$	0.006	\\
5.395	&	-0.045	$\pm$	0.017	$\pm$	0.004	&	0.133	$\pm$	0.029	$\pm$	0.013	&	0.006	$\pm$	0.030	$\pm$	0.004	\\
5.405	&	0.019	$\pm$	0.027	$\pm$	0.004	&	0.162	$\pm$	0.048	$\pm$	0.004	&	-0.017	$\pm$	0.017	$\pm$	0.003	\\
5.415	&	0.018	$\pm$	0.025	$\pm$	0.002	&	0.071	$\pm$	0.051	$\pm$	0.004	&	0.074	$\pm$	0.042	$\pm$	0.014	\\
5.425	&	0.069	$\pm$	0.037	$\pm$	0.018	&	0.143	$\pm$	0.059	$\pm$	0.005	&	0.045	$\pm$	0.037	$\pm$	0.005	\\
5.435	&	-0.015	$\pm$	0.014	$\pm$	0.003	&	0.145	$\pm$	0.035	$\pm$	0.017	&	-0.014	$\pm$	0.016	$\pm$	0.000	\\
5.445	&	0.014	$\pm$	0.017	$\pm$	0.001	&	0.087	$\pm$	0.042	$\pm$	0.007	&	0.045	$\pm$	0.031	$\pm$	0.006	\\
5.455	&	0.000	$\pm$	0.017	$\pm$	0.001	&	0.169	$\pm$	0.035	$\pm$	0.019	&	-0.020	$\pm$	0.013	$\pm$	0.002	\\
5.465	&	-0.014	$\pm$	0.011	$\pm$	0.002	&	0.057	$\pm$	0.017	$\pm$	0.003	&	-0.067	$\pm$	0.019	$\pm$	0.003	\\
5.475	&	0.071	$\pm$	0.032	$\pm$	0.013	&	0.100	$\pm$	0.042	$\pm$	0.006	&	0.028	$\pm$	0.029	$\pm$	0.010	\\
5.485	&	0.005	$\pm$	0.021	$\pm$	0.007	&	0.178	$\pm$	0.041	$\pm$	0.006	&	-0.02	$\pm$	0.015	$\pm$	0.002	\\
5.495	&	0.028	$\pm$	0.027	$\pm$	0.005	&	0.125	$\pm$	0.049	$\pm$	0.004	&	0.023	$\pm$	0.029	$\pm$	0.001	\\
5.505	&	0.044	$\pm$	0.029	$\pm$	0.003	&	0.064	$\pm$	0.047	$\pm$	0.003	&	0.019	$\pm$	0.020	$\pm$	0.001	\\
5.515	&	0.017	$\pm$	0.024	$\pm$	0.010	&	0.114	$\pm$	0.042	$\pm$	0.009	&	-0.006	$\pm$	0.024	$\pm$	0.012	\\
5.525	&	0.015	$\pm$	0.024	$\pm$	0.000	&	0.171	$\pm$	0.033	$\pm$	0.021	&	-0.020	$\pm$	0.009	$\pm$	0.002	\\
5.535	&	0.011	$\pm$	0.008	$\pm$	0.000	&	0.123	$\pm$	0.008	$\pm$	0.013	&	0.005	$\pm$	0.008	$\pm$	0.003	\\
5.545	&	-0.002	$\pm$	0.005	$\pm$	0.001	&	0.073	$\pm$	0.029	$\pm$	0.005	&	0.003	$\pm$	0.014	$\pm$	0.001	\\
5.555	&	0.023	$\pm$	0.015	$\pm$	0.007	&	0.072	$\pm$	0.022	$\pm$	0.011	&	0.003	$\pm$	0.007	$\pm$	0.004	\\
5.565	&	0.003	$\pm$	0.015	$\pm$	0.000	&	0.053	$\pm$	0.031	$\pm$	0.005	&	0.007	$\pm$	0.022	$\pm$	0.003	\\
5.575	&	0.003	$\pm$	0.010	$\pm$	0.000	&	0.078	$\pm$	0.036	$\pm$	0.008	&	0.022	$\pm$	0.026	$\pm$	0.004	\\
5.585	&	0.006	$\pm$	0.029	$\pm$	0.018	&	0.150	$\pm$	0.119	$\pm$	0.010	&	0.005	$\pm$	0.014	$\pm$	0.002	\\
5.595	&	0.047	$\pm$	0.028	$\pm$	0.004	&	0.095	$\pm$	0.041	$\pm$	0.001	&	0.020	$\pm$	0.023	$\pm$	0.000	\\
5.605	&	0.015	$\pm$	0.018	$\pm$	0.000	&	0.043	$\pm$	0.015	$\pm$	0.006	&	-0.014	$\pm$	0.006	$\pm$	0.002	\\
5.615	&	0.031	$\pm$	0.018	$\pm$	0.002	&	0.042	$\pm$	0.035	$\pm$	0.004	&	0.032	$\pm$	0.014	$\pm$	0.003	\\
5.625	&	-0.010	$\pm$	0.015	$\pm$	0.002	&	0.027	$\pm$	0.021	$\pm$	0.013	&	0.06	$\pm$	0.027	$\pm$	0.008	\\
5.635	&	0.053	$\pm$	0.040	$\pm$	0.004	&	0.056	$\pm$	0.062	$\pm$	0.006	&	0.017	$\pm$	0.032	$\pm$	0.004	\\
5.645	&	0.011	$\pm$	0.014	$\pm$	0.002	&	0.073	$\pm$	0.028	$\pm$	0.000	&	0.006	$\pm$	0.011	$\pm$	0.000	\\
5.655	&	0.033	$\pm$	0.024	$\pm$	0.000	&	0.039	$\pm$	0.036	$\pm$	0.013	&	0.014	$\pm$	0.015	$\pm$	0.001	\\
5.665	&	-0.012	$\pm$	0.027	$\pm$	0.008	&	0.091	$\pm$	0.024	$\pm$	0.029	&	0.028	$\pm$	0.018	$\pm$	0.006	\\
5.675	&	-0.047	$\pm$	0.012	$\pm$	0.006	&	0.111	$\pm$	0.015	$\pm$	0.013	&	-0.012	$\pm$	0.006	$\pm$	0.001	\\
5.685	&	0.019	$\pm$	0.019	$\pm$	0.003	&	0.083	$\pm$	0.031	$\pm$	0.004	&	-0.004	$\pm$	0.014	$\pm$	0.001	\\
5.695	&	0.030	$\pm$	0.019	$\pm$	0.003	&	0.045	$\pm$	0.026	$\pm$	0.005	&	0.003	$\pm$	0.014	$\pm$	0.001	\\
5.705	&	0.016	$\pm$	0.016	$\pm$	0.001	&	0.085	$\pm$	0.032	$\pm$	0.009	&	0.018	$\pm$	0.019	$\pm$	0.004	\\
5.715	&	0.037	$\pm$	0.026	$\pm$	0.001	&	0.052	$\pm$	0.039	$\pm$	0.000	&	0.012	$\pm$	0.025	$\pm$	0.002	\\
5.725	&	0.007	$\pm$	0.005	$\pm$	0.004	&	0.079	$\pm$	0.007	$\pm$	0.004	&	0.012	$\pm$	0.006	$\pm$	0.000	\\
5.735	&	0.028	$\pm$	0.044	$\pm$	0.006	&	0.058	$\pm$	0.06	$\pm$	0.013	&	0.033	$\pm$	0.049	$\pm$	0.005	\\
5.745	&	0.028	$\pm$	0.027	$\pm$	0.002	&	0.024	$\pm$	0.031	$\pm$	0.004	&	0.015	$\pm$	0.016	$\pm$	0.003	\\
5.755	&	-0.005	$\pm$	0.019	$\pm$	0.002	&	0.098	$\pm$	0.033	$\pm$	0.012	&	0.007	$\pm$	0.017	$\pm$	0.002	\\
5.765	&	-0.006	$\pm$	0.012	$\pm$	0.004	&	-0.001	$\pm$	0.01	$\pm$	0.005	&	0.062	$\pm$	0.023	$\pm$	0.008	\\
5.775	&	0.017	$\pm$	0.014	$\pm$	0.001	&	0.057	$\pm$	0.012	$\pm$	0.006	&	-0.020	$\pm$	0.005	$\pm$	0.002	\\
5.785	&	-0.015	$\pm$	0.005	$\pm$	0.002	&	0.069	$\pm$	0.024	$\pm$	0.008	&	0.021	$\pm$	0.020	$\pm$	0.004	\\
5.795	&	0.048	$\pm$	0.025	$\pm$	0.000	&	0.041	$\pm$	0.026	$\pm$	0.000	&	0.001	$\pm$	0.011	$\pm$	0.002	\\
5.805	&	-0.005	$\pm$	0.012	$\pm$	0.002	&	0.101	$\pm$	0.032	$\pm$	0.005	&	-0.015	$\pm$	0.013	$\pm$	0.003	\\
5.815	&	-0.009	$\pm$	0.013	$\pm$	0.008	&	0.110	$\pm$	0.03	$\pm$	0.007	&	-0.019	$\pm$	0.021	$\pm$	0.005	\\
5.825	&	0.052	$\pm$	0.026	$\pm$	0.004	&	0.019	$\pm$	0.024	$\pm$	0.003	&	0.010	$\pm$	0.012	$\pm$	0.002	\\
5.835	&	0.039	$\pm$	0.015	$\pm$	0.007	&	0.037	$\pm$	0.019	$\pm$	0.005	&	0.060	$\pm$	0.020	$\pm$	0.005	\\
5.845	&	-0.016	$\pm$	0.013	$\pm$	0.005	&	0.061	$\pm$	0.027	$\pm$	0.001	&	-0.013	$\pm$	0.020	$\pm$	0.004	\\
5.855	&	-0.012	$\pm$	0.004	$\pm$	0.001	&	0.115	$\pm$	0.03	$\pm$	0.013	&	0.002	$\pm$	0.023	$\pm$	0.002	\\
5.865	&	-0.04	$\pm$	0.009	$\pm$	0.004	&	0.107	$\pm$	0.014	$\pm$	0.012	&	-0.051	$\pm$	0.020	$\pm$	0.006	\\
5.875	&	-0.011	$\pm$	0.009	$\pm$	0.003	&	0.106	$\pm$	0.012	$\pm$	0.014	&	-0.085	$\pm$	0.011	$\pm$	0.011	\\
5.885	&	0.042	$\pm$	0.022	$\pm$	0.004	&	0.015	$\pm$	0.026	$\pm$	0.001	&	0.038	$\pm$	0.021	$\pm$	0.006	\\
5.895	&	-0.008	$\pm$	0.010	$\pm$	0.002	&	0.054	$\pm$	0.024	$\pm$	0.006	&	0.023	$\pm$	0.018	$\pm$	0.004	\\
5.905	&	0.002	$\pm$	0.011	$\pm$	0.002	&	0.050	$\pm$	0.028	$\pm$	0.009	&	0.002	$\pm$	0.009	$\pm$	0.001	\\
5.915	&	0.007	$\pm$	0.012	$\pm$	0.004	&	0.069	$\pm$	0.023	$\pm$	0.004	&	-0.007	$\pm$	0.008	$\pm$	0.000	\\
5.925	&	0.010	$\pm$	0.024	$\pm$	0.002	&	0.030	$\pm$	0.033	$\pm$	0.002	&	0.009	$\pm$	0.013	$\pm$	0.000	\\
5.935	&	0.007	$\pm$	0.014	$\pm$	0.000	&	0.049	$\pm$	0.011	$\pm$	0.006	&	-0.006	$\pm$	0.001	$\pm$	0.001	\\
5.945	&	-0.014	$\pm$	0.022	$\pm$	0.003	&	0.056	$\pm$	0.014	$\pm$	0.007	&	-0.007	$\pm$	0.002	$\pm$	0.001	\\
5.955	&	0.034	$\pm$	0.011	$\pm$	0.003	&	0.069	$\pm$	0.021	$\pm$	0.009	&	0.032	$\pm$	0.022	$\pm$	0.001	\\
5.965	&	-0.011	$\pm$	0.016	$\pm$	0.004	&	0.158	$\pm$	0.026	$\pm$	0.014	&	-0.046	$\pm$	0.011	$\pm$	0.008	\\
5.975	&	0.033	$\pm$	0.011	$\pm$	0.002	&	0.079	$\pm$	0.019	$\pm$	0.008	&	-0.006	$\pm$	0.003	$\pm$	0.002	\\
5.985	&	0.019	$\pm$	0.013	$\pm$	0.005	&	0.041	$\pm$	0.022	$\pm$	0.004	&	0.029	$\pm$	0.022	$\pm$	0.003	\\
5.995	&	0.052	$\pm$	0.012	$\pm$	0.001	&	0.009	$\pm$	0.019	$\pm$	0.002	&	0.013	$\pm$	0.009	$\pm$	0.000	\\

\end{tabularx}
\end{widetext}

\end{document}